\documentclass[12pt]{article}

\usepackage{amsfonts,amssymb}

\newcommand{\smin}{\,\raisebox{0.06em}{${\scriptstyle \in}$}\,}
\newcommand{\nsmin}{\,\raisebox{0.06em}{${\scriptstyle \notin}$}\,}
\newcommand{\ssmin}{\,\raisebox{0.06em}{${\scriptscriptstyle \in}$}\,}

\newcommand{\smcup}{\,\raisebox{0.06em}{${\scriptstyle \cup}$}\,}

\newcommand{\oktimes}{\,\raisebox{-0.2em}{$\stackrel{\otimes}{,}$}\,}

\newtheorem{propos}{Proposition}

\begin{document}
\title{From Dynamical to Numerical R-Matrices: \\
       A Case Study for the Calogero Models}
\author{Michael Forger and Axel Winterhalder
        \thanks{forger@ime.usp.br and winter@ime.usp.br}}
\date{Departamento de Matem\'atica Aplicada, \\
      Instituto de Matem\'atica e Estat\'{\i}stica, \\
      Universidade de S\~ao Paulo, \\
      Caixa Postal 66281, \\[2mm]
      BR--05311-970~ S\~ao Paulo, S.P., Brazil}
\maketitle
\renewcommand{\thefootnote}{\fnsymbol{footnote}}
\footnotetext[1]{Work supported by CNPq (Conselho Nacional de Desenvolvimento
                 Cient\'{\i}fico e Tecno\-l\'o\-gico), Brazil, by FAPESP
                 (Funda\c{c}\~ao de Amparo \`a Pesquisa do Estado de S\~ao
                 Paulo), Brazil.}
\renewcommand{\thefootnote}{\arabic{footnote}}
\setcounter{footnote}{0}
\thispagestyle{empty}
\begin{abstract}        
\noindent
Within the class of integrable Calogero models associated with (semi-)\,%
simple Lie algebras and with symmetric pairs of Lie algebras identified in a
previous paper, we analyze whether and to what extent it is possible to find
a gauge transformation that takes the traditional Lax pair with its dynamical
$R$-matrix to a new Lax pair with a numerical $R$-matrix.
\end{abstract}
\vspace{-5mm}
\begin{flushright}
 \parbox{12em}
 {\begin{center}
   Universidade de S\~ao Paulo \\
   RT-MAP-0202 \\
   December 2002
  \end{center}}
\end{flushright}

\newpage

\section{Introduction}

In a recent paper \cite{FW1}, we have performed a systematic analysis
of the Calogero-Moser-Sutherland models, or Calogero models, for short,
which constitute an important class of completely integrable
Hamiltonian systems. Our work follows the traditional Lie algebraic
approach outlined long ago by Olshanetsky and Perelomov~\cite{OP1,%
OP2,Pe} which is based on the use of \mbox{(semi-)}\,simple Lie
algebras and, more generally, of symmetric pairs, extending it so
as to encompass the construction not only of a Lax representation
for the equations of motion but also that of a dynamical $R$-matrix.
The existence of these structures was found to depend on the
possibility of solving a simple set of algebraic constraints
for a certain function $F$ or $K$ that assigns to each root $\alpha$
a generator $F_\alpha$ or $K_\alpha$ in the pertinent Cartan subalgebra
$\mathfrak{h}$\,:\footnote{In this paper, we adopt a slightly modified
notation: for reasons to become clear towards the end of the paper,
we shall in the case of symmetric pairs denote the generators
$F_\alpha$ of Ref.~\cite{FW1} by $K_\alpha$.} these state that
for any two roots $\alpha$ and $\beta$
\begin{equation} \label{eq:LAAC1}
 \mbox{\sl g}_\alpha \, \alpha(F_\beta) \, - \,
 \mbox{\sl g}_\beta  \, \beta(F_\alpha)~
 =~\mathit{\Gamma}_{\alpha,\beta}
\end{equation}
in the case of (semi-)\,simple Lie algebras $\mathfrak{g}$ and
\begin{equation} \label{eq:SPAC1}
 \mbox{\sl g}_\alpha \, \alpha(K_\beta) \, - \,
 \mbox{\sl g}_\beta  \, \beta(K_\alpha)~
 =~\mathit{\Gamma}_{\alpha,\beta}^\theta
\end{equation}
in the case of symmetric pairs $(\mathfrak{g},\theta)$, where the
coefficients $\mathit{\Gamma}_{\alpha,\beta}$ and $\mathit{\Gamma}_%
{\alpha,\beta}^\theta$ are defined in terms of the structure constants
$N_{\alpha,\beta}$ of $\mathfrak{g}$ and the coupling constants
$\mbox{\sl g}_\alpha$ of the model by
\begin{equation} \label{eq:LAAC2}
 \mathit{\Gamma}_{\alpha,\beta}~
 =~\mbox{\sl g}_{\alpha + \beta} \, N_{\alpha,\beta}
\end{equation}
and by
\begin{equation} \label{eq:SPAC2}
 \mathit{\Gamma}_{\alpha,\beta}^\theta~
 =~{\textstyle \frac{1}{4}} \,
   \Bigl( \mbox{\sl g}_{\alpha+\beta} \, N_{\alpha,\beta} \, + \,
          \mbox{\sl g}_{\theta\alpha+\beta} \,
          N_{\theta\alpha,\beta} \, + \,
          \mbox{\sl g}_{\alpha+\theta\beta} \,
          N_{\alpha,\theta\beta} \, + \,
          \mbox{\sl g}_{\theta\alpha+\theta\beta} \,
          N_{\theta\alpha,\theta\beta} \Bigr)
\end{equation}
respectively.\footnote{In the case of symmetric pairs, it is also assumed
that the root generators $E_\alpha$ in $\mathfrak{g}$ can be and have been
chosen so that $\, \theta E_\alpha = E_{\theta\alpha}$ for all $\, \alpha
\ssmin\, \Delta$, implying that $\, N_{\theta\alpha,\theta\beta} =
N_{\alpha,\beta} \,$ for all $\, \alpha,\beta \ssmin\, \Delta$.
As~explained in the erratum to Ref.~\cite{FW1}, this is not always
possible but is a necessary condition for our proof of integrability,
and it can always be arranged to hold for all roots $\alpha$ if it can
be made to hold for all real roots $\alpha$, that is, all roots $\alpha$
in $\Delta$ satisfying $\, \theta\alpha = -\alpha$.} In the first case,
it was found that a solution of these constraints exists only for
the Lie algebras $\mathfrak{sl}(n,\mathbb{C})$ of the $A$-series,
whereas in the second case, explicit solutions were found for the
complex Grassmannians \mbox{$\, SU(p+q) /$} \mbox{$S(U(p) \times
U(q)) \,$} of the $A\,III\/$-\,series when $\, |p-q| \leqslant 1$.

Following a somewhat different direction, several authors~\cite{HW,FP}
have recently shown by explicit matrix computations that the Calogero
models based on $\mathfrak{sl}(n,\mathbb{C})$, degenerate as well as
elliptic, admit a gauge transformation taking the dynamical $R$-matrix
into a numerical one: this is achieved by explicitly constructing a
group-valued function on the configuration space which is used to
conjugate the standard Lax pair and dynamical $R$-matrix of the
model into a new Lax pair and a numerical $R$-matrix.

In the present paper, we systematize the method of Feh\'er and Pusztai~%
\cite{FP}, adapting it to the formalism developed in our previous work~%
\cite{FW1}: this allows us to extend it from the degenerate to the
elliptic models as well as from the case of Lie algebras to that of
symmetric pairs. In all cases, we find that the existence of a gauge
transformation with the desired property can be reduced to a set of
purely algebraic constraints which are similar to but not identical
with the integrability constraints~(\ref{eq:LAAC1}) and~(\ref{eq:SPAC1})
found in Ref.~\cite{FW1}. In the case of Lie algebras, it turns out that
these various constraints all have one and the same solution, thus
confirming the previous results of other authors~\cite{HW,FP} that
the well-known dynamical $R$-matrices of the Calogero models based
on $\mathfrak{sl}(n,\mathbb{C})$ can be gauge transformed to numerical
$R$-matrices. In the case of symmetric pairs, however, we find extra
constraints on the root system, over and above those that guarantee
integrability. In particular, for the $A\,III\;\!$-\,series of complex
complex Grassmannians $\, SU(p+q) / S(U(p) \times U(q)) \,$ where
integrability has in Ref.~\cite{FW1} been shown to hold when
$\, |p-q| \leqslant 1$, these constraints exclude the case
$\, |p-q| = 1 \,$ but allow for a solution in the case $\, p=q$.
This means that the dynamical $R$-matrices for the Calogero models
associated with the classical root systems can be gauge transformed
to numerical $R$-matrices for the $C_n$ and $D_n$ systems but not for
the $B_n$ and $BC_n$ systems; remarkably, the latter are just the
ones containing an explicit dependence on the coupling constants.

The paper is organized as follows. In Sect.~2, we give a brief summary
of the method of gauge transforming Lax pairs and dynamical $R$-matrices
in integrable systems such as the Calogero models; moreover, we collect
a number of identities to be used repeatedly later on. In Sect.~3, we
present our calculations for the case of Lie algebras and in Sect.~4
those for symmetric pairs. Finally, in Sect.~5 we draw our conclusions
and comment on perspectives for further work.

\section{Gauge Transformations}

Consider an integrable model with a finite-dimensional phase space which
we assume to be the cotangent bundle $T^* Q$ of a configuration space $Q$.
Integrability is encoded into the existence of a Lax representation for
the equations of motion,
\begin{equation} \label{eq:LR}
 \dot{L}~=~[L\,,M]~,
\end{equation}
together with that of an $R$-matrix whose role is to control the Poisson
brackets between the components of the Lax matrix $L$, according to the
formula~\cite{BV}
\begin{equation} \label{eq:PB}
 \{ L_1 \>\! , L_2 \}~=~[R_{12}\,,L_1] - [R_{21}\,,L_2]~.
\end{equation}
Here, $L$ and $M$ are maps from $T^* Q$ into a given Lie algebra
$\mathfrak{g}$ whereas $R$ will in general be a map from $T^* Q$ into
the second tensor power $\, U(\mathfrak{g}) \otimes U(\mathfrak{g}) \,$
of the universal enveloping algebra $U(\mathfrak{g})$ of $\mathfrak{g}$;
as usual, $\, L_1 = L \otimes 1$, $L_2 = 1 \otimes L \,$ etc.. The choice
of $\mathfrak{g}$ is far from obvious; it reflects the hidden symmetries
that are present in the model. Moreover, even if one fixes $\mathfrak{g}$
and a connected Lie group $G$ that has $\mathfrak{g}$ as its Lie algebra,
$L$, $M$ and $R$ are not uniquely determined. In particular, we are free
to perform a \emph{gauge transformation} by an arbitrary function $g$ on
$T^* Q$ with values in $G$, as follows:
\begin{equation} \label{eq:GTLAXM}
 L^\prime~=~g \>\! L \>\! g^{-1}~,
\end{equation}
\begin{equation} \label{eq:GTRMAT}
 R_{12}^{\,\prime}~
 =~g_1^{} \, g_2^{} \biggl( R_{12}^{} \, + \,
                            g_1^{-1} \{ g_1^{} \>\!, L_2^{} \} \, + \,
                            {\textstyle \frac{1}{2}}
                            \left[ \, g_1^{-1} g_2^{-1} \,
                                      \{ g_1^{} \>\!, g_2^{} \} \,,
                                      L_2^{} \, \right] \biggr) \,
   g_1^{-1} g_2^{-1}~.
\end{equation}
%\begin{equation} \label{eq:GTLAXM2}
% L^\prime(u)~=~g(u) \, L(u) \, g^{-1}(u)~,
%\end{equation}
%\begin{equation} \label{eq:GTRMAT2}
% \begin{array}{rcl}
%  R^\prime(u,v) \!\!&=&\!\! (g(u) \otimes g(v)) \\[2mm]
%  & & \biggl( R(u,v) \, + \, (g^{-1}(u) \otimes 1) \{g(u) \oktimes L(v)\} \\
%  & & \hspace{1cm} + \,
%             \left[ (g^{-1}(u) \otimes g^{-1}(v)) \{g(u) \oktimes g(v)\} \, ,
%                    1 \otimes L(v) \right] \biggr) \\[3mm]
%  & & \hspace{6.4cm} (g^{-1}(u) \otimes g^{-1}(v))~.
% \end{array}
%\end{equation}
The second transformation law is dictated by the requirement that the
fundamental Poisson bracket relation (\ref{eq:PB}) should be preserved
under this transformation, which is easy to check. Note that in general,
$L$, $M$ and $g$ may depend on a spectral parameter~$u$, in which case
$R$ will depend on two spectral parameters $u$ and $v$.

In the case of the Calogero models of interest here, $\mathfrak{g}$ is a
simple complex Lie algebra, with Cartan subalgebra $\mathfrak{h}$ and
corresponding root system $\Delta$ fixed once and for all, $Q$ is an
open subset in a real subspace of $\mathfrak{h}$ in which we fix a
basis $\, \{ H_1,\ldots,H_r \}$, $L$ is of the form
\begin{equation} \label{eq:LAXM1}
 L(q,p\,;u)~=~\sum_{j=1}^r \, p_j \>\! H_j \, + \, 
              \sum_{\alpha \ssmin \Delta} L_\alpha(q,u) \, E_\alpha~,
\end{equation}
with functions $L_\alpha$ whose explicit form will be needed only later,
and $R$ is independent of the momentum variables. (For details, see
Ref.~\cite{FW1}.) Our aim in what follows will be to determine $g$
in such a way that $R^{\,\prime}$ becomes constant (as a function
on phase space). To this end, we shall assume that $g$ is also
independent of the momentum variables and introduce the
``gauge potentials''
\begin{equation} \label{eq:GPOT1}
 A_j(u)~=~g^{-1}(u) \, \partial_j \>\! g(u)~.
\end{equation}
Reverting to ordinary tensor notation, we get
\[
 (g^{-1}(u) \otimes 1) \, \{ g(u) \oktimes L(v) \}~
 = \; - \, \sum_{j=1}^r \, g^{-1}(u) \, \{p_j\,,g(u)\} \otimes H_j~
 = \; - \, \sum_{j=1}^r \, A_j(u) \otimes H_j~,
\]
so eqn~(\ref{eq:GTRMAT}) simplifies to
\[
 R^{\,\prime}(u,v)~
 =~(g(u) \otimes g(v))
   \biggl( R(u,v) \, - \, \sum_{j=1}^r \, A_j(u) \otimes H_j \biggr)
   (g^{-1}(u) \otimes g^{-1}(v))~.
\]
This implies
\begin{eqnarray*}
\lefteqn{(g^{-1}(u) \otimes g^{-1}(v)) \; \partial_k R^{\,\prime}(u,v) \;
         (g(u) \otimes g(v))}                              \hspace{1cm} \\[2mm]
 &=&\!\! \partial_k \Bigl( R(u,v) \, - \,
         \sum_{j=1}^r \, A_j(u) \otimes H_j \Bigr)                           \\
 & &\!\! + \,
         \Bigl[ \, (g^{-1}(u) \otimes g^{-1}(v)) \;
                   \partial_k \Bigl( g(u) \otimes g(v) \Bigr) \, , \,
                   R(u,v) \, - \, \sum_{j=1}^r \, A_j(u) \otimes H_j \,
         \Bigr]~,
\end{eqnarray*}
so the condition that the partial derivatives $\partial_k R^{\,\prime}(u,v)$
of $R^{\,\prime}(u,v)$ all vanish amounts to requiring
\begin{eqnarray*}
\lefteqn{\partial_k \Bigl( R(u,v) \, - \,
         \sum_{j=1}^r \, A_j(u) \otimes H_j \Bigr)} \hspace{1cm} \\
 &&- \, \Bigl[ \, R(u,v) \, - \, \sum_{j=1}^r \, A_j(u) \otimes H_j \; , \,
                  A_k(u) \otimes 1 \, + \, 1 \otimes A_k(v) \, \Bigr]~=~0~.
\end{eqnarray*}
Using the integrability condition
\begin{equation} \label{eq:GPOT2}
 \partial_k A_l(u) - \partial_l A_k(u) + [A_k(u),A_l(u)]~=~0
\end{equation}
that follows from eqn~(\ref{eq:GPOT1}), this can be rewritten in the form
\begin{equation} \label{eq:NRMAT}
 \begin{array}{l}
  {\displaystyle {\partial_k R(u,v) \, - \,
                  \sum_{j=1}^r \, \partial_j A_k(u) \otimes H_j}} \\
  {\displaystyle {~- \, \Bigl[ \, R(u,v) \, , \,
                  A_k(u) \otimes 1 \, + \, 1 \otimes A_k(v) \, \Bigr] \, + \,
                  \sum_{j=1}^r \, A_j(u) \otimes [H_j \, , A_k(v)]~=~0~.}}
 \end{array}
\end{equation}
In order to compute the content of eqns~(\ref{eq:GPOT2}) and~(\ref{eq:NRMAT}),
we shall in what follows expand the gauge potential according to
\begin{equation} \label{eq:GPOT3}
 A_j(u)~=~A_j^{\mathfrak{h}}(u) \, + \,
          \sum_{\alpha \ssmin \Delta} A_j^\alpha(u) \, E_\alpha~,
\end{equation}
which allows us, in particular, to decompose eqn~(\ref{eq:GPOT2}) into its
Cartan part
\begin{equation} \label{eq:GPOT4}
 \partial_k^{\vphantom{\mathfrak{h}}} A_l^{\mathfrak{h}}(u) \, - \,
 \partial_l^{\vphantom{\mathfrak{h}}} A_k^{\mathfrak{h}}(u) \, + \,
 \sum_{\alpha \ssmin \Delta} A_k^\alpha(u) A_l^{-\alpha}(u) \, H_\alpha^{}~=~0
\end{equation}
and its root part
\begin{equation} \label{eq:GPOT5}
 \begin{array}{rcl}
  \partial_k^{\vphantom{\alpha}} A_l^\alpha(u) \, - \,
  \partial_l^{\vphantom{\alpha}} A_k^\alpha(u) \!\! &+& \!\!
  \alpha(A_k^{\mathfrak{h}}(u))
  A_l^{\vphantom{\mathfrak{h}}\alpha}(u) \, - \,
  \alpha(A_l^{\mathfrak{h}}(u))
  A_k^{\vphantom{\mathfrak{h}}\alpha}(u) \\[4mm]
  &+& \!\!\! {\displaystyle
              \sum_{\beta,\gamma \ssmin \Delta \atop \beta + \gamma = \alpha}
              N_{\beta,\gamma}^{} A_k^\beta(u) A_l^\gamma(u)~=~0~.}
 \end{array}
\end{equation}
In what follows, we shall analyze under what conditions this system
of equations admits solutions when we insert the explicit expressions
for~$R$ given in Ref.~\cite{FW1} and evaluate the commutators in eqn~%
(\ref{eq:NRMAT}) using the usual abbreviation $\, \alpha_j = \alpha
(H_j) \,$ and the relations
\begin{equation} \label{eq:COMR1}
 \begin{array}{c}
  [ \, H_j \otimes H_j \,,\, E_\alpha \otimes 1 \, ]~
  =~\alpha_j \; E_\alpha \otimes H_j~, \\[2mm]
  [ \, H_j \otimes H_j \,,\, 1 \otimes E_\alpha \, ]~
  =~\alpha_j \; H_j \otimes E_\alpha~,
 \end{array}
\end{equation}
(no summation over $j$),
\begin{equation} \label{eq:COMR2}
 \begin{array}{c}
  [ \, F_\gamma \otimes E_\gamma \,,\, H_j \otimes 1 \, ]~=~0~, \\[2mm]
  [ \, F_\gamma \otimes E_\gamma \,,\, 1 \otimes H_j \, ]~
  =~- \, \gamma_j \; F_\gamma \otimes E_\gamma~, \\[4mm]
  [ \, F_\gamma \otimes E_\gamma \,,\, E_\delta \otimes 1 \, ]~
  =~\delta(F_\gamma) \; E_\delta \otimes E_\gamma~, \\[4mm]
  [ \, F_\gamma \otimes E_\gamma \,,\, 1 \otimes E_\gamma \, ]~=~0~, \\[2mm]
  [ \, F_\gamma \otimes E_\gamma \,,\, 1 \otimes E_{-\gamma} \, ]~
  =~F_\gamma \otimes H_\gamma~, \\[2mm]
  [ \, F_\gamma \otimes E_\gamma \,,\, 1 \otimes E_\delta \, ]~
  =~N_{\gamma,\delta} \, F_\gamma \otimes E_{\gamma+\delta} \quad
  \mbox{if $\, \gamma \pm \delta \neq 0$}~,
 \end{array}
\end{equation}
(valid for any set of generators $F_\gamma$ belonging to the Cartan subalgebra
$\mathfrak{h}$) and
\begin{equation} \label{eq:COMR3}
 \begin{array}{c}
  [ \, E_\gamma \otimes E_{-\gamma} \,,\, H_j \otimes 1 \, ]~
  =~- \, \gamma_j \; E_\gamma \otimes E_{-\gamma}~, \\[2mm]
  [ \, E_\gamma \otimes E_{-\gamma} \,,\, 1 \otimes H_j \, ]~
  =~\gamma_j \; E_\gamma \otimes E_{-\gamma}~,
 \end{array}
\end{equation}
\begin{equation} \label{eq:COMR4}
 \begin{array}{c}
  [ \, E_\gamma \otimes E_{-\gamma} \,,\, E_\gamma \otimes 1 \, ]~=~0~, \\[2mm]
  [ \, E_\gamma \otimes E_{-\gamma} \,,\, E_{-\gamma} \otimes 1 \, ]~
  =~H_\gamma \otimes E_{-\gamma}~, \\[2mm]
  [ \, E_\gamma \otimes E_{-\gamma} \,,\, E_\delta \otimes 1 \, ]~
  =~N_{\gamma,\delta} \, E_{\gamma+\delta} \otimes E_{-\gamma} \quad
  \mbox{if $\, \gamma \pm \delta \neq 0$}~,
 \end{array}
\end{equation}
\vspace*{3mm}
\begin{equation} \label{eq:COMR5}
 \begin{array}{c}
  [ \, E_\gamma \otimes E_{-\gamma} \,,\, 1 \otimes E_\gamma \, ]~
  =~- \, E_\gamma \otimes H_\gamma~, \\[2mm]
  [ \, E_\gamma \otimes E_{-\gamma} \,,\, 1 \otimes E_{-\gamma} \, ]~=~0~,
  \\[2mm]
  [ \, E_\gamma \otimes E_{-\gamma} \,,\, 1 \otimes E_\delta \, ]~
  =~N_{-\gamma,\delta} \, E_\gamma \otimes E_{-\gamma+\delta} \quad
  \mbox{if $\, \gamma \pm \delta \neq 0$}~,
 \end{array}
\vspace*{3mm}
\end{equation}
The second step would be to determine $g$ itself and, from there, find
$L^\prime$ and $R^{\,\prime}$: this question will be addressed elsewhere
in order not to overload our presentation here.

Concluding this section, let us for later use collect the functional
identities satisfied by the coefficient functions $L_\alpha$ that
appear in eqn~(\ref{eq:LAXM1}) above. For the degenerate models,
$\, L_\alpha(q,u) = \mathrm{i} \, \mbox{\sl g}_\alpha \, w(\alpha(q)) \,$
where $w$ is an odd function of its argument,
\begin{equation} \label{eq:ANSYMW}
 w(-t)~= \; - \, w(t)~,
\end{equation}
that satisfies the differential equation
\begin{equation} \label{eq:DIFEQW}
 \left( \frac{w^\prime}{w} \right)^{\!\prime}~=~w^2~,
\end{equation}
as well as the functional equation
\begin{equation} \label{eq:FUNEQW}
 \Bigl( \, \frac{w^\prime(s)}{w(s)} \, + \,
           \frac{w^\prime(t)}{w(t)} \, \Bigr) \, w(s+t) \,
 + \, w(s) \, w(t)~=~0~,
\end{equation}
already employed in Ref.~\cite{FW1}. For the elliptic models, $L_\alpha(q,u)
= \mathrm{i} \, \mbox{\sl g}_\alpha \, \Phi(\alpha(q),u) \,$ where $\Phi$ and
the closely related Weierstrass zeta function satisfy the symmetry properties
\begin{equation} \label{eq:EFUNEQ1}
 \Phi(-z_1,-z_2)~=~- \Phi(z_1,z_2)~~~,~~~
 \zeta(-z)~=~- \, \zeta(z)
\end{equation}
and the functional equations
\begin{eqnarray}
 &\Phi(s,u) \, \Phi(-s,u)~=~\zeta^\prime(s) - \zeta^\prime(u)~,&
 \label{eq:EFUNEQ2} \\[1mm]
 &\Phi(s,u) \, \Phi^\prime(-s,u) \, - \, \Phi^\prime(s,u) \, \Phi(-s,u)~
 = \; - \, \zeta^{\prime\prime}(s)~,&
 \label{eq:EFUNEQ3} \\[1mm]
 &\Phi(s,u) \, \Phi^\prime(t,u) \, - \, \Phi^\prime(s,u) \, \Phi(t,u)~
 = \; - \left( \zeta^\prime(s) - \zeta^\prime(t) \right) \Phi(s+t,u)~,&
 \label{eq:EFUNEQ4} \\[1mm]
 &\Phi(-s,v-u) \, \Phi(s+t,v) \, + \, \Phi(-t,u-v) \, \Phi(s+t,u)~
 = \; - \, \Phi(s,u) \, \Phi(t,v)~,&
 \label{eq:EFUNEQ5} \\[1mm]
 &\Phi(-s,u-v) \, \Phi(s,u) \, +
  \left( \zeta(v-u) + \zeta(u) \right) \Phi(s,v)~=~\Phi^\prime(s,v)~,&
 \label{eq:EFUNEQ6}
\end{eqnarray}
already employed in Ref.\ \cite{FW1}, as well as the additional functional
equations
\begin{eqnarray}
 &\Phi^\prime(s,u)~= \, \left( \zeta(s+u) - \zeta(s) \right) \Phi(s,u)~,&
 \label{eq:EFUNEQ7} \\[1mm]
 &\Phi(s,u) \, \Phi(t,u)~= \, \left( \zeta(s) + \zeta(t) + \zeta(u) -
                                     \zeta(s+t+u) \right) \Phi(s+t,u)~,
 \label{eq:EFUNEQ8}
\end{eqnarray}
where $\Phi^\prime$ denotes the derivative of $\Phi$ with respect to the
first argument; all of these can be derived from the representation of
$\Phi$ and $\zeta$ in terms of the Weierstrass $\sigma$ function:
\[
 \Phi(z_1,z_2)~=~\frac{\sigma(z_1 + z_2)}{\sigma(z_1) \, \sigma(z_2)}~~~,~~~
 \zeta(z)~=~\frac{\sigma^\prime(z)}{\sigma(z)}~.
\]
Note that in the degenerate case, the spectral parameter drops out.
In fact, all of the calculations to be presented in what follows can
be carried out for the degenerate case in exactly the same manner as
for the elliptic case, provided one performs the following substitutions:
\vspace{2mm}
\begin{equation}
 \begin{array}{ccc}
  \Phi(s,u) \,,\, \Phi(s,v) &\rightarrow& w(s) \\[2mm]
  \Phi(s,u-v) \,,\, \Phi(s,v-u) \,,\, \zeta(s)
  &\rightarrow& - \, {\displaystyle \frac{w^\prime(s)}{w(s)}} \\[4mm]
  \begin{array}{c}
   \zeta(u) \,,\, \zeta(v) \,,\, \zeta(u-v) \,,\, \zeta(v-u) \\[1mm]
   \zeta(s+u) \,,\, \zeta(s+t+u)
  \end{array}
  &\rightarrow& 0 \rule[-6mm]{0mm}{7mm}
 \end{array}
\end{equation}
Therefore, we shall suppress the calculations for the degenerate models,
except at the few points where substantial differences arise.

\pagebreak

\section{Calogero Models for Semisimple Lie Algebras}

According to Ref.~\cite{FW1}, the standard Lax matrix $L$ and the dynamical
$R$-matrix for the Calogero models associated with the root system $\Delta$
of a simple complex Lie algebra $\mathfrak{g}$ read
\begin{equation} \label{eq:LADLAXM}
 L~=~\sum_{j=1}^r \, p_j \>\! H_j \, + \,
     \sum_{\alpha \ssmin \Delta} \mathrm{i} \, \mbox{\sl g}_\alpha \,
     w(\alpha(q)) \, E_\alpha~,
\end{equation}
\begin{equation} \label{eq:LADRMAT}
 R~=~\sum_{\alpha \ssmin \Delta} w(\alpha(q)) \,
     F_\alpha \otimes E_\alpha \, + \,
     \sum_{\alpha \ssmin \Delta} \frac{w^\prime(\alpha(q))}{w(\alpha(q))} \,
     E_\alpha \otimes E_{-\alpha}~,
\vspace{2mm}
\end{equation}
for the degenerate model and
\begin{equation} \label{eq:LAELAXM}
 L(u)~=~\sum_{j=1}^r \, p_j \>\! H_j \, + \,
        \sum_{\alpha \ssmin \Delta} \mathrm{i} \, \mbox{\sl g}_\alpha \,
        \Phi(\alpha(q),u) \, E_\alpha~,
\vspace{-6mm}
\end{equation}
\begin{eqnarray} \label{eq:LAERMAT}
 R(u,v) \!\!
 &=&\!\! - \, \sum_{j=1}^r \, (\zeta(u-v) + \zeta(v)) \, H_j \otimes H_j
                                                                   \nonumber \\
 & &\!\! + \, \sum_{\alpha \ssmin \Delta} \Phi(\alpha(q),v) \,
              F_\alpha \otimes E_\alpha                                      \\
 & &\!\! - \, \sum_{\alpha \ssmin \Delta} \Phi(\alpha(q),u-v) \,
              E_\alpha \otimes E_{-\alpha}~,                       \nonumber
\end{eqnarray}
for the elliptic model. As has been shown in Ref.~\cite{FW1}, integrability
requires the generators $\, F_\alpha \smin\, \mathfrak{h}_{\mathbb{R}} \,$
appearing in eqns~(\ref{eq:LADRMAT}) and~(\ref{eq:LAERMAT}) to satisfy the
constraints (\ref{eq:LAAC1}). Moreover, writing
\begin{equation} \label{eq:LACOEF1}
 F_\alpha^\pm~=~{\textstyle \frac{1}{2}} \, (F_\alpha^{} \pm F_{-\alpha}^{})~,
\end{equation}
we also impose the condition
\begin{equation} \label{eq:LAAC3}
 \alpha(F_\alpha^+)~=~0~,
\end{equation}
which follows from eqn~(\ref{eq:LAAC1}) by setting $\, \beta = - \alpha \,$
when $\, \mbox{\sl g}_\alpha \neq 0 \,$ but turns out to be true in general,
independent of this hypothesis.

In order to compute the content of eqns~(\ref{eq:GPOT2}) and~(\ref{eq:NRMAT}),
we further expand the Cartan part of the gauge potential according to
\begin{equation} \label{eq:LAGPOT3}
 A_j^{\mathfrak{h}}(u)~=~\sum_{k=1}^r A_j^k(u) \, H_k~.
\end{equation}
Then inserting eqns~(\ref{eq:LAERMAT}), (\ref{eq:GPOT3}) and
(\ref{eq:LAGPOT3}) into eqn~(\ref{eq:NRMAT}), we obtain
\vspace{2mm}
\begin{eqnarray*}
 0 \!\!&=&\!\!    \sum_{\alpha \ssmin \Delta}
                  \alpha_k \, \Phi^\prime(\alpha(q),v) \;
                  F_\alpha \otimes E_\alpha \, - \,
                  \sum_{\alpha \ssmin \Delta}
                  \alpha_k \, \Phi^\prime(\alpha(q),u-v) \;
                  E_\alpha \otimes E_{-\alpha}                               \\
 & & \mbox{} - \, \sum_{j=1}^r \,
                  \partial_j^{\vphantom{\mathfrak{h}}}
                  A_k^{\mathfrak{h}}(q,u) \otimes H_j \, - \,
                  \sum_{j=1}^r \, \sum_{\alpha \ssmin \Delta}
                  \partial_j^{\vphantom{\mathfrak{h}}} A_k^\alpha(q,u) \;
                  E_\alpha \otimes H_j                                       \\
 & & \mbox{} + \, \sum_{j=1}^r \, \sum_{\alpha \ssmin \Delta} \,
                  \Bigl(\,  \zeta(u-v) \,+\, \zeta(v) \, \Bigr) \,
                  A^\alpha_k(q,u) \;
                  [ \, H_j \otimes H_j \,,\, E_{\alpha} \otimes 1 \, ]       \\
 & & \mbox{} + \, \sum_{j=1}^r \, \sum_{\alpha \ssmin \Delta} \,
                  \Bigl(\,  \zeta(u-v) \,+\, \zeta(v) \, \Bigr) \,
                  A^\alpha_k(q,v) \;
                  [ \, H_j \otimes H_j \,,\, 1 \otimes E_{\alpha} \, ]       \\
 & & \mbox{} - \, \sum_{j=1}^r \, \sum_{\gamma \ssmin \Delta}
                  \Phi(\gamma(q),v) \, A_k^j(q,u) \;
                  [ \, F_\gamma \otimes E_\gamma \,,\, H_j \otimes 1 \, ]    \\
 & & \mbox{} - \, \sum_{j=1}^r \, \sum_{\gamma \ssmin \Delta}
                  \Phi(\gamma(q),v) \, A_k^j(q,v) \;
                  [ \, F_\gamma \otimes E_\gamma \,,\, 1 \otimes H_j \, ]    \\
 & & \mbox{} - \, \sum_{\gamma,\delta \ssmin \Delta}
                  \Phi(\gamma(q),v) \, A_k^\delta(q,u) \;
                  [ \, F_\gamma \otimes E_\gamma \,,\,
                       E_\delta \otimes 1 \, ]                               \\
 & & \mbox{} - \, \sum_{\gamma,\delta \ssmin \Delta}
                  \Phi(\gamma(q),v) \, A_k^\delta(q,v) \;
                  [ \, F_\gamma \otimes E_\gamma \,,\,
                       1 \otimes E_\delta \, ]                               \\
 & & \mbox{} + \, \sum_{j=1}^r \, \sum_{\gamma \ssmin \Delta}
                  \Phi(\gamma(q),u-v) \, A_k^j(q,u) \;
                  [ \, E_\gamma \otimes E_{-\gamma} \,,\, H_j \otimes 1 \, ] \\
 & & \mbox{} + \, \sum_{j=1}^r \, \sum_{\gamma \ssmin \Delta}
                  \Phi(\gamma(q),u-v) \, A_k^j(q,v) \;
                  [ \, E_\gamma \otimes E_{-\gamma} \,,\, 1 \otimes H_j \, ] \\
 & & \mbox{} + \, \sum_{\gamma,\delta \ssmin \Delta}
                  \Phi(\gamma(q),u-v) \, A_k^\delta(q,u) \;
                  [ \, E_\gamma \otimes E_{-\gamma} \,,\,
                       E_\delta \otimes 1 \, ]                               \\
 & & \mbox{} + \, \sum_{\gamma,\delta \ssmin \Delta}
                  \Phi(\gamma(q),u-v) \, A_k^\delta(q,v) \;
                  [ \, E_\gamma \otimes E_{-\gamma} \,,\,
                       1 \otimes E_\delta \, ]                               \\
 & & \mbox{} + \, \sum_{j=1}^r \, \sum_{\alpha \ssmin \Delta}
                  \alpha_j^{} A_k^\alpha(q,v) \;
                  A_j^{\mathfrak{h}}(q,u) \otimes E_\alpha \, + \,
                  \sum_{j=1}^r \, \sum_{\alpha,\beta \ssmin \Delta}
                  \beta_j^{} A_j^\alpha(q,u) A_k^\beta(q,v) \;
                  E_\alpha \otimes E_\beta~.
\end{eqnarray*}
%in the elliptic case.

\noindent
Using eqns~(\ref{eq:COMR1})--(\ref{eq:COMR5}) to carry out the commutators,
together with the relation
\begin{equation} \label{eq:LACSA}
 \sum_{j=1}^r \, \alpha_j \>\! H_j~=~H_\alpha~,
\end{equation}
we can collect the terms to identify the components of eqn~(\ref{eq:NRMAT})
along the various subspaces of $\, \mathfrak{g} \otimes \mathfrak{g} \,$:
 \\[2mm]
those along $\, \mathfrak{h} \otimes H_j$ $(1 \leqslant j \leqslant r)$,
\begin{equation} \label{eq:LAEDCHH}
 \partial_j^{} A_k^{\mathfrak{h}}(q,u) \, + \,
 \sum_{\alpha \ssmin \Delta} \alpha_j^{} \, \Phi(\alpha(q),v) \,
                             A_k^{-\alpha}(q,v) \, F_\alpha^{}~=~0~,
\end{equation}
those along $\, \mathfrak{h} \otimes \mathfrak{g}_\alpha$
$(\alpha \smin\, \Delta)$,
\begin{eqnarray} \label{eq:LAEDCHE}
 \alpha_k^{} \, \Phi^\prime(\alpha(q),v) \, F_\alpha^{} \!\!&+&\!\!\!
 \Bigl( \zeta(u-v) + \zeta(v) \Bigr)
 A_k^\alpha(q,v) \, H_\alpha^{} \, + \,
 \Phi(\alpha(q),v) \, \alpha(A_k^{\mathfrak{h}}(q,v)) \, F_\alpha^{}
 \nonumber \\[3mm]
 &-&\!\!\! \sum_{\gamma,\delta \ssmin \Delta \atop
                 \gamma + \delta = \alpha}
           N_{\gamma,\delta}^{} \, \Phi(\gamma(q),v) \,
           A_k^\delta(q,v) \, F_{\gamma}^{} \\[-3mm]
 &-&\!\!   \Phi(-\alpha(q),u-v) \, A_k^\alpha(q,u) \, H_\alpha^{} \, + \,
           \sum_{j=1}^r \, \alpha_j^{} \, A_k^\alpha(q,v)
           A_j^{\mathfrak{h}}(q,u)~=~0~, \nonumber
\end{eqnarray}
those along $\, \mathfrak{g}_\alpha \otimes H_j$
$(\alpha \smin\, \Delta \,,\, 1 \leqslant j \leqslant r)$,
\begin{equation} \label{eq:LAEDCEH}
 \partial_j^{} A_k^\alpha(q,u) \, + \,
 \alpha_j^{} \, \Phi(\alpha(q),u-v) \, A_k^\alpha(q,v) \, - \,
 \alpha_j^{} \, (\zeta(u-v) + \zeta(v)) \, A_k^\alpha(q,u)~=~0~,
\end{equation}
those along $\, \mathfrak{g}_\alpha \otimes \mathfrak{g}_\alpha$
$(\alpha \smin\, \Delta)$,
\begin{equation} \label{eq:LAEDCEE1}
 \alpha(F_\alpha^{}) \, \Phi(\alpha(q),v) \, A_k^\alpha(q,u) \, - \,
 \sum_{j=1}^r \, \alpha_j^{} \,  A_j^\alpha(q,u) A_k^\alpha(q,v)~=~0~,
\end{equation}
those along $\, \mathfrak{g}_\alpha \otimes \mathfrak{g}_{-\alpha}$
$(\alpha \smin\, \Delta)$,
\vspace{2mm}
\begin{eqnarray} \label{eq:LAEDCEE2}
 \alpha_k^{} \, \Phi^\prime(\alpha(q),u-v) \!\!&+&\!\!
 \alpha(F_{-\alpha}^{}) \, \Phi(-\alpha(q),v) \, A_k^\alpha(q,u)
 \nonumber \\[2mm]
 &+&\!\! \Phi(\alpha(q),u-v) \,
         \Bigl( \, \alpha(A_k^{\mathfrak{h}}(q,u)) \, - \,
                   \alpha(A_k^{\mathfrak{h}}(q,v)) \Bigr)             \\
 &+&\!\! \sum_{j=1}^r \, \alpha_j^{} \,
         A_j^\alpha(q,u) A_k^{-\alpha}(q,v)~=~0~, \nonumber
\end{eqnarray}
and finally those along $\, \mathfrak{g}_\alpha \otimes \mathfrak{g}_\beta \,$
with $\, \alpha,\beta \smin\, \Delta~,~\alpha \pm \beta \neq 0$,
\vspace{2mm}
\begin{eqnarray} \label{eq:LAEDCEE3}
 && \alpha(F_\beta^{}) \, \Phi(\beta(q),v) \, A_k^\alpha(q,u)
 \nonumber \\[2mm]
 && - \; N_{\alpha,\beta}^{} \,
    \Bigl( \, \Phi(\alpha(q),u-v) \, A_k^{\alpha + \beta}(q,v) \, - \,
              \Phi(-\beta(q),u-v) \, A_k^{\alpha + \beta}(q,u) \, \Bigr)
 \\[-1mm]
 && - \, \sum_{j=1}^r \, \beta_j A_j^\alpha(q,u) A_k^\beta(q,v)~=~0~.
 \nonumber
\end{eqnarray}
%in the elliptic case.
This is a complicated set of equations which we shall solve in a series of
steps.

We~begin by considering the differential equation~(\ref{eq:LAEDCEH})
for the root part of the gauge potential, which by using the functional
equation~(\ref{eq:EFUNEQ6}) (with $u$ and $v$ interchanged) is seen to
have the simple solution
\begin{equation} \label{eq:LAGPOTR}
 A_k^\alpha(q,u)~=~\Phi(\alpha(q),u) \, a_k^\alpha~,
\end{equation}
where the coefficients $a_k^\alpha$ are constants that must be determined
from the remaining equations. For what follows, we shall find it convenient
to assemble these constants into a vector in $\mathfrak{h}_{\mathbb{R}}$
by writing, for any $\, \alpha \smin\, \Delta$,
\begin{equation} \label{eq:LACOEF2}
 a_\alpha^{}~=~\sum_{j=1}^r \, a_j^\alpha \>\! H_j^{}~,
\end{equation}
so that of course
\begin{equation} \label{eq:LACOEF3}
 a_k^\alpha~=~(H_k^{} \,, a_\alpha^{})~.
\end{equation}
In analogy with eqn~(\ref{eq:LACOEF1}), we also introduce the abbreviation
\begin{equation} \label{eq:LACOEF4}
 a_\alpha^\pm~=~{\textstyle \frac{1}{2}} \, (a_\alpha^{} \pm a_{-\alpha}^{})~.
\end{equation}

Now we are ready to state the first main result of this section.
\begin{propos}
 ~The integrable Calogero model associated with the root system of a simple
 complex Lie algebra $\mathfrak{g}$ admits a gauge transformation $g$ from the
 standard Lax pair of Olshanetsky and Perelomov and the dynamical $R$-matrix
 of Ref.~\cite{FW1}, as given by eqns~(\ref{eq:LADLAXM}-\ref{eq:LAERMAT}),
 to a new Lax pair with a numerical $R$-matrix~ if and only if the set of
 generators $\, F_\alpha \smin\, \mathfrak{h}_{\mathbb{R}} \,$ appearing
 in eqns~(\ref{eq:LADRMAT}) and~(\ref{eq:LAERMAT}) satisfies the algebraic
 constraints
 \begin{equation} \label{eq:LAAC4}
  \alpha(F_\alpha^+)~=~0~,
 \end{equation}
 \begin{equation} \label{eq:LAAC5}
  F_\alpha^-~=~\frac{\epsilon_\alpha}{\sqrt{2} \, |\alpha|} \, H_\alpha~,
 \vspace{2mm}
 \end{equation}
 \begin{equation} \label{eq:LAAC6}
  \begin{array}{c}
   \alpha(F_\beta^{}) \, F_\alpha^{} \, - \, \beta(F_\alpha^{}) \, F_\beta^{}~
   =~N_{\alpha,\beta}^{} \, F_{\alpha+\beta}^{} \\[2mm]
   \mbox{for $\, \alpha,\beta \smin\, \Delta \,$ such that
   $\, \beta \neq \pm \alpha$}~,
   \rule[-4mm]{0mm}{5mm}
  \end{array}
 \end{equation}
 as well as the additional algebraic constraints
 \begin{equation} \label{eq:LAEAC1}
  \sum_{\alpha \ssmin \Delta} H_\alpha \otimes F_\alpha \otimes F_{-\alpha}~
  =~0~,
 \end{equation}
 \begin{equation} \label{eq:LAEAC2}
  \sum_{\beta,\gamma \ssmin \Delta \atop \beta + \gamma = \alpha}
  N_{\beta,\gamma} \, F_\beta \otimes F_\gamma~
  =~H_\alpha \otimes F_\alpha \, - \, F_\alpha \otimes H_\alpha~,
 \end{equation}
 to be imposed in the case of the elliptic model, where $\, F_\alpha^\pm
 = \frac{1}{2} \, (F_\alpha^{} \pm F_{-\alpha}^{}) \,$ as above, with
 $\, \epsilon_\alpha = \pm 1$.
 In this case, the root part and the Cartan part of the gauge potential
 $\, A_k^{} = g^{-1} \, \partial_k^{} g \,$ associated with this gauge
 transformation $g$ are given by
 \begin{equation} \label{eq:LADGPOT1}
  A_k^\alpha(q)~=~w(\alpha(q)) \; (H_k \,, F_\alpha^{})~,
 \end{equation}
 and
 \begin{equation} \label{eq:LADGPOT2}
  A_k^{\mathfrak{h}}(q)~
  =~\sum_{\alpha \ssmin \Delta} \frac{w^\prime(\alpha(q))}{w(\alpha(q))}~
    (H_k^{} \,, F_{-\alpha}^{}) \, F_\alpha^{}~,
 \end{equation}
 for the degenerate model and by
 \begin{equation} \label{eq:LAEGPOT1}
  A_k^\alpha(q,u)~=~\Phi(\alpha(q),u) \; (H_k^{} \,, F_\alpha^{})~,
 \end{equation}
 and
 \begin{equation} \label{eq:LAEGPOT2}
  A_k^{\mathfrak{h}}(q,u)~
  = \; - \, \sum_{\alpha \ssmin \Delta} \, \zeta(\alpha(q)) \,
            (H_k^{} \,, F_{-\alpha}^{}) \, F_\alpha^{} \, - \,
            \zeta(u) \, H_k^{}~,
 \end{equation}
 for the elliptic model.
\end{propos}
\textbf{Note.}~~As we shall show after completing the proof of Proposition~1,
eqn~(\ref{eq:LAAC6}) forces all roots $\alpha$ in $\Delta$ to have the same
length (which by convention we fix to be $\sqrt{2}$) and also allows for a
choice of basis in which the signs $\epsilon_\alpha$ are independent of~%
$\alpha$. so that eqn~(\ref{eq:LAAC5}) can be simplified as follows:
\begin{equation} \label{eq:LAAC7}
 F_\alpha^-~=~{\textstyle \frac{1}{2}} \, \epsilon \, H_\alpha~.
\end{equation}

\noindent
\textbf{Proof.}~~With the vector notation introduce above, we can first
of all reduce eqn~(\ref{eq:LAEDCEE1}) to a single algebraic constraint:
\begin{equation} \label{eq:LAIAC01}
 \alpha(a_\alpha^{})~=~\alpha(F_\alpha^{})~.
\end{equation}
Note that replacing $\alpha$ by $-\alpha$ and adding/subtracting the two
equations, we get
\begin{eqnarray}
 &\alpha(F_\alpha^+)~=~\alpha(a_\alpha^+)~,&              \label{eq:LAIAC02} \\
 &\alpha(F_\alpha^-)~=~\alpha(a_\alpha^-)~.&              \label{eq:LAIAC03}
\end{eqnarray}
Using eqn~(\ref{eq:LAAC3}), the first of these can be sharpened as follows:
\begin{equation} \label{eq:LAIAC04}
 \alpha(F_\alpha^+)~=~0~=~\alpha(a_\alpha^+)~.
\end{equation}
Next, inserting eqn~(\ref{eq:LAGPOTR}) together with the functional equation~%
(\ref{eq:EFUNEQ2}) into the differential equation (\ref{eq:LAEDCHH}) for the
Cartan part of the gauge potential, we see that this equation can be solved
by setting
\begin{equation} \label{eq:LAGPOTC1}
 A_k^{\mathfrak{h}}(q,u)~
 = \; - \, \sum_{\alpha \ssmin \Delta} \, \zeta(\alpha(q)) \,
           a_k^{-\alpha} \, F_\alpha^{} \, - \, a_k^{\mathfrak{h}}(u)~,
\end{equation} 
where the $a_k^{\mathfrak{h}}(u)$ are constants that must be determined from
the remaining equations, provided we assume the coefficients $a_k^\alpha$ to
satisfy the relation
\begin{equation} \label{eq:LAIAC05}
 \sum_{\alpha \ssmin \Delta} \, \alpha_j^{} \, 
 a_k^{-\alpha} \, F_\alpha^{}~=~0 \qquad
 \mbox{for $\, 1 \leqslant j\,,k \leqslant r$}~.
\end{equation}
Converted into a tensor equation, it reads
\begin{equation} \label{eq:LAIAC06}
 \sum_{\alpha \ssmin \Delta} H_\alpha \otimes F_\alpha \otimes a_{-\alpha}~
 =~0~,
\end{equation}
which leads back to eqn~(\ref{eq:LAIAC05}) by taking the scalar product with
$H_j$ in the first and with $H_k$ in the third tensor factor. Note that in
the degenerate case, the same argument works, but eqn~(\ref{eq:LAIAC05}/%
\ref{eq:LAIAC06}) is not needed. Similarly, inserting eqn~(\ref{eq:LAGPOTR})
together with the functional equation~(\ref{eq:EFUNEQ6}) into eqn~%
(\ref{eq:LAEDCEE2}), we obtain
\[
 \begin{array}{l}
  \alpha_k^{} \,
  \Bigl( - \, \Phi(\alpha(q),u) \, \Phi(\alpha(q),-v) \, +
         \left( \zeta(u) - \zeta(v) \right) \Phi(\alpha(q),u-v) \Bigr) \\[4mm]
   - \; \alpha(F_{-\alpha}^{}) \, a_k^\alpha \,
  \Phi(\alpha(q),u) \, \Phi(\alpha(q),-v) \\[3mm]
  - \; \Phi(\alpha(q),u-v) \,
  \Bigl( \alpha(a_k^{\mathfrak{h}}(u)) \, - \,
         \alpha(a_k^{\mathfrak{h}}(v)) \Bigr) \\[4mm]
  - \; \alpha(a_\alpha^{}) \, a_k^{-\alpha} \,
  \Phi(\alpha(q),u) \, \Phi(\alpha(q),-v)~=~0~.
 \end{array}
\]
Obviously, the terms proportional to $\Phi(\alpha(q),u-v)$ cancel provided
we set
\begin{equation} \label{eq:LAGPOTC2}
 a_k^{\mathfrak{h}}(u)~=~\zeta(u) \, H_k~,
\end{equation}
and the remaining terms cancel if we impose the relation
\begin{equation} \label{eq:LAIAC07}
 \alpha_k^{} \, + \, \alpha(F_{-\alpha}) \, a_k^\alpha \, + \,
 \alpha(a_\alpha) \, a_k^{-\alpha}~=~0~.
\end{equation}
Converted to a vector equation in $\mathfrak{h}_{\mathbb{R}}$, it reads
\begin{equation} \label{eq:LAIAC08}
 H_\alpha \, + \, \alpha(F_{-\alpha}) \, a_\alpha \, + \,
 \alpha(a_\alpha) \, a_{-\alpha} ~=~ 0~,
\end{equation}
which leads back to eqn~(\ref{eq:LAIAC07}) by taking the scalar product with
$H_k$. Even simpler to handle is eqn~(\ref{eq:LAEDCEE3}), which by insertion
of the functional equation~(\ref{eq:EFUNEQ5}) reduces to the relation
\begin{equation} \label{eq:LAIAC09}
 \begin{array}{c}
  \alpha(F_\beta^{}) \, a_k^\alpha \, - \,
  N_{\alpha,\beta}^{} \, a_k^{\alpha+\beta} \, - \,
  \beta(a_\alpha^{}) \, a_k^\beta~=~0 \\[2mm]
  \mbox{for $\, \alpha,\beta \smin\, \Delta \,$ such that
  $\, \beta \neq \pm \alpha$}~.
 \end{array}
\end{equation}
Converted to a vector equation in $\mathfrak{h}_{\mathbb{R}}$, it reads
\begin{equation} \label{eq:LAIAC10}
 \begin{array}{c}
  \alpha(F_\beta) \, a_\alpha \, - \,
  N_{\alpha, \beta} \, a_{\alpha+\beta} \, - \,
  \beta(a_\alpha) \, a_\beta~=~0 \\[2mm]
  \mbox{for $\, \alpha,\beta \smin\, \Delta \,$ such that
  $\, \beta \neq \pm \alpha$}~,
 \end{array}
\end{equation} 
which leads back to eqn~(\ref{eq:LAIAC09}) by taking the scalar product with
$H_k$.

Before proceeding to the solution of the remaining equations, let us pause
to draw a few consequences of the algebraic constraints (\ref{eq:LAIAC01})--%
(\ref{eq:LAIAC04}) and (\ref{eq:LAIAC07}/\ref{eq:LAIAC08}) derived so far;
this will help us considerably to simplify our further work. First of all,
eqns~({\ref{eq:LAIAC01})--(\ref{eq:LAIAC04}) state that
\[
 \alpha(F_{-\alpha})~= \; - \, \alpha(F_\alpha)~
 = \; - \, \alpha(a_\alpha)~=~\alpha(a_{-\alpha})~,
\]
implying that eqn~(\ref{eq:LAIAC08}) can be reduced to
\[
 \alpha(a_\alpha) \, (a_\alpha - a_{-\alpha})~=~H_\alpha~.
\vspace{2mm}
\]
Applying $\alpha$ to this relation and using the previous equation again,
we conclude that
\begin{equation} \label{eq:LAIAC11}
 \alpha(a_\alpha)~=~\epsilon_\alpha \; \frac{|\alpha|}{\sqrt{2}}
 \qquad \mbox{and} \qquad
 a_\alpha^-~=~\frac{\epsilon_\alpha}{\sqrt{2} \, |\alpha|} \, H_\alpha
\end{equation}
where $\, \epsilon_\alpha = \epsilon_{-\alpha} \,$ is a sign factor ($\pm 1$).
Next, we simplify all these equations by showing that eqns~(\ref{eq:LAIAC09}/%
\ref{eq:LAIAC10}) and~(\ref{eq:LAIAC11}) in fact force the vectors $a_\alpha$
and $F_\alpha$ to be equal. To prove this, we begin by symmetrizing eqn~%
(\ref{eq:LAIAC10}) with respect to the exchange of $\alpha$ and $\beta$,
obtaining
\[
 \alpha(F_\beta - a_\beta) \, a_\alpha \, + \,
 \beta(F_\alpha - a_\alpha) \, a_\beta~=~0~.
\]
Symmetrizing with respect to the exchange of $\beta$ and $-\beta$ and
inserting eqn~(\ref{eq:LAIAC11}) gives
\begin{equation} \label{eq:LAIAC12}
 \alpha(F_\beta^+ - a_\beta^+) \, a_\alpha \, + \,
 \frac{\epsilon_\beta}{\sqrt{2} \, |\beta|} \,
 \beta(F_\alpha - a_\alpha) \, H_\beta~=~0~.
\end{equation}
Symmetrizing again with respect to the exchange of $\alpha$ and $-\alpha$
and inserting eqn~(\ref{eq:LAIAC11}) then leads to
\[
 \frac{\epsilon_\alpha}{\sqrt{2} \, |\alpha|} \,
 \alpha(F_\beta^+ - a_\beta^+) \, H_\alpha \, + \,
 \frac{\epsilon_\beta}{\sqrt{2} \, |\beta|} \,
 \beta(F_\alpha^+ - a_\alpha^+) \, H_\beta~=~0~.
\]
But $H_\alpha$ and $H_\beta$ are linearly independent since, as stated in
eqn~(\ref{eq:LAIAC10}), the roots $\alpha$ and $\beta$ are supposed to be
non-proportional, so the coefficients must vanish separately, that is,
for any two roots $\, \alpha,\beta \smin \Delta$, we have
\[
 \beta(F_\alpha^+ - a_\alpha^+)~=~0
\]
whenever $\beta$ is not proportional to $\alpha$ and, according to eqn~%
(\ref{eq:LAIAC04}), also when $\beta$ is proportional to $\alpha$. Since 
$\Delta$ generates $\mathfrak{h}_{\mathbb{R}}$, this simply means that
$\, a_\alpha^+ = F_\alpha^+$. Inserting this conclusion back into eqn~%
(\ref{eq:LAIAC12}) and applying once more the same argument, we arrive
at the result that $\, a_\alpha = F_\alpha \,$. With this result, eqns~%
(\ref{eq:LAIAC01})--(\ref{eq:LAIAC03}) and~(\ref{eq:LAIAC08}) reduce to
trivial identities whereas eqns~(\ref{eq:LAIAC04}), (\ref{eq:LAIAC11}),
(\ref{eq:LAIAC10}), (\ref{eq:LAIAC06}), (\ref{eq:LAGPOTR}) with~%
(\ref{eq:LACOEF3}) and~(\ref{eq:LAGPOTC1}) with (\ref{eq:LAGPOTC2})
assume the form given in eqns~(\ref{eq:LAAC4}), (\ref{eq:LAAC5}),
(\ref{eq:LAAC6}), (\ref{eq:LAEAC1}), (\ref{eq:LAEGPOT1}) and~%
(\ref{eq:LAEGPOT2}), respectively.

Let us summarize the results obtained so far. With the exception of
eqn~(\ref{eq:LAEDCHE}), the system of equations (\ref{eq:LAEDCHH})--%
(\ref{eq:LAEDCEE3}) has been completely solved in terms of the explicit
formulae (\ref{eq:LADGPOT1})--(\ref{eq:LAEGPOT2}) for the gauge potential
with the explicit formula (\ref{eq:LAAC5}) for the odd part $F_\alpha^-$
of the coefficient vectors $F_\alpha$ and the algebraic constraints
(\ref{eq:LAAC4}), (\ref{eq:LAAC6}) and (\ref{eq:LAEAC1}). Thus we are
left with the task of verifying the implications of eqns~(\ref{eq:GPOT4}),
(\ref{eq:GPOT5}) and (\ref{eq:LAEDCHE}).

Beginning with eqn~(\ref{eq:GPOT4}), we use the functional equation~%
(\ref{eq:EFUNEQ2}) to compute
\vspace{2mm}
\begin{eqnarray*}
\lefteqn{\partial_k^{\vphantom{\mathfrak{h}}} A_l^{\mathfrak{h}}(q,u) \, - \,
         \partial_l^{\vphantom{\mathfrak{h}}} A_k^{\mathfrak{h}}(q,u) \, + \,
         \sum_{\alpha \ssmin \Delta} A_k^\alpha(q,u) A_l^{-\alpha}(q,u)
         H_\alpha^{}}                                      \hspace{5mm} \\[1mm]
 &=&\!\! \sum_{\alpha \ssmin \Delta} \zeta^\prime(\alpha(q))
         \Bigl( \, \alpha_l^{} \, (H_k^{} \,, F_{-\alpha}^{}) \,
                   F_\alpha^{} \, - \,
                   \alpha_k^{} \, (H_l^{} \,, F_{-\alpha}^{}) \,
                   F_\alpha^{} \Bigr)                                   \\
 & & \mbox{} + \, \sum_{\alpha \ssmin \Delta} 
         \Phi(\alpha(q),u) \, \Phi(-\alpha(q),u) \;
         (H_k^{} \,, F_\alpha^{}) \, (H_l^{} \,, F_{-\alpha}^{}) \,
         H_\alpha^{} \hspace*{3cm}
\end{eqnarray*}
\begin{eqnarray*}
 \mbox{}
 &=&\!\! {\textstyle \frac{1}{2}} \;
         \sum_{\alpha \ssmin \Delta} \zeta^\prime(\alpha(q))
         \Bigl( \, + \; \alpha_l^{} \, (H_k^{} \,, F_{-\alpha}^{}) \,
                        F_\alpha^{} \,
                   - \, \alpha_l^{} \, (H_k^{} \,, F_\alpha^{}) \,
                        F_{-\alpha}^{}                                 \\[-3mm]
 & & \hspace{28mm} - \; \alpha_k^{} \, (H_l^{} \,, F_{-\alpha}^{}) \,
                        F_\alpha^{} \,
                   + \, \alpha_k^{} \, (H_l^{} \,, F_\alpha^{}) \,
                        F_{-\alpha}^{}                                  \\
 & & \hspace{28mm} + \; (H_k^{} \,, F_\alpha^{}) \,
                        (H_l^{} \,, F_{-\alpha}^{}) \, H_\alpha^{} \,
                   - \, (H_k^{} \,, F_{-\alpha}^{}) \,
                        (H_l^{} \,, F_\alpha^{}) \, H_\alpha^{} \Bigr)  \\[1mm]
 & & \mbox{} - \, \zeta^\prime(u) \, \sum_{\alpha \ssmin \Delta}
                  (H_k^{} \,, F_\alpha^{}) \,
                  (H_l^{} \,, F_{-\alpha}^{}) \, H_\alpha^{}            \\[4mm]
 &=&\!\! \sum_{\alpha \ssmin \Delta} \zeta^\prime(\alpha(q))
         \Bigl( \, - \; \alpha_l^{} \, (H_k^{} \,, F_\alpha^-) \, F_\alpha^+ \,
                   + \, \alpha_l^{} \, (H_k^{} \,, F_\alpha^+) \, F_\alpha^-
                                                                       \\[-3mm]
 & &\hspace{2.55cm}+ \; \alpha_k^{} \, (H_l^{} \,, F_\alpha^-) \, F_\alpha^+ \,
                   - \, \alpha_k^{} \, (H_l^{} \,, F_\alpha^+) \, F_\alpha^- \\
 & &\hspace{2.55cm}+ \; (H_k^{} \,, F_\alpha^-) \, (H_l^{} \,, F_\alpha^+) \,
                        H_\alpha^{} \,
                   - \, (H_k^{} \,, F_\alpha^+) \, (H_l^{} \,, F_\alpha^-) \,
                        H_\alpha^{} \Bigr)                              \\[1mm]
 & & \mbox{} - \, \zeta^\prime(u) \, \sum_{\alpha \ssmin \Delta}
                  (H_k^{} \,, F_\alpha) \, (H_l^{} \,, F_{-\alpha}) \,
                  H_\alpha^{}                                           \\
\end{eqnarray*}
and can use eqns~(\ref{eq:LAAC5}) and~(\ref{eq:LAEAC1}) to verify that
the terms under the first sum cancel \mbox{mutually} in pairs whereas the
second sum vanishes. Note that in the degenerate case, the same argument
works, but eqn~(\ref{eq:LAEAC1}) is not needed.

For the proof of eqn~(\ref{eq:GPOT5}), the trick is  to split the sum over
roots $\beta$ coming from the third and fourth term into contributions with
$\, \beta = \alpha$, which cancel mutually, contributions with $\, \beta
= - \alpha$, which combine with the contributions coming from the first and
second term (transformed using the functional equation~(\ref{eq:EFUNEQ7})),
and the remaining contributions with $\, \beta \neq \pm\alpha$: these can
be complemented by terms that also cancel mutually (marked by underlining)
and then be combined with the contributions from the fifth term (transformed
using the functional equation (\ref{eq:EFUNEQ8})):
\vspace{3mm}
\begin{eqnarray*}
\lefteqn{
 \begin{array}{rcl}
  \partial_k^{\vphantom{\alpha}} A_l^\alpha(q,u) \, - \,
  \partial_l^{\vphantom{\alpha}} A_k^\alpha(q,u) \!\! &+& \!\!
  \alpha(A_k^{\mathfrak{h}}(q,u)) \,
  A_l^{\vphantom{\mathfrak{h}}\alpha}(q,u) \, - \,
  \alpha(A_l^{\mathfrak{h}}(q,u)) \,
  A_k^{\vphantom{\mathfrak{h}}\alpha}(q,u) \\[4mm]
  &+& \!\!\! {\displaystyle
              \sum_{\beta,\gamma \ssmin \Delta \atop \beta + \gamma = \alpha}
              N_{\beta,\gamma}^{} A_k^\beta(q,u) A_l^\gamma(q,u)}
 \end{array}
}                                                          \hspace{3mm} \\[5mm]
 &=&\!\! \Phi^\prime(\alpha(q),u) \,
         \alpha_k^{} \, (H_l^{} \,, F_\alpha^{}) \, - \,
         \Phi^\prime(\alpha(q),u) \,
         \alpha_l^{} \, (H_k^{} \,, F_\alpha^{})                        \\[4mm]
 & &\!\! \mbox{} + \, \Phi(\alpha(q),u)
         \Bigl( \, \sum_{\beta \ssmin \Delta}
                   \zeta(\beta(q)) \; \alpha(F_{-\beta}^{}) \,
                   (H_k^{} \,, F_\beta^{}) \, (H_l^{} \,, F_\alpha^{}) \, - \,
                   \zeta(u) \; \alpha_k^{} \, (H_l^{} \,, F_\alpha^{})
                   \Bigr)                                               \\
 & &\!\! \mbox{} - \, \Phi(\alpha(q),u)
         \Bigl( \, \sum_{\beta \ssmin \Delta}
                   \zeta(\beta(q)) \, \alpha(F_{-\beta}^{}) \,
                   (H_l^{} \,, F_\beta^{}) \, (H_k^{} \,, F_\alpha^{}) \, - \,
                   \zeta(u) \; \alpha_l^{} \, (H_k^{} \,, F_\alpha^{})
                   \Bigr)                                               \\
 & &\!\! \mbox{} + \,
         \sum_{\gamma,\delta \ssmin \Delta \atop \gamma + \delta = \alpha}
         N_{\gamma,\delta}^{} \, \Phi(\gamma(q),u) \, \Phi(\delta(q),u) \,
         (H_k^{} \,, F_\gamma^{}) \, (H_l^{} \,, F_\delta^{})           \\[2cm]
 &=&\!\! \Bigl( \Phi^\prime(\alpha(q),u) \, - \,
                \Phi(\alpha(q),u) \, \zeta(u) \Bigr) \,
         \Bigl( \, (H_k \,, H_\alpha) \, (H_l \,, F_\alpha) \, - \,
                   (H_l \,, H_\alpha) \, (H_k \,, F_\alpha) \Bigr)      \\[2mm]
 & &\!\! \mbox{} + \, \Phi(\alpha(q),u) \, \zeta(\alpha(q))             \\[1mm]
 & &\!\! \mbox{} \hspace{5mm} \times
         \Bigl( \mbox{} + \, \alpha(F_{-\alpha}) \,
                   (H_k \,, F_\alpha) \, (H_l \,, F_\alpha) \, - \,
                   \alpha(F_\alpha) \,
                   (H_k \,, F_{-\alpha}) \, (H_l \,, F_\alpha)         \\[-2mm]
 & &\!\! \mbox{} \hspace{7mm} \phantom{\times \Bigl(}
                \mbox{} - \, \alpha(F_{-\alpha}) \,
                   (H_l \,, F_\alpha) \, (H_k \,, F_\alpha) \, + \,
                   \alpha(F_\alpha) \,
                   (H_l \,, F_{-\alpha}) \, (H_k \,, F_\alpha) \, \Bigr)\\[1mm]
 & &\!\! \mbox{} + \, \Phi(\alpha(q),u)
         \sum_{\beta \ssmin \Delta \atop \beta \neq \pm\alpha}
         \zeta(\beta(q)) \, (H_k^{} \,, F_\beta^{}) \,
         \Bigl( \, \alpha(F_{-\beta}^{}) \, (H_l^{} \,, F_\alpha^{}) \, + \,
                   \underline{\beta(F_\alpha^{}) \, (H_l^{} \,, F_{-\beta}^{})}
                \, \Bigr)                                               \\
 & &\!\! \mbox{} - \, \Phi(\alpha(q),u) 
         \sum_{\beta \ssmin \Delta \atop \beta \neq \pm\alpha}
         \zeta(\beta(q)) \, (H_l^{} \,, F_\beta^{}) \,
         \Bigl( \, \alpha(F_{-\beta}^{}) \, (H_k^{} \,, F_\alpha^{}) \, + \,
                   \underline{\beta(F_\alpha^{}) \, (H_k^{} \,, F_{-\beta}^{})}
                \, \Bigr)                                               \\
 & &\!\! \mbox{} + \, \Phi(\alpha(q),u)
         \sum_{\gamma,\delta \ssmin \Delta \atop \gamma + \delta = \alpha}
         \Bigl( \, \zeta(\gamma(q)) \, + \, \zeta(\delta(q)) \Bigr) \,
         N_{\gamma,\delta} \,
         (H_k \,, F_\gamma) \, (H_l \,, F_\delta)                       \\
 & &\!\! \mbox{} - \, \Phi(\alpha(q),u)
         \Bigl( \, \zeta(\alpha(q) + u) \, - \, \zeta(u) \Bigr)
         \sum_{\gamma,\delta \ssmin \Delta \atop \gamma + \delta = \alpha}
         N_{\gamma,\delta} \,
         (H_k \,, F_\gamma) \, (H_l \,, F_\delta)                       \\[3mm]
 &=&\!\! \Phi(\alpha(q),u) \,
         \Bigl( \, \zeta(\alpha(q)+u) \, - \, \zeta(\alpha(q)) \, - \,
                   \zeta(u) \Bigr)                                      \\
 & &\!\! \mbox{} \times
         \Bigl( \, (H_k^{} \,, H_\alpha^{}) \, (H_l^{} \,, F_\alpha^+) \, - \,
                   (H_k^{} \,, F_\alpha^+) \, (H_l^{} \,, H_\alpha^{}) \\[-1mm]
 & &\!\! \mbox{} \hspace{1cm}
         - \, 2 \>\! \alpha(F_\alpha^-) \,
              (H_k^{} \,, F_\alpha^-) \, (H_l^{} \,, F_\alpha^+) \,
         + \, 2 \>\! \alpha(F_\alpha^-) \,
              (H_k^{} \,, F_\alpha^+) \, (H_l^{} \,, F_\alpha^-) \Bigr) \\[2mm]
 & &\!\! \mbox{} + \, \Phi(\alpha(q),u) \,
         \Bigl( \, \zeta(\alpha(q)+u) \, - \, \zeta(u) \Bigr)           \\
 & &\!\! \mbox{} \hspace{6mm} \times
         \Bigl( \, 2 \>\! \alpha(F_\alpha^-) \,
                   (H_k^{} \,, F_\alpha^-) \, (H_l^{} \,, F_\alpha^+) \, - \,
                   2 \>\! \alpha(F_\alpha^-) \,
                   (H_k^{} \,, F_\alpha^+) \, (H_l^{} \,, F_\alpha^-)   \\
 & &\!\! \mbox{} \hspace{3cm} - \,
         \sum_{\beta,\gamma \ssmin \Delta \atop \beta + \gamma = \alpha}
         N_{\beta,\gamma} \,
         (H_k \,, F_\beta) \, (H_l \,, F_\gamma) \, \Bigr)              \\
 & &\!\! \mbox{} + \, \Phi(\alpha(q),u)
         \sum_{\beta \ssmin \Delta \atop \beta \neq \pm\alpha}
         \zeta(\beta(q)) \, (H_k^{} \,, F_\beta^{})                    \\[-5mm]
 & & \hspace*{35mm} \times
         \Bigl( H_l^{} \,, \alpha(F_{-\beta}^{}) \, F_\alpha^{} \, + \,
                           \beta(F_\alpha^{}) \, F_{-\beta}^{} \, + \,
                           N_{\beta,\alpha-\beta}^{} \,
                           F_{\alpha-\beta}^{} \Bigr)                   \\[2mm]
 & &\!\! \mbox{} - \, \Phi(\alpha(q),u)
         \sum_{\beta \ssmin \Delta \atop \beta \neq \pm\alpha}
         \zeta(\beta(q)) \, (H_l^{} \,, F_\beta^{})                    \\[-5mm]
 & & \hspace*{35mm} \times
         \Bigl( H_k^{} \,, \alpha(F_{-\beta}^{}) \, F_\alpha^{} \, + \,
                           \beta(F_\alpha^{}) \, F_{-\beta}^{} \, - \,
                           N_{\alpha-\beta,\beta}^{} \,
                           F_{\alpha-\beta}^{} \Bigr)~.
\end{eqnarray*}
The last two terms vanish due to eqn~(\ref{eq:LAAC6}), while the first term
vanishes due to eqn~(\ref{eq:LAAC5}). The same reasoning shows that the
second term will vanish provided we impose the condition~(\ref{eq:LAEAC2}).
Note that in the degenerate case, the same argument works, but eqn~%
(\ref{eq:LAEAC2}) is not needed.

\pagebreak

The proof of eqn~(\ref{eq:LAEDCHE}) proceeds along similar lines, using the
functional equations~(\ref{eq:EFUNEQ6})--(\ref{eq:EFUNEQ8}):
\vspace{2mm}
\begin{eqnarray*}
\lefteqn{ 
 \begin{array}{rcl}
  \alpha_k^{} \, \Phi^\prime(\alpha(q),v) \, F_\alpha^{} \!\!&+&\!\!
  \bigl( \zeta(u-v) + \zeta(v) \bigr) \, A_k^\alpha(q,v) \, H_\alpha^{} \, + \,
  \Phi(\alpha(q),v) \; \alpha(A_k^{\mathfrak{h}}(q,v)) \, F_\alpha^{} \\[5mm]
  &-&\!\!\!
  {\displaystyle
   \sum_{\gamma,\delta \ssmin \Delta \atop \gamma + \delta = \alpha}
   N_{\gamma,\delta}^{} \, \Phi(\gamma(q),v) \,
   A_k^\delta(q,v) \, F_\gamma^{}} \\[-2mm]
  &-&\!\!\!
  {\displaystyle
   \Phi(-\alpha(q),u-v) \, A_k^\alpha(q,u) \, H_\alpha^{} \, + \,
   \sum_{j=1}^r \, \alpha_j^{} \, A_k^\alpha(q,v) A_j^{\mathfrak{h}}(q,u)}
 \end{array}}                                              \hspace{5mm} \\[4mm]
 &=&\!\! \Phi^\prime(\alpha(q),v) \; \alpha_k^{} \, F_\alpha^{} \, + \,
         \bigl( \zeta(u-v) + \zeta(v) \bigr) \;
         \Phi(\alpha(q),v) \; (H_k^{} \,, F_\alpha^{}) \, H_\alpha^{}   \\[5mm]
 & & \mbox{} + \,
         \Phi(\alpha(q),v) \sum_{\beta \ssmin \Delta} \zeta(\beta(q)) \,
         \alpha(F_{-\beta}^{}) \,
         (H_k^{} \,, F_\beta^{}) \, F_\alpha^{} \, - \,
         \Phi(\alpha(q),v) \, \zeta(v) \, \alpha_k^{} \, F_\alpha^{}    \\[1mm]
 & & \mbox{} - \,
         \sum_{\gamma,\delta \ssmin \Delta \atop \gamma + \delta = \alpha}
         N_{\gamma,\delta}^{} \, \Phi(\gamma(q),v) \, \Phi(\delta(q),v) \,
         (H_k^{} \,, F_\delta^{}) \, F_\gamma^{}                        \\[1mm]
 & & \mbox{} - \, \Phi(-\alpha(q),u-v) \, \Phi(\alpha(q),u) \,
         (H_k^{} \,, F_\alpha^{}) \, H_\alpha^{}                        \\[3mm]
 & & \mbox{} - \, \Phi(\alpha(q),v) \,
         \sum_{j=1}^r \sum_{\beta \ssmin \Delta}
         \zeta(\beta(q)) \; \alpha_j^{} \, (H_j^{} \,, F_{-\beta}^{}) \,
         (H_k^{} \,, F_\alpha^{}) \, F_\beta^{}                         \\
 & & \mbox{} - \, \Phi(\alpha(q),v) \, \zeta(u) \,
         \sum_{j=1}^r \alpha_j^{} \, (H_k^{} \,, F_\alpha^{}) \, H_j^{} \\[4mm]
 &=&\!\! \Bigl( \Phi^\prime(\alpha(q),v) \, - \,
                \Phi(\alpha(q),v) \, \zeta(v) \Bigr) \,
         (H_k^{} \,, H_\alpha^{}) \, F_\alpha^{}                        \\[2mm]
 & & \mbox{} + \,
         \Bigl( \bigl( \zeta(u-v) + \zeta(v) - \zeta(u) \bigr) \,
                \Phi(\alpha(q),v)                                      \\[-1mm]
 & & \mbox{} \qquad - \,
         \Phi(-\alpha(q),u-v) \, \Phi(\alpha(q),u) \Bigr) \,
         (H_k^{} \,, F_\alpha^{}) \, H_\alpha^{}                        \\[2mm]
 & & \mbox{} + \, \Phi(\alpha(q),v) \, \zeta(\alpha(q)) \,
         \Bigl( \mbox{} + \, \alpha(F_{-\alpha}^{}) \,
                (H_k^{} \,, F_\alpha^{}) \, F_\alpha^{} \, - \,
                \alpha(F_\alpha^{}) \, (H_k^{} \,, F_{-\alpha}^{}) \,
                F_\alpha^{}                                            \\[-1mm]
 & & \mbox{} \hspace{41mm} - \, \alpha(F_{-\alpha}^{}) \,
                (H_k^{} \,, F_\alpha^{}) \, F_\alpha^{} \, + \,
                \alpha(F_\alpha^{}) \, (H_k^{} \,, F_\alpha^{}) \,
                F_{-\alpha}^{} \Bigr)                                   \\[2mm]
 & & \mbox{} + \, \Phi(\alpha(q),v)
         \sum_{\beta \ssmin \Delta \atop \beta \neq \pm\alpha}
         \zeta(\beta(q)) \; (H_k^{} \,, F_\beta^{}) \,
         \Bigl( \alpha(F_{-\beta}^{}) F_\alpha^{} \, + \,
                \underline{\beta(F_\alpha^{}) F_{-\beta}^{}} \Bigr)     \\
 & & \mbox{} - \, \Phi(\alpha(q),v)
         \sum_{\beta \ssmin \Delta \atop \beta \neq \pm\alpha}
         \zeta(\beta(q)) \;
         \Bigl( H_k^{} \,, \alpha(F_{-\beta}^{}) F_\alpha^{} \, + \,
                \underline{\beta(F_\alpha^{}) F_{-\beta}^{}} \Bigr) \,
         F_\beta^{}                                                     \\
 & & \mbox{} - \, \Phi(\alpha(q),v)
         \sum_{\gamma,\delta \ssmin \Delta \atop \gamma + \delta = \alpha}
         \Bigl( \, \zeta(\gamma(q)) \, + \, \zeta(\delta(q)) \, \Bigr)
         N_{\gamma,\delta} \, (H_k \,, F_\delta) \, F_\gamma            \\
 & & \mbox{} + \, \Phi(\alpha(q),v)
         \Bigl( \, \zeta(\alpha(q) + v) \, - \, \zeta(v) \, \Bigr)
         \sum_{\gamma,\delta \ssmin \Delta \atop \gamma + \delta = \alpha}
         N_{\gamma,\delta} \, (H_k \,, F_\delta) \, F_\gamma            \\[3mm]
 &=&\!\! \Phi(\alpha(q),v) \,
         \Bigl( \, \zeta(\alpha(q)+v) \, - \, \zeta(\alpha(q)) \, - \,
                   \zeta(v) \Bigr) \,
         \Bigl( \, (H_k^{} \,, H_\alpha^{}) \, F_\alpha^{} \, - \,
                   (H_k^{} \,, F_\alpha^{}) \, H_\alpha^{} \, \Bigr)    \\[2mm]
 & & \mbox{} + \,
         \Phi(\alpha(q),v) \; \zeta(\alpha(q)) \; 2 \>\! \alpha(F_\alpha^-) \,
         \Bigl( \, (H_k^{} \,, F_\alpha^-) \, F_\alpha^+ \, - \,
                   (H_k^{} \,, F_\alpha^+) \, F_\alpha^- \, \Bigr)      \\[2mm]
 & & \mbox{} + \, \Phi(\alpha(q),v)
         \sum_{\beta \ssmin \Delta \atop \beta \neq \pm\alpha}
         \zeta(\beta(q))                                               \\[-5mm]
 & & \hspace*{35mm} \times \,
          (H_k^{} \,, F_\beta^{}) \;
          \Bigl( \alpha(F_{-\beta}^{}) \, F_\alpha^{} \, + \,
                 \beta(F_\alpha^{}) \, F_{-\beta}^{} \, - \,
                 N_{\alpha-\beta,\beta}^{} \,
                 F_{\alpha-\beta}^{} \Bigr)                             \\[2mm]
 & & \mbox{} - \, \Phi(\alpha(q),v)
         \sum_{\beta \ssmin \Delta \atop \beta \neq \pm\alpha}
         \zeta(\beta(q))                                               \\[-5mm]
 & & \hspace*{35mm} \times
         \Bigl( H_k^{} \,, \alpha(F_{-\beta}^{}) \, F_\alpha^{} \, + \,
                           \beta(F_\alpha^{}) \, F_{-\beta}^{} \, + \,
                           N_{\beta,\alpha-\beta}^{} \,
                           F_{\alpha-\beta}^{} \Bigr) \; F_\beta^{}     \\[3mm]
 & & \mbox{} + \, \Phi(\alpha(q),v)
         \Bigl( \, \zeta(\alpha(q) + v) \, - \, \zeta(v) \, \Bigr)
         \sum_{\gamma,\delta \ssmin \Delta \atop \gamma + \delta = \alpha}
         N_{\gamma,\delta} \, (H_k \,, F_\delta) \, F_\gamma            \\[4mm]
 &=&\!\! \Phi(\alpha(q),v)
         \Bigl( \, \zeta(\alpha(q) + v) \, - \, \zeta(v) \Bigr)         \\[2mm]
 & & \mbox{} \times
         \Bigl( \, (H_k^{} \,, H_\alpha^{}) \, F_\alpha^{} \, - \,
                   (H_k^{} \,, F_\alpha^{}) \, H_\alpha^{} \, - \,
         \sum_{\beta,\gamma \ssmin \Delta \atop \beta + \gamma = \alpha}
         N_{\beta,\gamma} \, (H_k \,, F_\beta) \, F_\gamma \, \Bigr)~,
\end{eqnarray*}
where in the last step, we have used eqns~(\ref{eq:LAAC6}) and~%
(\ref{eq:LAAC5}). Again, this expression will vanish provided we
impose the condition~(\ref{eq:LAEAC2}). Note that in the degenerate
case, the same argument works, but eqn~(\ref{eq:LAEAC2}) is not needed.
\hspace*{\fill} \raisebox{-2mm}{$\Box$} \vspace{4mm}

Having concluded the proof of Proposition~1, we pass to analyzing the
implications of the algebraic constraints that we have derived. As it
turns out, the conditions stated in Proposition~1 are sufficiently
strong to allow for a complete classification of all possible solutions.
As a by-product, we shall be able to reduce eqn~(\ref{eq:LAAC5}) to the
form given in eqn~(\ref{eq:LAAC7}).

A first step in this direction is taken by observing that the signs
$\epsilon_\alpha$ that appear in eqn~(\ref{eq:LAAC5}) may without loss
of generality be assumed to be independent of $\alpha$:
\begin{equation} \label{eq:LAIAC13}
 \epsilon_\alpha~=~\epsilon \qquad \mbox{for all $\, \alpha \smin \Delta$}~.
\end{equation}
This freedom of choice follows from the possibility of performing a
transformation that changes the signs of the root generators without
modifying any of the relations between generators and structure constants
used in the preceding calculations: it is given by
\begin{eqnarray*}
 &E_\alpha~~\longrightarrow~~
  E_\alpha^\prime~=~\epsilon \, \epsilon_\alpha \, E_\alpha~,& \\[2mm]
 &H_\alpha~~\longrightarrow~~H_\alpha^\prime~=~H_\alpha~,& \\[2mm]
 &N_{\alpha,\beta}~~\longrightarrow~~
  N_{\alpha,\beta}^\prime~
  =~{\displaystyle \frac{\epsilon \, \epsilon_\alpha \epsilon_\beta}
                        {\epsilon_{\alpha+\beta}}} \; N_{\alpha,\beta}~,\\[1mm]
 &F_\alpha~~\longrightarrow~~
  F_\alpha^\prime~=~\epsilon \, \epsilon_\alpha \, F_\alpha~.&
\end{eqnarray*}
and, omitting the primes, brings eqn~(\ref{eq:LAAC5}) into the form
\begin{equation} \label{eq:LAAC8}
 F_\alpha^-~=~\frac{\epsilon}{\sqrt{2} \, |\alpha|} \, H_\alpha^{}~.
\end{equation}
Next, let us write down the system obtained from eqn~(\ref{eq:LAAC6}) upon
replacing $\alpha$ by $-\alpha$ and $\beta$ by $-\beta$:
\begin{equation} \label{eq:LAIAC14}
 \begin{array}{c}
  \alpha(F_\beta) \, F_\alpha \, - \, \beta(F_\alpha) \, F_\beta~
  =~N_{\alpha,\beta} \, F_{\alpha+\beta}~, \\[1mm]
  \alpha(F_{-\beta}) \, F_\alpha \, + \, \beta(F_\alpha) \, F_{-\beta}~
  =~N_{\alpha,-\beta} \, F_{\alpha-\beta}~, \\[1mm]
  - \, \alpha(F_\beta) \, F_{-\alpha} \, - \, \beta(F_{-\alpha}) \, F_\beta~
  =~N_{-\alpha,\beta} \, F_{-\alpha+\beta}~, \\[1mm]
  - \, \alpha(F_{-\beta}) \, F_{-\alpha} \,
  + \, \beta(F_{-\alpha}) \, F_{-\beta}~
  =~N_{-\alpha,-\beta} \, F_{-\alpha-\beta}~.
 \end{array}
\end{equation}
Adding these four equations gives
\[
 \alpha(F_\beta^+) \, F_\alpha^- \, - \, \beta(F_\alpha^+) \, F_\beta^-~
 =~{\textstyle \frac{1}{2}} \, N_{\alpha,\beta} \, F_{\alpha+\beta}^- \, + \,
   {\textstyle \frac{1}{2}} \, N_{\alpha,-\beta} \, F_{\alpha-\beta}^-~.
\]
Inserting eqn~(\ref{eq:LAAC5}) and separating the coefficients of
$H_\alpha$ and $H_\beta$, we conclude that
\begin{equation} \label{eq:LAIAC15}
 \frac{1}{|\beta|} \, 2 \>\! \beta(F_\alpha^+)~
 =~\frac{1}{|\alpha-\beta|} \, N_{\alpha,-\beta} \, - \,
   \frac{1}{|\alpha+\beta|} \, N_{\alpha,\beta}~,
\end{equation}
plus the same equation with $\alpha$ and $\beta$ interchanged. It is to be
noted that this derivation is only valid when $\, \beta \neq \pm \alpha$,
as stated in eqn~(\ref{eq:LAAC6}): this supplementary condition is also
needed in order to guarantee that $H_\alpha$ and $H_\beta$ are linearly
independent but can in fact be eliminated from eqn~(\ref{eq:LAIAC15})
since this formula is automatically satisfied when $\, \beta = \pm \alpha$.
(Indeed, for $\, \beta = \pm \alpha$ the rhs is understood to vanish since
$2 \>\! \alpha$ and $0$ do not belong to the root system~$\Delta$, whereas
the lhs vanishes as a consequence of eqn~(\ref{eq:LAAC4}).)

The algebraic equation (\ref{eq:LAIAC15}) is identical with a special case
of eqn~(42) of Ref.~\cite{FW1}, obtained by replacing the coupling constants
$\mbox{\sl g}_\gamma$ by $\, 1 / |\gamma|$. As has been shown in Sect.~2.2
of Ref.~\cite{FW1}, there is only one type of simple complex Lie algebra
$\mathfrak{g}$ for which there exists a solution, namely those of the
$A$-series. In particular, $\mathfrak{g}$ is simply laced, and all its
roots have the same length, which according to our convention equals
$\sqrt{2}$, and eqn~(\ref{eq:LAAC8}) simplifies to the form given in
eqn~(\ref{eq:LAAC7}). Explicitly, if $\mathfrak{g} = \mathfrak{sl}%
(n,\mathbb{C}) \,$ with $\mathfrak{h}_{\mathbb{R}}$ consisting of
the real diagonal $(n \times n)$-matrices, we have $\, \Delta =
\{ \alpha_{ab} \, / \, 1 \leqslant a \neq b \leqslant n \} \,$
with $\, \alpha_{ab}(H) = H_{aa} - H_{bb} \,$ for $\, H \smin\,
\mathfrak{h}_{\mathbb{R}} \,$ and take $\, E_{\alpha_{ab}} = E_{ab} \,$
where $E_{ab}$ is the matrix whose entry in the $a^{\mathrm{th}}$ row and
$b^{\,\mathrm{th}}$ column is $1$ while all other entries are $0\,$; then
the structure constants $\, N_{ab,cd} = N_{\alpha_{ab},\alpha_{cd}} \,$
are given by
\begin{equation}
 N_{ab,cd}~=~\delta_{bc} - \delta_{ad}~,
\end{equation}
and writing $\, F_{ab}^\pm = F_{\alpha_{ab}}^\pm$, we have
\begin{equation} \label{eq:LASOL1}
 F_{ab}^+~= \; - \, \frac{1}{2} \, ( E_{aa} + E_{bb} ) \,
               + \, \frac{1}{n} \, \mathbf{1}_n~,
\end{equation}
and
\begin{equation} \label{eq:LASOL2}
 F_{ab}^-~=~\frac{\epsilon}{2} \, ( E_{aa} - E_{bb} )~,
\end{equation}
implying that for $\, \epsilon = +1$,
\begin{equation} \label{eq:LASOL3}
 F_{ab}^{}~= \; - \, E_{bb} \, + \, \frac{1}{n} \, \mathbf{1}_n~,
\end{equation}
while for $\, \epsilon = -1$,
\begin{equation} \label{eq:LASOL4}
 F_{ab}^{}~= \; - \, E_{aa} \, + \, \frac{1}{n} \, \mathbf{1}_n~.
\end{equation}
It is then easy to check that $F$, as defined by eqns~(\ref{eq:LASOL1})--%
(\ref{eq:LASOL4}), satisfies all the conditions stated in Proposition~1.
To see this, assume for simplicity that $\, \epsilon = +1 \,$ (noting that
the case $\, \epsilon = -1 \,$ can be obtained from this one by replacing
$F_\alpha$ by $F_{-\alpha}$, which does not affect the validity of any of
the equations (\ref{eq:LAAC6})--(\ref{eq:LAEAC2})). Then assuming, for
example, that $\, \alpha = \alpha_{ab} \,$ and $\, \beta = \alpha_{cd}$,
the lhs $\, \alpha_{ab}(F_{cd}) F_{ab} \, - \, \alpha_{cd}(F_{ab}) F_{cd} \,$
and rhs $\, N_{ab,cd} F_{\alpha_{ab}+\alpha_{cd}} \,$ of eqn~(\ref{eq:LAAC6})
are both equal to
\[
% \begin{array}{cc}
%  \mbox{} + \,
  \delta_{ad} (E_{bb} - {\displaystyle \frac{1}{n}} \, \mathbf{1}_n) \, - \,
  \delta_{bc} (E_{dd} - {\displaystyle \frac{1}{n}} \, \mathbf{1}_n)
%  & \mbox{if~$\epsilon = + 1 $}~, \\[4mm]
%  \mbox{} - \,
%  \delta_{bc} (E_{aa} - {\displaystyle \frac{1}{n}} \, \mathbf{1}_n) \, + \,
%  \delta_{ad} (E_{cc} - {\displaystyle \frac{1}{n}} \, \mathbf{1}_n)
%  & \mbox{if~$\epsilon = - 1 $}~,
% \end{array}
\]
except when $\, a=d \,$ and $\, b=c \,$ $(\beta = -\alpha)$, where the rhs
is understood to vanish while the lhs does not. Similarly, the formula
\[
 \sum_{a=1}^n \, (E_{aa} - {\displaystyle \frac{1}{n}} \, \mathbf{1}_n)~=~0
\]
allow us to verify eqns~(\ref{eq:LAEAC1}) and~(\ref{eq:LAEAC2}): the lhs of
eqn~(\ref{eq:LAEAC1}) becomes
\begin{eqnarray*}
\lefteqn{\sum_{1 \leqslant a \neq b \leqslant n} \, (E_{aa} - E_{bb}) \otimes
         (E_{bb} - {\displaystyle \frac{1}{n}} \, \mathbf{1}_n) \otimes
         (E_{aa} - {\displaystyle \frac{1}{n}} \, \mathbf{1}_n)}
                                                           \hspace{5mm} \\
 &=&\!\! \mbox{} - \, \sum_{a=1}^n \, E_{aa} \otimes
         (E_{aa} - {\displaystyle \frac{1}{n}} \, \mathbf{1}_n) \otimes
         (E_{aa} - {\displaystyle \frac{1}{n}} \, \mathbf{1}_n)         \\
 & &\!\! \mbox{} + \, \sum_{b=1}^n \, E_{b\>\!b\;\!} \otimes
         (E_{b\>\!b\,} - {\displaystyle \frac{1}{n}} \, \mathbf{1}_n) \otimes
         (E_{b\>\!b\,} - {\displaystyle \frac{1}{n}} \, \mathbf{1}_n)~,
\end{eqnarray*}

\pagebreak

\noindent
which vanishes as required, while that of eqn~(\ref{eq:LAEAC2}),
for $\, \alpha = \alpha_{ab}$, becomes
\vspace{1mm}
\begin{eqnarray*}
\lefteqn{\sum_{1 \leqslant c \leqslant n \atop c \neq a \,,\, c \neq b} \,
 ( F_{ac} \otimes F_{cb} \, - \, F_{cb} \otimes F_{ac} )}
                                                          \hspace{5mm} \\[-1mm]
 &=&\!\! \sum_{1 \leqslant c \leqslant n \atop c \neq a \,,\, c \neq b} \,
         (E_{cc} - {\displaystyle \frac{1}{n}} \, \mathbf{1}_n) \otimes
         (E_{bb} - {\displaystyle \frac{1}{n}} \, \mathbf{1}_n) \, - \,
         (E_{bb} - {\displaystyle \frac{1}{n}} \, \mathbf{1}_n) \otimes
         (E_{cc} - {\displaystyle \frac{1}{n}} \, \mathbf{1}_n)        \\[-1mm]
 &=&\!\! \mbox{} - \,
         (E_{aa} + E_{bb} - {\displaystyle \frac{2}{n}} \, \mathbf{1}_n)
         \otimes
         (E_{bb} - {\displaystyle \frac{1}{n}} \, \mathbf{1}_n) \, + \,
         (E_{bb} - {\displaystyle \frac{1}{n}} \, \mathbf{1}_n)
         \otimes
         (E_{aa} + E_{bb} - {\displaystyle \frac{2}{n}} \, \mathbf{1}_n)\\[2mm]
 &=&\!\! \mbox{} - \,
         (E_{aa} - E_{bb}) \otimes
         (E_{bb} - {\displaystyle \frac{1}{n}} \, \mathbf{1}_n) \, + \,
         (E_{bb} - {\displaystyle \frac{1}{n}} \, \mathbf{1}_n) \otimes
         (E_{aa} - E_{bb}) \\[4mm]
 &=&\!\! H_{ab} \otimes F_{ab} \, -  \, F_{ab} \otimes H_{ab}~,
\vspace{1mm}
\end{eqnarray*}
as required. In this way, we have rederived the main result of Refs~\cite{HW}
and~\cite{FP}, which states that the dynamical $R$-matrix of the integrable
Calogero model associated with the root system of the simple Lie algebra
$\, \mathfrak{g} = \mathfrak{sl}(n,\mathbb{C}) \,$ of the $A$-series can
be gauge transformed to a numerical $R$-matrix.

\section{Calogero Models for Symmetric Pairs}

According to Ref.~\cite{FW1}, the standard Lax matrix $L$ and the dynamical
$R$-matrix for the Calogero models associated with the root system $\Delta$
of a symmetric pair $(\mathfrak{g},\theta)$ read
\begin{equation} \label{eq:SPDLAXM}
 L~=~\sum_{j=1}^r \, p_j H_j \, + \,
     \sum_{\alpha \ssmin \tilde{\Delta}} i \, \mbox{\sl g}_\alpha \,
     w(\alpha(q)) \, E_\alpha~,
\end{equation}
\begin{equation} \label{eq:SPDRMAT}
 R~=~\sum_{\alpha \ssmin \tilde{\Delta}} w(\alpha(q)) \,
     K_\alpha \otimes E_\alpha \, + \,
     {\textstyle \frac{1}{2}} \,
     \sum_{\alpha \ssmin \tilde{\Delta}}
          \frac{w^\prime(\alpha(q))}{w(\alpha(q))} \,
     \Bigl( E_\alpha \otimes E_{-\alpha} \, + \,
            E_{\theta\alpha} \otimes E_{-\alpha} \Bigr)~,
\vspace{2mm}
\end{equation}
for the degenerate model and
\begin{equation} \label{eq:SPELAXM}
 L(u)~=~\sum_{j=1}^r \, p_j H_j \, + \,
        \sum_{\alpha \ssmin \tilde{\Delta}} i \, \mbox{\sl g}_\alpha \,
        \Phi(\alpha(q),u) \, E_\alpha~,
\vspace{-5mm}
\end{equation}
\begin{eqnarray} \label{eq:SPERMAT}
 R(u,v) \!\!
 &=&\!\! - \, {\textstyle {1 \over 2}} \, \sum_{j=1}^r \,
              (\zeta(u-v) + \zeta(u+v)) \, H_j \otimes H_j    \nonumber \\[1mm]
 & &\!\! - \, {\textstyle {1 \over 2}} \,
              (\zeta(u-v) - \zeta(u+v) + 2 \>\! \zeta(v)) \,
              C_{\mathfrak{z}}                                \nonumber \\[3mm]
 & &\!\! + \, \sum_{\alpha \ssmin \tilde{\Delta}}
              \Phi(\alpha(q),v) \, K_\alpha \otimes E_\alpha                 \\
 & &\!\! - \, {\textstyle {1 \over 2}} \, \sum_{\alpha \ssmin \tilde{\Delta}}
              \Bigl( \Phi(\alpha(q),u-v) \,
                     E_\alpha \otimes E_{-\alpha} \, + \,
                     \Phi(\alpha(q),-u-v) \,
                     E_{\theta\alpha} \otimes E_{-\alpha} \Bigr)~,
                                                        \quad \nonumber
\end{eqnarray}
for the elliptic model, where $\, \Delta = \Delta_0 \smcup \tilde{\Delta} \,$
with
\[
 \begin{array}{ccc}
  \Delta_0 \!\!&=&\!\! \{ \, \alpha \smin\, \Delta \, / \,
                             \theta\alpha = \alpha \, \}~, \\[2mm]
  \tilde{\Delta} \!\!&=&\!\! \{ \, \alpha \smin\, \Delta \, / \,
                                   \theta\alpha \neq \alpha \, \}~.
 \end{array}
\]
As has been shown in Ref.~\cite{FW1}, integrability requires the generators
$\, K_\alpha \smin\, \mathrm{i} \mathfrak{b}_0 \,$ appearing in eqns~%
(\ref{eq:SPDRMAT}) and~(\ref{eq:SPERMAT}) to satisfy the constraints
(\ref{eq:SPAC1}). Moreover, writing
\begin{equation} \label{eq:SPCOEF1}
 K_\alpha^\pm~=~{\textstyle \frac{1}{2}} \, (K_\alpha^{} \pm K_{-\alpha}^{})~,
\end{equation}
we also impose the condition
\begin{equation} \label{eq:SPAC3}
 \alpha(K_\alpha^+)~=~0 \qquad
 \mbox{for $\, \alpha \smin \tilde{\Delta} \,$ such that
  $\, \theta\alpha - \alpha \,\nsmin\, \Delta$}~,
\end{equation}
which follows from eqn~(\ref{eq:SPAC1}) by setting $\, \beta = - \alpha \,$
when $\, \mbox{\sl g}_\alpha \neq 0 \,$ but turns out to be true in general,
independent of this hypothesis.

Before proceeding with the calculations, we pause to note that the $R$-matrices
given by eqns~(\ref{eq:SPDRMAT}) and~(\ref{eq:SPERMAT}) have certain symmetry
properties with respect to the auto\-morphism $\theta$ that we want to be
preserved under the gauge transformation to $R^{\,\prime}(u,v)$: in the
degenerate case, $R$ takes values in $\, \mathfrak{k} \otimes \mathfrak{m} \,$
whereas in the elliptic case, $R(u,v)$ is even under the action of $\, \theta
\otimes 1 \,$ and odd under the action of $\, 1 \otimes \theta$ when these
are combined with a change of sign in the corresponding spectral parameter.
This can be achieved by imposing a restriction on the action of $\theta$
on $g$ or, equivalently, on the gauge potentials $A_j$: in the degenerate
case, $g$ should take values in~$K$ and the $A_j$ should take values in
$\mathfrak{k}$, whereas in the elliptic case, we require that
\begin{equation} \label{eq:SPGPOT1}
 \theta(g(-u))~=~g(u)~~,~~\theta(A_j(-u))~= \; A_j(u)~,
\end{equation}
or in terms of the components of the gauge potentials in the expansion
(\ref{eq:GPOT3}),
\begin{equation} \label{eq:SPGPOT2}
 \theta(A_j^{\mathfrak{h}}(-u))~= \; A_j^{\mathfrak{h}}(u)~~,~~
 A_j^{\theta\alpha}(-u)~= \; A_j^\alpha(u)~.
\end{equation}

In order to compute the content of eqns~(\ref{eq:GPOT2}) and~(\ref{eq:NRMAT}),
we further expand the Cartan part of the gauge potential according to
\begin{equation} \label{eq:SPGPOT3}
 A_j^{\mathfrak{h}}(u)~=~\sum_{k=1}^{r+s} A_j^k(u) \, H_k~.
\end{equation}
Then inserting eqns~(\ref{eq:SPERMAT}), (\ref{eq:GPOT3}) and
(\ref{eq:SPGPOT3}) into eqn~(\ref{eq:NRMAT}), we obtain
\vspace{2mm}
\begin{eqnarray*}
 0 \!\!&=&\!\!    \sum_{\alpha \ssmin \tilde{\Delta}}
                  \alpha_k \, \Phi^\prime(\alpha(q),v) \;
                  K_\alpha \otimes E_\alpha                             \\[1mm]
 & & \mbox{} - \, {\textstyle \frac{1}{2}} 
                  \sum_{\alpha \ssmin \tilde{\Delta}}
                  \alpha_k \, \Phi^\prime(\alpha(q),u-v) \;
                  E_\alpha \otimes E_{-\alpha} \, - \,
                  {\textstyle \frac{1}{2}} 
                  \sum_{\alpha \ssmin \tilde{\Delta}}
                  \alpha_k \, \Phi^\prime(\alpha(q),-u-v) \;
                  E_{\theta\alpha} \otimes E_{-\alpha}                       \\
 & & \mbox{} - \, \sum_{j=1}^r \,
                  \partial_j^{\vphantom{\mathfrak{h}}}
                  A_k^{\mathfrak{h}}(q,u) \otimes H_j \, - \,
                  \sum_{j=1}^r \, \sum_{\alpha \ssmin \Delta}
                  \partial_j^{\vphantom{\mathfrak{h}}} A_k^\alpha(q,u) \;
                  E_\alpha \otimes H_j                                       \\
 & & \mbox{} + \, {\textstyle \frac{1}{2}} \,
                  \sum_{j=1}^r \, \sum_{\alpha \ssmin \Delta} \,
                  \Bigl( \, \zeta(u-v) \,+\, \zeta(u+v) \Bigr) \,
                  A^\alpha_k(q,u) \;
                  [ \, H_j \otimes H_j \,,\, E_{\alpha} \otimes 1 \, ]       \\
 & & \mbox{} + \, {\textstyle \frac{1}{2}} \,
                  \sum_{j=1}^r \, \sum_{\alpha \ssmin \Delta} \,
                  \Bigl( \, \zeta(u-v) \,+\, \zeta(u+v) \Bigr) \,
                  A^\alpha_k(q,v) \;
                  [ \, H_j \otimes H_j \,,\, 1 \otimes E_{\alpha} \, ]       \\
 & & \mbox{} + \, {\textstyle \frac{1}{2}} \,
                  \sum_{j=1}^{r+s} \,
                  \Bigl( \, \zeta(u-v) \,-\, \zeta(u+v) \,+\,
                            2 \>\! \zeta(v) \Bigr) \,
                  A^j_k(q,u) \;
                  [ \, C_{\mathfrak{z}} \,,\, H_j \otimes 1 \, ]       \\[-1mm]
 & & \mbox{} + \, {\textstyle \frac{1}{2}} \,
                  \sum_{j=1}^{r+s} \,
                  \Bigl( \, \zeta(u-v) \,-\, \zeta(u+v) \,+\,
                            2 \>\! \zeta(v) \Bigr) \,
                  A^j_k(q,v) \;
                  [ \, C_{\mathfrak{z}} \,,\, 1 \otimes H_j \, ]        \\[1mm]
 & & \mbox{} + \, {\textstyle \frac{1}{2}}
                  \sum_{\alpha \ssmin \Delta} \,
                  \Bigl( \, \zeta(u-v) \,-\, \zeta(u+v) \,+\,
                            2 \>\! \zeta(v) \Bigr) \,
                  A^\alpha_k(q,u) \;
                  [ \, C_{\mathfrak{z}} \,,\, E_{\alpha} \otimes 1 \, ] \\[1mm]
 & & \mbox{} + \, {\textstyle \frac{1}{2}}
                  \sum_{\alpha \ssmin \Delta} \,
                  \Bigl( \, \zeta(u-v) \,-\, \zeta(u+v) \,+\,
                            2 \>\! \zeta(v) \Bigr) \,
                  A^\alpha_k(q,v) \;
                  [ \, C_{\mathfrak{z}} \,,\, 1 \otimes E_{\alpha} \, ]      \\
 & & \mbox{} - \, \sum_{j=1}^{r+s} \, \sum_{\gamma \ssmin \tilde{\Delta}}
                  \Phi(\gamma(q),v) \, A_k^j(q,u) \;
                  [ \, K_\gamma \otimes E_\gamma \,,\, H_j \otimes 1 \, ]    \\
 & & \mbox{} - \, \sum_{j=1}^{r+s} \, \sum_{\gamma \ssmin \tilde{\Delta}}
                  \Phi(\gamma(q),v) \, A_k^j(q,v) \;
                  [ \, K_\gamma \otimes E_\gamma \,,\, 1 \otimes H_j \, ]
                                                                        \\[1mm]
 & & \mbox{} - \, \sum_{\gamma \ssmin \tilde{\Delta} \atop
                        \delta \ssmin \Delta}
                  \Phi(\gamma(q),v) \, A_k^\delta(q,u) \;
                  [ \, K_\gamma \otimes E_\gamma \,,\,
                       E_\delta \otimes 1 \, ]                               \\
 & & \mbox{} - \, \sum_{\gamma \ssmin \tilde{\Delta} \atop
                        \delta \ssmin \Delta}
                  \Phi(\gamma(q),v) \, A_k^\delta(q,v) \;
                  [ \, K_\gamma \otimes E_\gamma \,,\,
                       1 \otimes E_\delta \, ]                               \\
 & & \mbox{} + \, {\textstyle \frac{1}{2}} \,
                  \sum_{j=1}^{r+s} \, \sum_{\gamma \ssmin \tilde{\Delta}}
                  \Phi(\gamma(q),u-v) \, A_k^j(q,u) \;
                  [ \, E_\gamma \otimes E_{-\gamma} \,,\, H_j \otimes 1 \, ] \\
 & & \mbox{} + \, {\textstyle \frac{1}{2}} \,
                  \sum_{j=1}^{r+s} \, \sum_{\gamma \ssmin \tilde{\Delta}}
                  \Phi(\gamma(q),u-v) \, A_k^j(q,v) \;
                  [ \, E_\gamma \otimes E_{-\gamma} \,,\, 1 \otimes H_j \, ] \\
 & & \mbox{} + \, {\textstyle \frac{1}{2}} \,
                  \sum_{j=1}^{r+s} \, \sum_{\gamma \ssmin \tilde{\Delta}}
                  \Phi(\gamma(q),-u-v) \, A_k^j(q,u) \;
                  [ \, E_{\theta\gamma} \otimes E_{-\gamma} \,,\,
                       H_j \otimes 1 \, ]                                    \\
 & & \mbox{} + \, {\textstyle \frac{1}{2}} \,
                  \sum_{j=1}^{r+s} \, \sum_{\gamma \ssmin \tilde{\Delta}}
                  \Phi(\gamma(q),-u-v) \, A_k^j(q,v) \;
                  [ \, E_{\theta\gamma} \otimes E_{-\gamma} \,,\,
                       1 \otimes H_j \, ]                               \\[2mm]
 & & \mbox{} + \, {\textstyle \frac{1}{2}} \,
                  \sum_{\gamma \ssmin \tilde{\Delta} \atop
                        \delta \ssmin \Delta}
                  \Phi(\gamma(q),u-v) \, A_k^\delta(q,u) \;
                  [ \, E_\gamma \otimes E_{-\gamma} \,,\,
                       E_\delta \otimes 1 \, ]                               \\
 & & \mbox{} + \, {\textstyle \frac{1}{2}} \,
                  \sum_{\gamma \ssmin \tilde{\Delta} \atop
                        \delta \ssmin \Delta}
                  \Phi(\gamma(q),u-v) \, A_k^\delta(q,v) \;
                  [ \, E_\gamma \otimes E_{-\gamma} \,,\,
                       1 \otimes E_\delta \, ]                               \\
 & & \mbox{} + \, {\textstyle \frac{1}{2}} \,
                  \sum_{\gamma \ssmin \tilde{\Delta} \atop
                        \delta \ssmin \Delta}
                  \Phi(\gamma(q),-u-v) \, A_k^\delta(q,u) \;
                  [ \, E_{\theta\gamma} \otimes E_{-\gamma} \,,\,
                       E_\delta \otimes 1 \, ]                               \\
 & & \mbox{} + \, {\textstyle \frac{1}{2}} \,
                  \sum_{\gamma \ssmin \tilde{\Delta} \atop
                        \delta \ssmin \Delta}
                  \Phi(\gamma(q),-u-v) \, A_k^\delta(q,v) \;
                  [ \, E_{\theta\gamma} \otimes E_{-\gamma} \,,\,
                       1 \otimes E_\delta \, ]                               \\
 & & \mbox{} + \, \sum_{j=1}^r \, \sum_{\alpha \ssmin \Delta}
                  \alpha_j^{} A_k^\alpha(q,v) \;
                  A_j^{\mathfrak{h}}(q,u) \otimes E_\alpha \, + \,
                  \sum_{j=1}^r \, \sum_{\alpha,\beta \ssmin \Delta}
                  \beta_j^{} A_j^\alpha(q,u) A_k^\beta(q,v) \;
                  E_\alpha \otimes E_\beta~.
\end{eqnarray*}

\noindent
Using the definition of $C_{\mathfrak{z}}\,$,
\begin{equation} \label{eq:CASIM}
 C_{\mathfrak{z}}~
 =~\sum_{j=r+1}^{r+s} H_j \otimes H_j \, + \,
   \sum_{\alpha \ssmin \Delta_0} E_\alpha \otimes E_{-\alpha}~,
\end{equation}
and eqns~(\ref{eq:COMR1})--(\ref{eq:COMR5}) to carry out the commutators,
together with the relation
\begin{equation} \label{eq:SPCSA}
 \sum_{j=1}^{r+s} \alpha_j \>\! H_j~=~H_{\alpha}~~,~~
 \sum_{j=1}^r \alpha_j \>\! H_j~=~(H_{\alpha})_{\mathfrak{a}}~~,~~
 \sum_{j=r+1}^{r+s} \alpha_j \>\! H_j~=~(H_{\alpha})_{\mathfrak{b}}~,
\end{equation}
we arrive at
\vspace{2mm}
\begin{eqnarray*}
 0 \!\!&=&\!\!    \sum_{\alpha \ssmin \tilde{\Delta}}
                  \alpha_k \, \Phi^\prime(\alpha(q),v) \;
                  K_\alpha \otimes E_\alpha                             \\[2mm]
 & & \mbox{} - \, {\textstyle \frac{1}{2}} 
                  \sum_{\alpha \ssmin \tilde{\Delta}}
                  \alpha_k \, \Phi^\prime(\alpha(q),u-v) \;
                  E_\alpha \otimes E_{-\alpha} \, - \,
                  {\textstyle \frac{1}{2}} 
                  \sum_{\alpha \ssmin \tilde{\Delta}}
                  \alpha_k \, \Phi^\prime(\alpha(q),-u-v) \;
                  \theta E_\alpha \otimes E_{-\alpha}                        \\
 & & \mbox{} - \, \sum_{j=1}^r \,
                  \partial_j^{\vphantom{\mathfrak{h}}}
                  A_k^{\mathfrak{h}}(q,u) \otimes H_j \, - \,
                  \sum_{j=1}^r \, \sum_{\alpha \ssmin \Delta}
                  \partial_j^{\vphantom{\mathfrak{h}}} A_k^\alpha(q,u) \;
                  E_\alpha \otimes H_j                                       \\
 & & \mbox{} + \, {\textstyle \frac{1}{2}} \,
                  \sum_{j=1}^r \, \sum_{\alpha \ssmin \tilde{\Delta}} \,
                  \alpha_j \, \Bigl( \, \zeta(u-v) \,+\, \zeta(u+v) \Bigr) \,
                  A^\alpha_k(q,u) \;
                  E_{\alpha} \otimes H_j                                     \\
 & & \mbox{} + \, {\textstyle \frac{1}{2}} \,
                  \sum_{j=1}^r \, \sum_{\alpha \ssmin \tilde{\Delta}} \,
                  \alpha_j \, \Bigl( \, \zeta(u-v) \,+\, \zeta(u+v) \Bigr) \,
                  A^\alpha_k(q,v) \;
                  H_j \otimes E_\alpha                                       \\
 & & \mbox{} - \, {\textstyle \frac{1}{2}} \,
                  \sum_{j=1}^{r+s} \, \sum_{\alpha \smin \Delta_0} \,
                  \alpha_j \, \Bigl( \, \zeta(u-v) \,-\, \zeta(u+v) \,+\,
                                        2 \>\! \zeta(v) \Bigr) \,
                  A^j_k(q,u) \;
                  E_\alpha \otimes E_{-\alpha}                               \\
 & & \mbox{} + \, {\textstyle \frac{1}{2}} \,
                  \sum_{j=1}^{r+s} \, \sum_{\alpha \smin \Delta_0} \,
                  \alpha_j \, \Bigl( \, \zeta(u-v) \,-\, \zeta(u+v) \,+\,
                                        2 \>\! \zeta(v) \Bigr) \,
                  A^j_k(q,v) \;
                  E_\alpha \otimes E_{-\alpha}                               \\
 & & \mbox{} + \, {\textstyle \frac{1}{2}} \,
                  \sum_{j=r+1}^{r+s} \, \sum_{\alpha \ssmin \Delta} \,
                  \alpha_j \, \Bigl( \, \zeta(u-v) \,-\, \zeta(u+v) \,+\,
                                        2 \>\! \zeta(v) \Bigr) \,
                  A^\alpha_k(q,u) \;
                  E_{\alpha} \otimes H_j                                     \\
 & & \mbox{} + \, {\textstyle \frac{1}{2}} \,
                  \sum_{j=r+1}^{r+s} \, \sum_{\alpha \ssmin \Delta} \,
                  \alpha_j \, \Bigl( \, \zeta(u-v) \,-\, \zeta(u+v) \,+\,
                                        2 \>\! \zeta(v) \Bigr) \,
                  A^\alpha_k(q,v) \;
                  H_j \otimes E_\alpha                                  \\[2mm]
 & & \mbox{} + \, {\textstyle \frac{1}{2}}
                  \sum_{\gamma \ssmin \Delta_0} \,
                  \Bigl( \, \zeta(u-v) \,-\, \zeta(u+v) \,+\,
                            2 \>\! \zeta(v) \Bigr) \,
                  A_k^{-\gamma}(q,u) \; H_\gamma \otimes E_{-\gamma}    \\[2mm]
 & & \mbox{} + \, {\textstyle \frac{1}{2}}
                  \sum_{\gamma \ssmin \Delta_0} \,
                  \Bigl( \, \zeta(u-v) \,-\, \zeta(u+v) \,+\,
                            2 \>\! \zeta(v) \Bigr) \,
                  A_k^\gamma(q,v) \; E_\gamma \otimes H_{-\gamma}       \\[2mm]
 & & \mbox{} + \, {\textstyle \frac{1}{2}}
                  \sum_{\alpha \ssmin \Delta \,, \gamma \ssmin \Delta_0 \atop
                        \alpha+\gamma \ssmin \Delta} N_{\gamma,\alpha}^{} \,
                  \Bigl( \, \zeta(u-v) \,-\, \zeta(u+v) \,+\,
                            2 \>\! \zeta(v) \Bigr) \,
                  A^\alpha_k(q,u) \;
                  E_{\gamma+\alpha} \otimes E_{-\gamma}                      \\
 & & \mbox{} + \, {\textstyle \frac{1}{2}}
                  \sum_{\alpha \ssmin \Delta \,, \gamma \ssmin \Delta_0 \atop
                        \alpha-\gamma \ssmin \Delta} N_{-\gamma,\alpha}^{} \,
                  \Bigl( \, \zeta(u-v) \,-\, \zeta(u+v) \,+\,
                            2 \>\! \zeta(v) \Bigr) \,
                  A^\alpha_k(q,v) \;
                  E_\gamma \otimes E_{-\gamma+\alpha}                        \\
 & & \mbox{} + \, \sum_{j=1}^{r+s} \, \sum_{\gamma \ssmin \tilde{\Delta}}
                  \gamma_j \, \Phi(\gamma(q),v) \, A_k^j(q,v) \;
                  K_\gamma \otimes E_\gamma                                  \\
 & & \mbox{} - \, \sum_{\gamma \ssmin \tilde{\Delta} \atop
                        \delta \ssmin \Delta}
                  \delta(K_\gamma^{}) \, \Phi(\gamma(q),v) \,
                  A_k^\delta(q,u) \; E_\delta \otimes E_\gamma               \\
 & & \mbox{} - \, \sum_{\gamma \ssmin \tilde{\Delta}}
                  \Phi(\gamma(q),v) \, A_k^{-\gamma}(q,v) \;
                  K_\gamma \otimes H_\gamma                                  \\
 & & \mbox{} - \, \sum_{\gamma \ssmin \tilde{\Delta} \,,\,
                        \delta \ssmin \Delta \atop
                        \gamma+\delta \ssmin \Delta}
                  N_{\gamma,\delta} \, \Phi(\gamma(q),v) \, A_k^\delta(q,v) \;
                  K_\gamma \otimes E_{\gamma+\delta}                         \\
 & & \mbox{} - \, {\textstyle \frac{1}{2}} \,
                  \sum_{j=1}^{r+s} \, \sum_{\gamma \ssmin \tilde{\Delta}}
                  \gamma_j \, \Phi(\gamma(q),u-v) \, A_k^j(q,u) \;
                  E_\gamma \otimes E_{-\gamma}                               \\
 & & \mbox{} + \, {\textstyle \frac{1}{2}} \,
                  \sum_{j=1}^{r+s} \, \sum_{\gamma \ssmin \tilde{\Delta}}
                  \gamma_j \, \Phi(\gamma(q),u-v) \, A_k^j(q,v) \;
                  E_\gamma \otimes E_{-\gamma}                               \\
 & & \mbox{} - \, {\textstyle \frac{1}{2}} \,
                  \sum_{j=1}^{r+s} \, \sum_{\gamma \ssmin \tilde{\Delta}}
                  (\theta\gamma)_j \, \Phi(\gamma(q),-u-v) \, A_k^j(q,u) \;
                  \theta E_\gamma \otimes E_{-\gamma}                        \\
 & & \mbox{} + \, {\textstyle \frac{1}{2}} \,
                  \sum_{j=1}^{r+s} \, \sum_{\gamma \ssmin \tilde{\Delta}}
                  \gamma_j \, \Phi(\gamma(q),-u-v) \, A_k^j(q,v) \;
                  \theta E_\gamma \otimes E_{-\gamma}                   \\[1mm]
 & & \mbox{} + \, {\textstyle \frac{1}{2}} \,
                  \sum_{\gamma \ssmin \tilde{\Delta}}
                  \Phi(\gamma(q),u-v) \, A_k^{-\gamma}(q,u) \;
                  H_\gamma \otimes E_{-\gamma}                               \\
 & & \mbox{} + \, {\textstyle \frac{1}{2}} \,
                  \sum_{\gamma \ssmin \tilde{\Delta} \,,\,
                        \delta \ssmin \Delta \atop
                        \gamma+\delta \ssmin \Delta}
                  N_{\gamma,\delta} \, \Phi(\gamma(q),u-v) \,
                  A_k^\delta(q,u) \; E_{\gamma+\delta} \otimes E_{-\gamma}   \\
 & & \mbox{} + \, {\textstyle \frac{1}{2}} \,
                  \sum_{\gamma \ssmin \tilde{\Delta}}
                  \Phi(\gamma(q),u-v) \, A_k^\gamma(q,v) \;
                  E_\gamma \otimes H_{-\gamma}                               \\
 & & \mbox{} + \, {\textstyle \frac{1}{2}} \,
                  \sum_{\gamma \ssmin \tilde{\Delta} \,,\,
                        \delta \ssmin \Delta \atop
                        \gamma-\delta \ssmin \Delta}
                  N_{-\gamma,\delta}^{} \, \Phi(\gamma(q),u-v) \,
                  A_k^\delta(q,v) \; E_\gamma \otimes E_{-\gamma+\delta}     \\
 & & \mbox{} + \, {\textstyle \frac{1}{2}} \,
                  \sum_{\gamma \ssmin \tilde{\Delta}}
                  \Phi(\gamma(q),-u-v) \, A_k^{-\theta\gamma}(q,u) \;
                  \theta H_\gamma \otimes E_{-\gamma}                        \\
 & & \mbox{} + \, {\textstyle \frac{1}{2}} \,
                  \sum_{\gamma \ssmin \tilde{\Delta} \,,\,
                        \delta \ssmin \Delta \atop
                        \theta\gamma+\delta \ssmin \Delta}
                  N_{\theta\gamma,\delta}^{} \, \Phi(\gamma(q),-u-v) \,
                  A_k^\delta(q,u) \;
                  E_{\theta\gamma+\delta} \otimes E_{-\gamma}                \\
 & & \mbox{} + \, {\textstyle \frac{1}{2}} \,
                  \sum_{\gamma \ssmin \tilde{\Delta}}
                  \Phi(\gamma(q),-u-v) \, A_k^\gamma(q,v) \;
                  E_{\theta\gamma} \otimes H_{-\gamma}                       \\
 & & \mbox{} + \, {\textstyle \frac{1}{2}} \,
                  \sum_{\gamma \ssmin \tilde{\Delta} \,,\,
                        \delta \ssmin \Delta \atop
                        \gamma-\delta \ssmin \Delta}
                  N_{-\gamma,\delta}^{} \, \Phi(\gamma(q),-u-v) \,
                  A_k^\delta(q,v) \;
                  E_{\theta\gamma} \otimes E_{-\gamma+\delta}                \\
 & & \mbox{} + \, \sum_{j=1}^r \, \sum_{\alpha \ssmin \Delta}
                  \alpha_j^{} A_k^\alpha(q,v) \;
                  A_j^{\mathfrak{h}}(q,u) \otimes E_\alpha \, + \,
                  \sum_{j=1}^r \, \sum_{\alpha,\beta \ssmin \Delta}
                  \beta_j^{} A_j^\alpha(q,u) A_k^\beta(q,v) \;
                  E_\alpha \otimes E_\beta~.
\end{eqnarray*}

\noindent
Relabelling summation indices and using cyclicity and antisymmetry of the
structure constants, we can bring this expression into the following form:
\vspace{4mm}
\begin{eqnarray*}
 0 \!\!&=&\!\!    \sum_{\alpha \ssmin \tilde{\Delta}}
                  \alpha_k \, \Phi^\prime(\alpha(q),v) \;
                  K_\alpha \otimes E_\alpha                             \\[2mm]
 & & \mbox{} - \, {\textstyle \frac{1}{2}} 
                  \sum_{\alpha \ssmin \tilde{\Delta}}
                  \alpha_k \, \Phi^\prime(\alpha(q),u-v) \;
                  E_\alpha \otimes E_{-\alpha} \, - \,
                  {\textstyle \frac{1}{2}} 
                  \sum_{\alpha \ssmin \tilde{\Delta}}
                  \alpha_k \, \Phi^\prime(\alpha(q),-u-v) \;
                  \theta E_\alpha \otimes E_{-\alpha}                        \\
 & & \mbox{} - \, \sum_{j=1}^r \,
                  \partial_j^{\vphantom{\mathfrak{h}}}
                  A_k^{\mathfrak{h}}(q,u) \otimes H_j \, - \,
                  \sum_{j=1}^r \, \sum_{\alpha \ssmin \Delta}
                  \partial_j^{\vphantom{\mathfrak{h}}} A_k^\alpha(q,u) \;
                  E_\alpha \otimes H_j                                       \\
 & & \mbox{} + \, {\textstyle \frac{1}{2}} \,
                  \sum_{j=1}^r \, \sum_{\alpha \ssmin \tilde{\Delta}} \,
                  \alpha_j \, \Bigl( \, \zeta(u-v) \,+\, \zeta(u+v) \Bigr) \,
                  A_k^\alpha(q,u) \;
                  E_{\alpha} \otimes H_j                                     \\
 & & \mbox{} + \, {\textstyle \frac{1}{2}} \,
                  \sum_{j=1}^r \, \sum_{\alpha \ssmin \tilde{\Delta}} \,
                  \alpha_j \, \Bigl( \, \zeta(u-v) \,+\, \zeta(u+v) \Bigr) \,
                  A_k^\alpha(q,v) \;
                  H_j \otimes E_\alpha                                       \\
 & & \mbox{} - \, {\textstyle \frac{1}{2}} \,
                  \sum_{j=1}^{r+s} \, \sum_{\alpha \smin \Delta_0} \,
                  \alpha_j \, \Bigl( \, \zeta(u-v) \,-\, \zeta(u+v) \,+\,
                                        2 \>\! \zeta(v) \Bigr) \,
                  A^j_k(q,u) \;
                  E_\alpha \otimes E_{-\alpha}                               \\
 & & \mbox{} + \, {\textstyle \frac{1}{2}} \,
                  \sum_{j=1}^{r+s} \, \sum_{\alpha \smin \Delta_0} \,
                  \alpha_j \, \Bigl( \, \zeta(u-v) \,-\, \zeta(u+v) \,+\,
                                        2 \>\! \zeta(v) \Bigr) \,
                  A^j_k(q,v) \;
                  E_\alpha \otimes E_{-\alpha}                               \\
 & & \mbox{} + \, {\textstyle \frac{1}{2}} \,
                  \sum_{j=r+1}^{r+s} \, \sum_{\alpha \ssmin \Delta} \,
                  \alpha_j \, \Bigl( \, \zeta(u-v) \,-\, \zeta(u+v) \,+\,
                                        2 \>\! \zeta(v) \Bigr) \,
                  A_k^\alpha(q,u) \;
                  E_{\alpha} \otimes H_j                                     \\
 & & \mbox{} + \, {\textstyle \frac{1}{2}} \,
                  \sum_{j=r+1}^{r+s} \, \sum_{\alpha \ssmin \Delta} \,
                  \alpha_j \, \Bigl( \, \zeta(u-v) \,-\, \zeta(u+v) \,+\,
                                        2 \>\! \zeta(v) \Bigr) \,
                  A_k^\alpha(q,v) \;
                  H_j \otimes E_\alpha                                  \\[2mm]
 & & \mbox{} - \, {\textstyle \frac{1}{2}}
                  \sum_{\alpha \ssmin \Delta_0} \,
                  \Bigl( \, \zeta(u-v) \,-\, \zeta(u+v) \,+\,
                            2 \>\! \zeta(v) \Bigr) \,
                  A_k^\alpha(q,u) \; H_\alpha \otimes E_\alpha          \\[2mm]
 & & \mbox{} - \, {\textstyle \frac{1}{2}}
                  \sum_{\alpha \ssmin \Delta_0} \,
                  \Bigl( \, \zeta(u-v) \,-\, \zeta(u+v) \,+\,
                            2 \>\! \zeta(v) \Bigr) \,
                  A_k^\alpha(q,v) \; E_\alpha \otimes H_\alpha          \\[2mm]
 & & \mbox{} - \, {\textstyle \frac{1}{2}}
                  \sum_{\alpha \ssmin \Delta \,,\, \beta \ssmin \Delta_0 \atop
                        \alpha+\beta \ssmin \Delta} N_{\alpha,\beta}^{} \,
                  \Bigl( \, \zeta(u-v) \,-\, \zeta(u+v) \,+\,
                            2 \>\! \zeta(v) \Bigr) \,
                  A_k^{\alpha+\beta}(q,u) \;
                  E_\alpha \otimes E_\beta                              \\[1mm]
 & & \mbox{} + \, {\textstyle \frac{1}{2}}
                  \sum_{\alpha \ssmin \Delta_0 \,,\, \beta \ssmin \Delta \atop
                        \alpha+\beta \ssmin \Delta} N_{\alpha,\beta}^{} \,
                  \Bigl( \, \zeta(u-v) \,-\, \zeta(u+v) \,+\,
                            2 \>\! \zeta(v) \Bigr) \,
                  A_k^{\alpha+\beta}(q,v) \;
                  E_\alpha \otimes E_\beta                                   \\
 & & \mbox{} + \, \sum_{j=1}^{r+s} \, \sum_{\alpha \ssmin \tilde{\Delta}}
                  \alpha_j \, \Phi(\alpha(q),v) \, A_k^j(q,v) \;
                  K_\alpha \otimes E_\alpha                             \\[1mm]
 & & \mbox{} - \, \sum_{\alpha \ssmin \Delta \atop
                        \beta \ssmin \tilde{\Delta}}
                  \alpha(K_\beta^{}) \, \Phi(\beta(q),v) \,
                  A_k^\alpha(q,u) \; E_\alpha \otimes E_\beta                \\
 & & \mbox{} - \, \sum_{\alpha \ssmin \tilde{\Delta}}
                  \Phi(\alpha(q),v) \, A_k^{-\alpha}(q,v) \;
                  K_\alpha \otimes H_\alpha                                  \\
 & & \mbox{} - \, \sum_{\gamma \ssmin \tilde{\Delta} \,,\,
                        \delta \ssmin \Delta \atop
                        \gamma+\delta \ssmin \Delta}
                  N_{\gamma,\delta} \, \Phi(\gamma(q),v) \, A_k^\delta(q,v) \;
                  K_\gamma \otimes E_{\gamma+\delta}                         \\
 & & \mbox{} - \, {\textstyle \frac{1}{2}} \,
                  \sum_{j=1}^{r+s} \, \sum_{\alpha \ssmin \tilde{\Delta}}
                  \alpha_j \, \Phi(\alpha(q),u-v) \, A_k^j(q,u) \;
                  E_\alpha \otimes E_{-\alpha}                               \\
 & & \mbox{} + \, {\textstyle \frac{1}{2}} \,
                  \sum_{j=1}^{r+s} \, \sum_{\alpha \ssmin \tilde{\Delta}}
                  \alpha_j \, \Phi(\alpha(q),u-v) \, A_k^j(q,v) \;
                  E_\alpha \otimes E_{-\alpha}                               \\
 & & \mbox{} - \, {\textstyle \frac{1}{2}} \,
                  \sum_{j=1}^{r+s} \, \sum_{\alpha \ssmin \tilde{\Delta}}
                  (\theta\alpha)_j \, \Phi(\alpha(q),-u-v) \, A_k^j(q,u) \;
                  \theta E_\alpha \otimes E_{-\alpha}                        \\
 & & \mbox{} + \, {\textstyle \frac{1}{2}} \,
                  \sum_{j=1}^{r+s} \, \sum_{\alpha \ssmin \tilde{\Delta}}
                  \alpha_j \, \Phi(\alpha(q),-u-v) \, A_k^j(q,v) \;
                  \theta E_\alpha \otimes E_{-\alpha}                   \\[2mm]
 & & \mbox{} - \, {\textstyle \frac{1}{2}} \,
                  \sum_{\alpha \ssmin \tilde{\Delta}}
                  \Phi(-\alpha(q),u-v) \, A_k^{\alpha}(q,u) \;
                  H_\alpha \otimes E_\alpha                                  \\
 & & \mbox{} - \, {\textstyle \frac{1}{2}} \,
                  \sum_{\alpha \ssmin \Delta \,,\,
                        \beta \ssmin \tilde{\Delta} \atop
                        \alpha+\beta \ssmin \Delta}
                  N_{\alpha,\beta} \, \Phi(-\beta(q),u-v) \,
                  A_k^{\alpha+\beta}(q,u) \;
                  E_\alpha \otimes E_\beta                                   \\
 & & \mbox{} - \, {\textstyle \frac{1}{2}} \,
                  \sum_{\alpha \ssmin \tilde{\Delta}}
                  \Phi(\alpha(q),u-v) \, A_k^\alpha(q,v) \;
                  E_\alpha \otimes H_\alpha                                  \\
 & & \mbox{} + \, {\textstyle \frac{1}{2}} \,
                  \sum_{\alpha \ssmin \tilde{\Delta} \,,\,
                        \beta \ssmin \Delta \atop
                        \alpha+\beta \ssmin \Delta}
                  N_{\alpha,\beta}^{} \, \Phi(\alpha(q),u-v) \,
                  A_k^{\alpha+\beta}(q,v) \;
                  E_\alpha \otimes E_\beta                                   \\
 & & \mbox{} - \, {\textstyle \frac{1}{2}} \,
                  \sum_{\alpha \ssmin \tilde{\Delta}}
                  \Phi(-\alpha(q),-u-v) \, A_k^\alpha(q,-u) \;
                  \theta H_\alpha \otimes E_\alpha                           \\
 & & \mbox{} - \, {\textstyle \frac{1}{2}} \,
                  \sum_{\alpha \ssmin \Delta  \,,\,
                        \beta \ssmin \tilde{\Delta} \atop
                        \alpha+\beta \ssmin \Delta}
                  N_{\alpha,\beta}^{} \, \Phi(-\beta(q),-u-v) \,
                  A_k^{\alpha+\beta}(q,-u) \;
                  \theta E_\alpha \otimes E_\beta                            \\
 & & \mbox{} - \, {\textstyle \frac{1}{2}} \,
                  \sum_{\alpha \ssmin \tilde{\Delta}}
                  \Phi(-\alpha(q),-u-v) \, A_k^\alpha(q,-v) \;
                  E_\alpha \otimes \theta H_\alpha                           \\
 & & \mbox{} + \, {\textstyle \frac{1}{2}} \,
                  \sum_{\alpha \ssmin \tilde{\Delta} \,,\,
                        \beta \ssmin \Delta \atop
                        \alpha+\beta \ssmin \Delta}
                  N_{\alpha,\beta}^{} \, \Phi(\alpha(q),-u-v) \,
                  A_k^{\alpha+\beta}(q,v) \;
                  \theta E_\alpha \otimes E_\beta                            \\
 & & \mbox{} + \, \sum_{j=1}^r \, \sum_{\alpha \ssmin \Delta}
                  \alpha_j^{} A_k^\alpha(q,v) \;
                  A_j^{\mathfrak{h}}(q,u) \otimes E_\alpha \, + \,
                  \sum_{j=1}^r \, \sum_{\alpha,\beta \ssmin \Delta}
                  \beta_j^{} A_j^\alpha(q,u) A_k^\beta(q,v) \;
                  E_\alpha \otimes E_\beta~.
\end{eqnarray*}

\noindent
Noting that for $\, \alpha \smin \Delta_0$, we have $\, \alpha_j = 0$ for
$1 \leqslant j \leqslant r \,$ and $\, (H_\alpha^{})_{\mathfrak{a}}^{} = 0$,
we can now identify the components of eqn~(\ref{eq:NRMAT}) along the various
subspaces of $\, \mathfrak{g} \otimes \mathfrak{g} \,$:
\vspace{3mm}
\begin{itemize}
 \item The components along $\, \mathfrak{h} \otimes H_j \,$ lead to the
       following system of equations. \\[2mm]
       For $\, 1 \leqslant j \leqslant r \,$:
       \begin{equation} \label{eq:SPEDCHH1}
        \partial_j^{} A_k^{\mathfrak{h}}(q,u) \, + \,
        \sum_{\alpha \ssmin \tilde{\Delta}}
        \alpha_j^{} \, \Phi(\alpha(q),v) \,
        A_k^{-\alpha}(q,v) \, K_\alpha^{}~=~0~.
       \end{equation}
       For $\, r+1 \leqslant j \leqslant r+s \,$:
       \begin{equation} \label{eq:SPEDCHH2}
        \sum_{\alpha \ssmin \tilde{\Delta}}
        \alpha_j^{} \, \Phi(\alpha(q),v) \,
        A_k^{-\alpha}(q,v) \, K_\alpha^{}~=~0~.
       \end{equation}

 \pagebreak

 \item The components along $\, \mathfrak{h} \otimes \mathfrak{g}_\alpha \,$
       lead to the following system of equations. \\[2mm]
       For $\, \alpha \smin\, \tilde{\Delta} \,$:
       \begin{equation} \label{eq:SPEDCHE1}
        \begin{array}{l}
         \alpha_k^{} \, \Phi^\prime(\alpha(q),v) \, K_\alpha^{} \, + \,
         {\textstyle \frac{1}{2}} \,
         \Bigl( \, \zeta(u-v) + \zeta(u+v) \Bigr) \,
         A_k^\alpha(q,v) \, (H_\alpha^{})_{\mathfrak{a}}^{} \\[3mm]
         \phantom{\alpha_k^{} \, \Phi^\prime(\alpha(q),v) \, K_\alpha^{} \,}
         + \, {\textstyle \frac{1}{2}} \,
         \Bigl( \, \zeta(u-v) - \zeta(u+v) + 2 \>\! \zeta(v) \Bigr) \,
         A_k^\alpha(q,v) \, (H_\alpha^{})_{\mathfrak{b}}^{} \\[6mm]
         \mbox{} + \, \Phi(\alpha(q),v) \,
                      \alpha(A_k^{\mathfrak{h}}(q,v)) \, K_\alpha^{} \, -
                      {\displaystyle \sum_{\gamma \ssmin \tilde{\Delta} ,
                                           \delta \ssmin \Delta \atop
                                           \gamma+\delta=\alpha}}
                      N_{\gamma,\delta}^{} \, \Phi(\gamma(q),v) \,
                      A_k^\delta(q,v) \, K_{\gamma}^{} \\[8mm]
         \mbox{} + \, {\textstyle \frac{1}{2}} \, \Phi(\alpha(q),v-u) \, 
                      A_k^\alpha(q,u) \, H_\alpha^{} \,
                 + \, {\textstyle \frac{1}{2}} \, \Phi(\alpha(q),v+u) \,
                      A_k^\alpha(q,-u) \, \theta H_\alpha^{} \\[4mm]
         \mbox{} + \, {\displaystyle \sum_{j=1}^r} \, \alpha_j^{} \,
                      A_k^\alpha(q,v) A_j^{\mathfrak{h}}(q,u)~
         =~0~.
        \end{array}
       \end{equation}
       For $\, \alpha \smin\, \Delta_0 \,$:
       \begin{equation} \label{eq:SPEDCHE2}
        \begin{array}{l}
         {\textstyle \frac{1}{2}} \,
         \Bigl( \, \zeta(u-v) - \zeta(u+v) + 2 \>\! \zeta(v) \Bigr)
         \Bigl( A_k^\alpha(q,v) \, - \, A_k^\alpha(q,u) \Bigr) \,
         H_\alpha^{} \\[4mm]
         \mbox{}~~ - {\displaystyle \sum_{\gamma,\delta \ssmin \tilde{\Delta}
                                          \atop \gamma+\delta=\alpha}}
           N_{\gamma,\delta}^{} \, \Phi(\gamma(q),v) \,
           A_k^\delta(q,v) \, K_{\gamma}^{}~=~0~.
        \end{array}
       \end{equation}
 \item The components along $\, \mathfrak{g}_\alpha \otimes H_j \,$ lead to
       the following system of equations. \\[2mm]
       For $\, \alpha \smin\, \tilde{\Delta} \,,\,
       1 \leqslant j \leqslant r \,$:
       \begin{equation} \label{eq:SPEDCEH1}
        \begin{array}{l}
         \partial_j^{} A_k^\alpha(q,u) \, - \,
         {\textstyle \frac{1}{2}} \, \alpha_j^{} \,
         \Bigl( \, \zeta(u-v) + \zeta(u+v) \Bigr) \, A_k^\alpha(q,u) \\[2mm]
         \mbox{} + \, {\textstyle \frac{1}{2}} \, \alpha_j^{} \,
         \Bigl( \, \Phi(\alpha(q),u-v) \, A_k^\alpha(q,v) \, + \,
                   \Phi(\alpha(q),u+v) \, A_k^\alpha(q,-v) \Bigr)~
         =~0~.
        \end{array}
       \end{equation}
       For $\, \alpha \smin\, \Delta_0$, $1 \leqslant j \leqslant r \,$:
       \begin{equation} \label{eq:SPEDCEH2}
        \partial_j^{} A_k^\alpha(q,u)~=~0~.
       \end{equation}
       For $\, \alpha \smin\, \tilde{\Delta} \,,\,
       r+1 \leqslant j \leqslant r+s \,$:
       \begin{equation} \label{eq:SPEDCEH3}
        \begin{array}{l}
         \Bigl( \, \zeta(u-v) - \zeta(u+v) + 2 \>\! \zeta(v) \Bigr) \,
         A_k^\alpha(q,u) \\[2mm]
         \mbox{} - \, \Bigl( \, \Phi(\alpha(q),u-v) \, A_k^\alpha(q,v) \, - \,
                                \Phi(\alpha(q),u+v) \, A_k^\alpha(q,-v) \Bigr)~
         =~0~.
        \end{array}
       \end{equation}
       For $\, \alpha \smin\, \Delta_0 \,,\,
       r+1 \leqslant j \leqslant r+s \,$:
       \begin{equation} \label{eq:SPEDCEH4}
        A_k^\alpha(q,u) \, - \, A_k^\alpha(q,v)~=~0~.
       \end{equation}

 \pagebreak

 \item The components along $\, \mathfrak{g}_\alpha \otimes
       \mathfrak{g}_\beta \,$ lead to the following system of
       equations. \\[2mm]
       For $\, \alpha \smin\, \Delta_0 \,$ and $\, \beta = -\alpha \,$:
       \begin{equation} \label{eq:SPEDCEE1}
        \alpha \left( A_k^{\mathfrak{h}}(q,u) \, - \,
                      A_k^{\mathfrak{h}}(q,v) \right) \,=~0~.
       \end{equation}
       For $\, \alpha,\beta \smin\, \Delta_0 \,$ with $\, \alpha + \beta
       \neq 0 \,$, no new condition arises,
       due to eqn~(\ref{eq:SPEDCEH4}). \\[1mm]
       For $\, \alpha \smin\, \Delta_0 \,,\, \beta \smin\, \tilde{\Delta} \,$:
       \begin{equation} \label{eq:SPEDCEE2}
        \begin{array}{l}
         {\textstyle \frac{1}{2}} \, N_{\alpha,\beta}^{} \,
         \Bigl( \, \zeta(u-v) \,-\, \zeta(u+v) \,+\,
                   2 \>\! \zeta(v) \Bigr) \, A_k^{\alpha+\beta}(q,v) \\[4mm]
         \mbox{} - \, \alpha(K_\beta) \, \Phi(\beta(q),v) \,
                      A_k^\alpha(q,u) \\[3mm]
         \mbox{} + \, {\textstyle \frac{1}{2}} \, N_{\alpha,\beta}^{} \,
                      \Bigl( \Phi(\beta(q),v-u) \,
                             A_k^{\alpha+\beta}(q,u) \, + \,
                             \Phi(\beta(q),v+u) \,
                             A_k^{\alpha+\beta}(q,-u) \Bigr) \\[2mm]
         \mbox{} + \, {\displaystyle \sum_{j=1}^r} \, \beta_j^{} \,
                      A_j^\alpha(q,u) A_k^\beta(q,v)~=~0~.
        \end{array}
       \end{equation}
       For $\, \alpha \smin\, \tilde{\Delta} \,,\, \beta \smin\, \Delta_0 \,$:
       \begin{equation} \label{eq:SPEDCEE3}
        \begin{array}{l}
         N_{\alpha,\beta}^{} \,
         \Bigl( \, \zeta(u-v) \,-\, \zeta(u+v) \,+\,
                   2 \>\! \zeta(v) \Bigr) \, A_k^{\alpha+\beta}(q,u) \\[4mm]
         \mbox{} - \, N_{\alpha,\beta}^{} \,
         \Bigl( \Phi(\alpha(q),u-v) \, A_k^{\alpha+\beta}(q,v) \, - \,
                \Phi(\alpha(q),u+v) \, A_k^{\alpha+\beta}(q,-v) \Bigr) \\[5mm]
         \mbox{} =~0~.
        \end{array}
       \end{equation}
       For $\, \alpha,\beta \smin\, \tilde{\Delta} \,$: \\[1mm]
       if $\, \beta = \alpha \neq - \theta\alpha \,$:
       \begin{equation} \label{eq:SPEDCEE4}
        \begin{array}{l}
         \alpha(K_\alpha^{}) \, \Phi(\alpha(q),v) \, A_k^\alpha(q,u) \\[4mm]
         \mbox{} - \, {\textstyle \frac{1}{2}} \, N_{\theta\alpha,\alpha}^{} \,
         \Phi(\alpha(q),u+v) \, \Bigl( A_k^{\alpha+\theta\alpha}(q,-u) \, - \,
                                       A_k^{\alpha+\theta\alpha}(q,v) \Bigr)
         \\[4mm]
         \mbox{} - \, {\displaystyle \sum_{j=1}^r} \, \alpha_j^{} \,
                      A_j^\alpha(q,u) A_k^\alpha(q,v)~=~0~.
        \end{array}
       \end{equation}
       if $\, \beta = - \theta\alpha \,$ (independently of whether
       $\, - \theta\alpha \neq \alpha \,$ or $\, - \theta\alpha = \alpha$):
       \begin{eqnarray} \label{eq:SPEDCEE5}
        \begin{array}{l}
         {\textstyle \frac{1}{2}} \, \alpha_k^{} \,
         \Phi^\prime(\alpha(q),u+v) \, - \, \alpha(K_{-\alpha}^{}) \,
         \Phi(\alpha(q),v) \, A_k^\alpha(q,u) \\[3mm]
         \mbox{} + \, {\textstyle \frac{1}{2}} \, \Phi(\alpha(q),u+v) \,
         \Bigl( \alpha (A_k^{\mathfrak{h}}(q,u)) \, - \,
                (\theta\alpha)(A_k^{\mathfrak{h}}(q,v)) \Bigr) \\[3mm]
         \mbox{} - \, {\textstyle \frac{1}{2}} \,
         N_{\alpha,-\theta\alpha}^{} \,
         \Bigl( \Phi(-\alpha(q),u-v) \, A_k^{\alpha-\theta\alpha}(q,u) \, - \,
                \Phi(\alpha(q),u-v) \, A_k^{\alpha-\theta\alpha}(q,v) \Bigr)~
         \\[2mm]
         \mbox{} + \, {\displaystyle \sum_{j=1}^r} \, \alpha_j^{} \,
                      A_j^\alpha(q,u) A_k^{-\theta\alpha}(q,v)~=~0~.
        \end{array}
       \end{eqnarray}
       if $\, \beta = \theta\alpha \neq - \alpha \,$:
       \begin{equation} \label{eq:SPEDCEE6}
        \begin{array}{l}
         \alpha(K_\alpha^{}) \, \Phi(-\alpha(q),v) \, A_k^\alpha(q,u) \\[3mm]
         \mbox{} + \, {\textstyle \frac{1}{2}} \, N_{\alpha,\theta\alpha}^{} \,
         \Phi(\alpha(q),u-v) \, \Bigl( A_k^{\alpha+\theta\alpha}(q,u) \, - \,
                                       A_k^{\alpha+\theta\alpha}(q,v) \Bigr)
         \\[2mm]
         \mbox{} + \, {\displaystyle \sum_{j=1}^r} \, \alpha_j^{} \,
                      A_j^\alpha(q,u) A_k^{\theta\alpha}(q,v)~=~0~.
        \end{array}
       \end{equation}
       if $\, \beta = - \alpha \,$ (independently of whether
       $\, \theta\alpha \neq -\alpha \,$ or $\, \theta\alpha = -\alpha$):
       \begin{eqnarray} \label{eq:SPEDCEE7}
        \begin{array}{l}
         {\textstyle \frac{1}{2}} \, \alpha_k^{} \,
         \Phi^\prime(\alpha(q),u-v) \, + \, \alpha(K_{-\alpha}^{}) \,
         \Phi(-\alpha(q),v) \, A_k^\alpha(q,u) \\[3mm]
         \mbox{} + \, {\textstyle \frac{1}{2}} \, \Phi(\alpha(q),u-v) \,
         \Bigl( \alpha(A_k^{\mathfrak{h}}(q,u)) \, - \,
                \alpha(A_k^{\mathfrak{h}}(q,v)) \Bigr) \\[3mm]
         \mbox{} - \, {\textstyle \frac{1}{2}} \,
         N_{\theta\alpha,-\alpha}^{} \,
         \Bigl( \Phi(-\alpha(q),u+v) \, A_k^{\theta\alpha-\alpha}(q,-u) \, - \,
                \Phi(\alpha(q),u+v) \, A_k^{\theta\alpha-\alpha}(q,v) \Bigr)~
         \\[2mm]
         \mbox{} + \, {\displaystyle \sum_{j=1}^r} \, \alpha_j^{} \,
                      A_j^\alpha(q,u) A_k^{-\alpha}(q,v)~=~0~.
        \end{array}
       \end{eqnarray}
       if $\, \beta \neq \pm\alpha \,$ and $\, \beta \neq \pm\theta\alpha \,$:
       \begin{eqnarray} \label{eq:SPEDCEE8}
        \begin{array}{l}
         \alpha(K_\beta^{}) \, \Phi(\beta(q),v) \, A_k^\alpha(q,u) \\[3mm]
         \mbox{} - \, {\textstyle \frac{1}{2}} \, N_{\alpha,\beta}^{} \,
         \Bigl( \Phi(\beta(q),v-u) \, A_k^{\alpha+\beta}(q,u) \, + \,
                \Phi(\alpha(q),u-v) \, A_k^{\alpha+\beta}(q,v) \Bigr)~~\\[3mm]
         \mbox{} - \, {\textstyle \frac{1}{2}} \, N_{\theta\alpha,\beta}^{} \,
         \Bigl( \Phi(\beta(q),u+v) \, A_k^{\theta\alpha+\beta}(q,-u) \, - \,
                \Phi(\alpha(q),u+v) \, A_k^{\theta\alpha+\beta}(q,v) \Bigr)~~
         \\[2mm]
         \mbox{} - \, {\displaystyle \sum_{j=1}^r} \, \beta_j^{} \,
                      A_j^\alpha(q,u) A_k^{\beta}(q,v)~=~0~.
        \end{array}
       \end{eqnarray}
\end{itemize}
This is a complicated set of equations which we shall solve in a series
of steps.

We~begin by considering the algebro-differential equations~%
(\ref{eq:SPEDCEH1})-(\ref{eq:SPEDCEH4}) for the root part of
the gauge potential, which we claim to have the simple solution
\begin{equation} \label{eq:SPGPOTR1}
 A_k^\alpha(q,u)~
 =~\left\{ \begin{array}{cc}
            \Phi(\alpha(q),u) \, M_k^\alpha &
            \mbox{for $\, \alpha \smin\, \tilde{\Delta}$} \\[2mm]
            M_k^\alpha & \mbox{for $\, \alpha \smin\, \Delta_0$}
           \end{array} \right\}~,
\end{equation}
where the coefficients $M_k^\alpha$ are constants that must be determined
from the remaining equations, subject to the constraint
\begin{equation} \label{eq:SPGPOTR2}
 M_k^{\theta\alpha}~= \; - \, M_k^\alpha \qquad
 \mbox{for $\, \alpha \smin\, \tilde{\Delta}$}~,
\end{equation}
imposed in order to guarantee the validity of eqn~(\ref{eq:SPGPOT2}).
(The corresponding constraint for $\, \alpha \smin \Delta_0 \,$ is empty.)
Indeed, the statement of eqn~(\ref{eq:SPGPOTR1}) for $\, \alpha \smin\,
\Delta_0$ follows directly from eqns~(\ref{eq:SPEDCEH2}) and~%
(\ref{eq:SPEDCEH4}). Similarly, the statement of eqn~(\ref{eq:SPGPOTR1})
for $\, \alpha \smin\, \tilde{\Delta}$ is an immediate consequence of eqn~%
(\ref{eq:SPEDCEH1}) in the degenerate case but is somewhat harder to prove
in the elliptic case. To this end, we recast the functional equation~%
(\ref{eq:EFUNEQ6}) into the form
\[
 \Phi(\alpha(q),u \mp v) \; \Phi(\alpha(q),\pm\, v)~
 =~\Bigl( \, \zeta(u \mp v) \, \pm \, \zeta(v) \Bigr) \,
   \Phi(\alpha(q),u) \, - \, \Phi^\prime(\alpha(q),u)
\]
and use the Ansatz
\[
 A_k^\alpha(q,u)~
 =~\Phi(\alpha(q),u) \, M_k^\alpha(q,u) \qquad
 \mbox{for $\, \alpha \smin\, \tilde{\Delta}$}
\]
to rewrite the differential equation (\ref{eq:SPEDCEH1}) in the form
\begin{eqnarray*}
\lefteqn{\Phi(\alpha(q),u) \, \partial_j^{} M_k^\alpha(q,u) \, + \,
         \alpha_j^{} \, \Phi^\prime(\alpha(q),u) \, M_k^\alpha(q,u)}    \\[1mm]
 & & \mbox{} - \, {\textstyle \frac{1}{2}} \, \alpha_j^{} \,
                  \Bigl( \, \zeta(u-v) \,+\, \zeta(u+v) \Bigr) \,
                  \Phi(\alpha(q),u) \, M_k^\alpha(q,u)                  \\
 & & \mbox{} + \, {\textstyle \frac{1}{2}} \, \alpha_j^{} \,
                  \Bigl( \Bigl( \, \zeta(u-v) \,+\, \zeta(v) \Bigr) \,
                         \Phi(\alpha(q),u) \, - \,
                         \Phi^\prime(\alpha(q),u) \Bigr) \,
                  M_k^\alpha(q,v)                                       \\
 & & \mbox{} + \, {\textstyle \frac{1}{2}} \, \alpha_j^{} \,
                  \Bigl( \Bigl( \, \zeta(u+v) \,-\, \zeta(v) \Bigr) \,
                         \Phi(\alpha(q),u) \, - \,
                         \Phi^\prime(\alpha(q),u) \Bigr) \,
                  M_k^\alpha(q,-v)                                      \\[3mm]
 &=&\! 0~,
\end{eqnarray*}
while the algebraic equation (\ref{eq:SPEDCEH3}) takes the form
\begin{eqnarray*}
\lefteqn{\Bigl( \, \zeta(u-v) \,-\, \zeta(u+v) \,+\, 2 \>\! \zeta(v) \Bigr) \,
         \Phi(\alpha(q),u) \, M_k^\alpha(q,u)}                          \\
 & & \mbox{} - \, \Bigl( \Bigl( \, \zeta(u-v) \,+\, \zeta(v) \Bigr) \,
                         \Phi(\alpha(q),u) \, - \,
                         \Phi^\prime(\alpha(q),u) \Bigr) \,
                  M_k^\alpha(q,v)                                       \\
 & & \mbox{} + \, \Bigl( \Bigl( \, \zeta(u+v) \,-\, \zeta(v) \Bigr) \,
                         \Phi(\alpha(q),u) \, - \,
                         \Phi^\prime(\alpha(q),u) \Bigr) \,
                  M_k^\alpha(q,-v)                                      \\[3mm]
 &=&\! 0~.
\end{eqnarray*}
Both of these equations can be simplified by using the functional equation~%
(\ref{eq:EFUNEQ7}) and subtracting $\, \frac{1}{2} \, \alpha_j$ times the
second from the first, with the result that
\begin{equation} \label{eq:SPGPOTR3}
 \begin{array}{rcl}
  \partial_j^{} M_k^\alpha(q,u) \!
  &=&\! \alpha_j^{} \,
        \Bigl( \, \zeta(u-v) \,+\, \zeta(v) \,+\,
                  \zeta(\alpha(q)) \,-\, \zeta(\alpha(q)+u) \Bigr) \\[2mm]
  & &   \times \, \Bigl( M_k^\alpha(q,u) - M_k^\alpha(q,v) \Bigr)~,
 \end{array}
\end{equation}
while
\begin{equation} \label{eq:SPGPOTR4}
 \begin{array}{rcl}
 \lefteqn{\Bigl( \, \zeta(u-v) \,-\, \zeta(u+v) \,+\, 2 \>\! \zeta(v) \Bigr) \,
          M_k^\alpha(q,u)}                                              \\[3mm]
  & & \mbox{} - \, \Bigl( \, \zeta(u-v) \,+\, \zeta(v) \,-\,
                          \zeta(\alpha(q)+u) \,+\, \zeta(\alpha(q)) \Bigr) \,
                   M_k^\alpha(q,v)                                      \\[3mm]
  & & \mbox{} + \, \Bigl( \, \zeta(u+v) \,-\, \zeta(v) \,-\,
                          \zeta(\alpha(q)+u) \,+\, \zeta(\alpha(q)) \Bigr) \,
                   M_k^\alpha(q,-v)                                     \\[4mm]
  &=&\! 0~.
 \end{array}
\end{equation}
Antisymmetrizing the last equation with respect to the exchange of $u$
and~$-u$ gives
\begin{equation} \label{eq:SPGPOTR5}
 \begin{array}{rcl}
  \lefteqn{\Bigl( \, \zeta(u-v) \,-\, \zeta(u+v) \,+\, 2 \>\! \zeta(v) \Bigr)
           \Bigl( M_k^\alpha(q,u) \, - \, M_k^\alpha(q,-u) \Bigr)}
                                                           \hspace{5mm} \\[3mm]
  &=&\!\! \Bigl( \, \zeta(u-v) \,+\, \zeta(u+v) \,+\,
                    \zeta(\alpha(q)-u) \,-\, \zeta(\alpha(q)+u) \Bigr)  \\[2mm]
  & &     \times \, \Bigl( M_k^\alpha(q,v) \, - \, M_k^\alpha(q,-v) \Bigr)~.
 \end{array}
\end{equation}
In order to analyze the consequences of this relation, we shall use the
identity
\[
 \zeta(x+y) - \zeta(x) - \zeta(y)~
 =~\frac{1}{2} \; \frac{\wp^\prime(x) - \wp^\prime(y)}{\wp(x) - \wp(y)}~,
\]
which can also be written as
\[
 \zeta(u-v) \,-\, \zeta(u+v) \,+\, 2 \>\! \zeta(v)~
  =~\frac{\wp^\prime(v)}{\wp(u) - \wp(v)}~,
\]
implying that
\begin{eqnarray*}
\lefteqn{\zeta(u-v) \,+\, \zeta(u+v) \,+\, \zeta(s-u) \,-\, \zeta(s+u)}
                                                           \hspace{5mm} \\[2mm]
 &=&\!\! \frac{\wp^\prime(u)}{\wp(u) - \wp(v)} \, + \,
         \frac{\wp^\prime(u)}{\wp(s) - \wp(u)}~
  =~     \frac{\wp^\prime(u) \, (\wp(s) - \wp(v))}
              {(\wp(s) - \wp(u)) \, (\wp(u) - \wp(v))}~.
\end{eqnarray*}
After rearranging coefficients, we conclude that eqn~(\ref{eq:SPGPOTR5}) can
be reformulated as stating that the function
\[
 \frac{\wp(\alpha(q)) - \wp(u)}{\wp^\prime(u)} \,
 \Bigl( M_k^\alpha(q,u) \, - \, M_k^\alpha(q,-u) \Bigr)
\]
is independent of $u$. Putting $\, u = \alpha(q)$, we see that it must actually
vanish identically, which is only possible if
\[
 M_k^\alpha(q,u) \, - \, M_k^\alpha(q,-u)~=~0~.
\]
Inserting this result back into eqn~(\ref{eq:SPGPOTR4}), we get
\[
 M_k^\alpha(q,u) \, - \, M_k^\alpha(q,v)~=~0~.
\]
Now eqn~(\ref{eq:SPGPOTR3}) implies that $M_k^\alpha$ is in fact a constant.

For what follows, we shall find it convenient to assemble the constants
introduced in eqn~(\ref{eq:SPGPOTR1}) above into a vector in $\mathfrak{a}_0$
by writing, for any $\, \alpha \smin\, \Delta$,
\begin{equation} \label{eq:SPCOEF2}
 M_\alpha^{}~=~\sum_{j=1}^r M_j^\alpha \>\! H_j^{}~,
\end{equation}
so that of course
\begin{equation} \label{eq:SPCOEF3}
 M_k^\alpha~=~(H_k^{} \,, M_\alpha^{})~.
\end{equation}
In analogy with eqn~(\ref{eq:SPCOEF1}), we also introduce the abbreviation
\begin{equation} \label{eq:SPCOEF4}
 M_\alpha^\pm~=~{\textstyle \frac{1}{2}} \, (M_\alpha^{} \pm M_{-\alpha}^{})~.
\end{equation}
Note that, for $\, \alpha \smin\, \tilde{\Delta}$, the generators
$\, M_\alpha^{} \smin\, \mathfrak{a}_0 \,$ are in a sense complementary
to the generators $\, K_\alpha^{} \smin\, \mathrm{i} \mathfrak{b}_0$.
(This observation will come to play an important role later on.)
In particular, eqn~(\ref{eq:SPGPOTR2}) amounts to the condition
\begin{equation} \label{eq:SPCOEF5}
 M_{\theta\alpha}^{}~= \; - \, M_\alpha^{} \qquad
 \mbox{for $\, \alpha \smin\, \tilde{\Delta}$}
\end{equation}
which is analogous to the condition
\begin{equation} \label{eq:SPCOEF6}
 K_{\theta\alpha}^{}~=~K_\alpha^{} \qquad
 \mbox{for $\, \alpha \smin\, \tilde{\Delta}$}
\end{equation}
of Ref.~\cite{FW1}.

Now we are ready to state the first main result of this section.
Our terminology will follow that of Ref.~\cite{Kn}, according to
which roots $\, \alpha \smin\, \Delta \,$ are called \emph{imaginary} if
$\, \theta\alpha = \alpha$, \emph{real} if $\, \theta\alpha = -\alpha \,$
and \emph{complex} if $\theta \alpha$ and $\alpha$ are linearly independent,
whereas two roots $\alpha$ and $\beta$ are called \emph{strongly orthogonal}
if both $\alpha+\beta$ and $\alpha-\beta$ are not roots (as is well known,
this implies that $\alpha$ and $\beta$ are orthogonal in the usual sense).
\begin{propos}
 The integrable Calogero model associated with the root system of a \linebreak
 symmetric pair ($\mathfrak{g},\theta)$ admits a gauge transformation $g$
 from the standard Lax pair of \linebreak Olshanetsky and Perelomov and the
 dynamical $R$-matrix of Ref.~\cite{FW1} to a new Lax pair with a numerical
 $R$-matrix~ if and only if\/ a) the automorphism $\theta$ acts on the root
 system~$\Delta$ in such a way that
 \begin{itemize}
  \item $\Delta$ contains no imaginary roots, i.e., $\Delta_0 = \emptyset \,$
        and $\, \tilde{\Delta} = \Delta$,
  \item for any complex root $\alpha$ in $\Delta$, $\theta\alpha$ and
        $\alpha$ are strongly orthogonal, i.e., $\theta\alpha \pm \alpha
        \nsmin \Delta$,
 \end{itemize}
 and b) the set of generators $\, K_\alpha^{} \smin\, \mathrm{i}
 \mathfrak{b}_0 \,$ appearing in eqns~(\ref{eq:SPDRMAT}) and~%
 (\ref{eq:SPERMAT}) can be complemented by a set of generators
 $\, M_\alpha \smin\, \mathfrak{a}_0 \,$ which, taken together,
 satisfy the algebraic constraints
 \vspace{2mm}
 \begin{equation} \label{eq:SPAC4}
  \alpha(K_\alpha^+)~=~0~~,~~
  \alpha(M_\alpha^+)~=~0~,
 \vspace{2mm}
 \end{equation}
 \begin{equation} \label{eq:SPAC5}
  K_\alpha^-~=~\frac{\epsilon_\alpha}{\sqrt{2} \, |\alpha|} \,
               (H_\alpha^{})_{\mathfrak{b}}^{}~~,~~
  M_\alpha^-~=~\frac{\epsilon_\alpha}{\sqrt{2} \, |\alpha|} \,
               (H_\alpha^{})_{\mathfrak{a}}^{}~,
 \vspace{2mm}
 \end{equation}
 \begin{equation} \label{eq:SPAC6}
  \begin{array}{c}
   \alpha(K_\beta^{}) \, K_\alpha^{} \, - \, \beta(K_\alpha^{}) \, K_\beta^{}~
   =~{\textstyle \frac{1}{2}} \,
     \Bigl( N_{\alpha,\beta}^{} \, K_{\alpha+\beta}^{} \, + \,
            N_{\theta\alpha,\beta}^{} \, K_{\theta\alpha+\beta}^{} \Bigr)
   \\[2mm]
   \alpha(K_\beta^{}) \, M_\alpha^{} \, - \, \beta(M_\alpha^{}) \, M_\beta^{}~
   =~{\textstyle \frac{1}{2}} \,
     \Bigl( N_{\alpha,\beta}^{} \, M_{\alpha+\beta}^{} \, - \,
            N_{\theta\alpha,\beta}^{} \, M_{\theta\alpha+\beta}^{} \Bigr)
   \\[4mm]
   \mbox{for $\, \alpha,\beta \smin\, \Delta \,$ such that
   $\, \beta \neq \pm \alpha \,,\, \beta \neq \pm \theta\alpha$}~,
   \rule[-4mm]{0mm}{5mm}
  \end{array}
 \end{equation}
 as well as the additional algebraic constraints
 \begin{equation} \label{eq:SPEAC1}
  \sum_{\alpha \ssmin \Delta} (H_\alpha^{})_{\mathfrak{a}}^{}
  \otimes K_\alpha^{} \otimes M_{-\alpha}^{}~=~0~~,~~
  \sum_{\alpha \ssmin \Delta} (H_\alpha^{})_{\mathfrak{b}}^{}
  \otimes M_\alpha^{} \otimes M_{-\alpha}^{}~=~0~,
 \end{equation}
 \begin{equation} \label{eq:SPEAC2}
  \begin{array}{ccc}
   {\displaystyle
    \sum_{\beta,\gamma \ssmin \Delta \atop \beta + \gamma = \alpha}
    N_{\beta,\gamma}^{} \, M_\beta^{} \otimes K_\gamma^{}} \!\!
    &=&\!\! (H_\alpha^{})_{\mathfrak{a}}^{} \otimes K_\alpha^{} \, - \,
            M_\alpha^{} \otimes (H_\alpha^{})_{\mathfrak{b}}^{}~,       \\[7mm]
   {\displaystyle
    \sum_{\beta,\gamma \ssmin \Delta \atop \beta + \gamma = \alpha}
    N_{\beta,\gamma}^{} \, M_\beta^{} \otimes M_\gamma^{}} \!\!
    &=&\!\! (H_\alpha^{})_{\mathfrak{a}}^{} \otimes M_\alpha^{} \, - \,
            M_\alpha^{} \otimes (H_\alpha^{})_{\mathfrak{a}}^{}~,
  \end{array}
 \end{equation}
 to be imposed in the case of the elliptic model, where $K_\alpha^\pm
 = \frac{1}{2} \, (K_\alpha^{} \pm K_{-\alpha}^{}) \,$ and \linebreak
 $M_\alpha^\pm = \frac{1}{2} \, (M_\alpha^{} \pm M_{-\alpha}^{}) \,$
 as above, with $\, \epsilon_\alpha = \pm 1$.
 In this case, the root part and the Cartan part of the potential
 $\, A_k^{} = g^{-1} \, \partial_k^{} g \,$ associated with this
 gauge transformation $g$ are given by
 \begin{equation} \label{eq:SPDGPOT1}
  A_k^\alpha(q)~=~w(\alpha(q)) \; (H_k^{} \,, M_\alpha^{})
 \end{equation}
 and
 \begin{equation} \label{eq:SPDGPOT2}
  A_k^{\mathfrak{h}}(q)~
  =~\sum_{\alpha \ssmin \Delta} \, \frac{w^\prime(\alpha(q))}{w(\alpha(q))} \;
            (H_k^{} \,, M_{-\alpha}^{}) \, K_\alpha^{}
 \end{equation}
 for the degenerate model and by
 \begin{equation} \label{eq:SPEGPOT1}
  A_k^\alpha(q,u)~=~\Phi(\alpha(q),u) \; (H_k^{} \,, M_\alpha^{})
 \end{equation}
 and
 \begin{equation} \label{eq:SPEGPOT2}
  A_k^{\mathfrak{h}}(q,u)~
  = \; - \, \sum_{\alpha \ssmin \Delta} \, \zeta(\alpha(q)) \,
            (H_k^{} \,, M_{-\alpha}^{}) \, K_\alpha^{} \, - \,
            \zeta(u) \, H_k^{}
 \end{equation}
 for the elliptic model.
\end{propos}
\textbf{Note.}~~As we shall show after completing the proof of Proposition~2,
eqn~(\ref{eq:SPAC6}) forces all roots $\alpha$ in $\Delta$ to have the same
length (which by convention we fix to be $\sqrt{2}$) and also allows for a
choice of basis in which the signs $\epsilon_\alpha$ are independent of~%
$\alpha$. so that eqn~(\ref{eq:SPAC5}) can be simplified as follows:
\begin{equation} \label{eq:SPAC7}
 K_\alpha^-~=~{\textstyle \frac{1}{2}} \, \epsilon \,
              (H_\alpha^{})_{\mathfrak{b}}^{}~~,~~
 M_\alpha^-~=~{\textstyle \frac{1}{2}} \, \epsilon \,
              (H_\alpha^{})_{\mathfrak{a}}^{}~.
\end{equation}

\noindent
\textbf{Proof.}~~With the vector notation introduced above, we can first
of all use eqn~(\ref{eq:SPEDCEH4}), with $\alpha$ replaced by $\, \alpha +
\theta\alpha $, to conclude that the middle terms in eqns~(\ref{eq:SPEDCEE4})
and~(\ref{eq:SPEDCEE6}) vanish identically and thus reduce both of them to a
single algebraic constraint:
\begin{equation} \label{eq:SPIAC01}
 \alpha(K_\alpha^{})~=~\alpha(M_\alpha^{}) \qquad
 \mbox{for $\, \alpha \smin \tilde{\Delta} \,$ such that
  $\, \theta\alpha \neq -\alpha$}~.
\end{equation}
Note that replacing $\alpha$ by $-\alpha$ and adding/subtracting the two
equations, we get
\begin{eqnarray}
 &\alpha(K_\alpha^+)~=~\alpha(M_\alpha^+) \qquad
  \mbox{for $\, \alpha \smin \tilde{\Delta} \,$ such that
   $\, \theta\alpha \neq -\alpha$}~,&                     \label{eq:SPIAC02} \\
 &\alpha(K_\alpha^-)~=~\alpha(M_\alpha^-) \qquad
  \mbox{for $\, \alpha \smin \tilde{\Delta} \,$ such that
   $\, \theta\alpha \neq -\alpha$}~.&                     \label{eq:SPIAC03}
\end{eqnarray}
Using eqn~(\ref{eq:SPAC3}), the first of these can be sharpened as follows:
\begin{equation} \label{eq:SPIAC04}
 \alpha(K_\alpha^+)~=~0~=~\alpha(M_\alpha^+) \qquad
 \mbox{for $\, \alpha \smin \tilde{\Delta} \,$ such that
  $\, \theta\alpha - \alpha \,\nsmin\, \Delta$}~.
\end{equation}
(Indeed, if $\, \theta\alpha = -\alpha \,$ so that eqn~(\ref{eq:SPIAC01})
no longer applies, eqn~(\ref{eq:SPIAC04}) remains correct because in this
case $\, M_\alpha^+ = 0$, according to eqns~(\ref{eq:SPCOEF4}) and~%
(\ref{eq:SPCOEF5}).)
Next, inserting eqn~(\ref{eq:SPGPOTR1}) together with the functional equation~%
(\ref{eq:EFUNEQ2}) into the differential equation (\ref{eq:SPEDCHH1}) for the
Cartan part of the gauge potential, we see that this equation can be solved
by setting
\begin{equation} \label{eq:SPGPOTC1}
 A_k^{\mathfrak{h}}(q,u)~
 = \; - \, \sum_{\alpha \ssmin \tilde{\Delta}} \, \zeta(\alpha(q)) \,
           M_k^{-\alpha} \, K_\alpha^{} \, - \, M_k^{\mathfrak{h}}(u)~,
\end{equation} 
where the $M_k^{\mathfrak{h}}(u)$ are constants that must be determined from
the remaining equations, provided we assume the coefficients $M_k^\alpha$ to
satisfy the relation
\begin{equation} \label{eq:SPIAC05}
 \sum_{\alpha \ssmin \tilde{\Delta}} \, \alpha_j^{} \, 
 M_k^{-\alpha} \, K_\alpha^{}~=~0 \qquad
 \mbox{for $\, 1 \leqslant j\,,k \leqslant r$}~.
\end{equation}
Converted into a tensor equation, it reads
\begin{equation} \label{eq:SPIAC06}
 \sum_{\alpha \ssmin \tilde{\Delta}}
 (H_\alpha^{})_{\mathfrak{a}}^{} \otimes K_\alpha^{} \otimes M_{-\alpha}^{}~
 =~0~,
\end{equation}
which leads back to eqn~(\ref{eq:SPIAC05}) by taking the scalar product with
$H_j$ in the first and with $H_k$ in the third tensor factor. Note that in
the degenerate case, the same argument works, but eqn~(\ref{eq:SPIAC05}/%
\ref{eq:SPIAC06}) is not needed. Moreover, eqn~(\ref{eq:SPEDCHH2}) is
satisfied as a consequence of the identity
\[
 \sum_{\alpha \ssmin \tilde{\Delta}} \, \alpha_j^{} \,
 f(\alpha(q)) \, M_k^{-\alpha} \, K_\alpha^{}~=~0 \qquad
 \mbox{for $\, r+1 \leqslant j \leqslant r+s$}~,
\]
which is valid for any even function $f$ (such as $\, \zeta^\prime$):
this is easily shown by replacing $\alpha$ by $\theta\alpha$ in the sum
and noting that $\, (\theta\alpha)_j = \alpha_j \,$ for $\, r+1 \leqslant
 j \leqslant r+s$, $(\theta\alpha)(q) = - \alpha(q)$, $M_k^{-\theta\alpha}
= - M_k^{-\alpha} \,$ and $\, K_{\theta\alpha} = K_\alpha \,$, so that
\[
  \sum_{\alpha \ssmin \tilde{\Delta}} \, \alpha_j^{} \,
  f(\alpha(q)) \, M_k^{-\alpha} \, K_\alpha^{}~
  = \; - \sum_{\alpha \ssmin \tilde{\Delta}} \, \alpha_j^{} \,
                    f(\alpha(q)) \, M_k^{-\alpha} \, K_\alpha^{}
 \qquad \mbox{for $\, r+1 \leqslant j \leqslant r+s$}~.
\]
Similarly, inserting eqn~(\ref{eq:SPGPOTR1}) together with the functional
equation~(\ref{eq:EFUNEQ6}) into eqns~(\ref{eq:SPEDCEE5}) and~%
(\ref{eq:SPEDCEE7}), we obtain
\[
 \begin{array}{l}
  {\textstyle \frac{1}{2}} \, \alpha_k^{} \,
  \Bigl( - \, \Phi(\alpha(q),u) \, \Phi(\alpha(q),v) \, +
         \left( \zeta(u) + \zeta(v) \right) \Phi(\alpha(q),u+v) \Bigr) \\[4mm]
  - \; \alpha(K_{-\alpha}^{}) \, M_k^\alpha \,
  \Phi(\alpha(q),u) \, \Phi(\alpha(q),v) \\[3mm]
  - \; {\textstyle \frac{1}{2}} \, \Phi(\alpha(q),u+v) \,
  \Bigl( \alpha(M_k^{\mathfrak{h}}(u)) \, - \,
         (\theta\alpha)(M_k^{\mathfrak{h}}(v)) \Bigr) \\[3mm]
  + \; {\textstyle \frac{1}{2}} \,
  N_{\alpha,-\theta\alpha}^{} \, M_k^{\alpha-\theta\alpha}
  \Bigl( - \, \Phi(-\alpha(q),u-v) \, \Phi(2 \>\! \alpha(q),u) \, - \,
              \Phi(-\alpha(q),v-u) \, \Phi(2 \>\! \alpha(q),v) \Bigr) \\[4mm]
  - \; \alpha(M_\alpha^{}) \, M_k^{-\alpha} \,
  \Phi(\alpha(q),u) \, \Phi(\alpha(q),v)~=~0
 \end{array}
\]
and
\[
 \begin{array}{l}
  {\textstyle \frac{1}{2}} \, \alpha_k^{} \,
  \Bigl( - \, \Phi(\alpha(q),u) \, \Phi(\alpha(q),-v) \, +
         \left( \zeta(u) - \zeta(v) \right) \Phi(\alpha(q),u-v) \Bigr) \\[4mm]
   - \; \alpha(K_{-\alpha}^{}) \, M_k^\alpha \,
  \Phi(\alpha(q),u) \, \Phi(\alpha(q),-v) \\[3mm]
  - \; {\textstyle \frac{1}{2}} \, \Phi(\alpha(q),u-v) \,
  \Bigl( \alpha(a_k^{\mathfrak{h}}(u)) \, - \,
         \alpha(a_k^{\mathfrak{h}}(v)) \Bigr) \\[3mm]
   - \; {\textstyle \frac{1}{2}} \,
  N_{\theta\alpha,-\alpha}^{} \, M_k^{\theta\alpha-\alpha}
  \Bigl( - \, \Phi(-\alpha(q),u+v) \, \Phi(2 \>\! \alpha(q),u) -
              \Phi(-\alpha(q),-u\!-\!v) \, \Phi(2 \>\! \alpha(q),-v) \Bigr)
  \\[4mm]
  - \; \alpha(M_\alpha^{}) \, M_k^{-\alpha} \,
  \Phi(\alpha(q),u) \, \Phi(\alpha(q),-v)~=~0
 \end{array}
\]
respectively. In both cases, the terms proportional to $\Phi(\alpha(q),
u \pm v)$ cancel provided we set
\begin{equation} \label{eq:SPGPOTC2}
 M_k^{\mathfrak{h}}(u)~=~\zeta(u) \, H_k~,
\end{equation}
which also guarantees that eqn~(\ref{eq:SPEDCEE1}) is valid, and
due to the functional equation~(\ref{eq:EFUNEQ5}), the remaining
terms then cancel if we impose the relation
\begin{equation} \label{eq:SPIAC07}
 {\textstyle \frac{1}{2}} \, \alpha_k^{} \, + \,
 \alpha(K_{-\alpha}^{}) \, M_k^\alpha \, + \,
 \alpha(M_\alpha^{}) \, M_k^{-\alpha} \, + \,
 {\textstyle \frac{1}{2}} \,
 N_{\theta\alpha,-\alpha}^{} \, M_k^{\theta\alpha-\alpha}~=~0 \qquad
 \mbox{for $\, \alpha \smin \tilde{\Delta}$}~.
\end{equation}
Converted to a vector equation in $\mathfrak{a}_0$, it reads
\begin{equation} \label{eq:SPIAC08}
 {\textstyle \frac{1}{2}} \, (H_\alpha^{})_{\mathfrak{a}}^{} \, + \,
 \alpha(K_{-\alpha}^{}) \, M_\alpha^{} \, + \,
 \alpha(M_\alpha^{}) \, M_{-\alpha}^{} \, + \,
 {\textstyle \frac{1}{2}} \,
 N_{\theta\alpha,-\alpha}^{} \, M_{\theta\alpha-\alpha}^{}~=~0 \qquad
 \mbox{for $\, \alpha \smin \tilde{\Delta}$}~,
\end{equation}
which leads back to eqn~(\ref{eq:SPIAC07}) by taking the scalar product
with $H_k$. Even simpler to handle is eqn~(\ref{eq:SPEDCEE8}), which by
insertion of eqn~(\ref{eq:SPGPOTR1}) together with the functional
equation~(\ref{eq:EFUNEQ5}) reduces to the relation
\begin{equation} \label{eq:SPIAC09}
 \begin{array}{c}
  \alpha(K_\beta^{}) \, M_k^\alpha \, - \, {\textstyle \frac{1}{2}} \,
  \Bigl( N_{\alpha,\beta}^{} \, M_k^{\alpha+\beta} \, - \,
         N_{\theta\alpha,\beta}^{} \, M_k^{\theta\alpha+\beta} \Bigr) \, - \,
  \beta(M_\alpha^{}) \, M_k^\beta~=~0 \\[3mm]
  \mbox{for $\, \alpha,\beta \smin \tilde{\Delta} \,$ such that
  $\, \beta \neq \pm \alpha \,,\, \beta \neq \pm \theta\alpha$}~.
 \end{array}
\end{equation}
Converted to a vector equation in $\mathfrak{a}_0$, it reads
\begin{equation} \label{eq:SPIAC10}
 \begin{array}{c}
  \alpha(K_\beta^{}) \, M_\alpha^{} \, - \, {\textstyle \frac{1}{2}} \,
  \Bigl( N_{\alpha,\beta}^{} \, M_{\alpha+\beta}^{} \, - \,
         N_{\theta\alpha,\beta}^{} \, M_{\theta\alpha+\beta}^{} \Bigr) \, - \,
  \beta(M_\alpha^{}) \, M_\beta^{}~=~0 \\[3mm]
  \mbox{for $\, \alpha,\beta \smin \tilde{\Delta} \,$ such that
  $\, \beta \neq \pm \alpha \,,\, \beta \neq \pm \theta\alpha$}~,
 \end{array}
\end{equation}
which leads back to eqn~(\ref{eq:SPIAC09}) by taking the scalar product with
$H_k$. Finally, inserting eqn~(\ref{eq:SPGPOTR1}) together with the functional
equation~(\ref{eq:EFUNEQ6}) (more precisely, the difference of two copies of
eqn~(\ref{eq:EFUNEQ6}): one with $\, v \rightarrow u \,,\, u \rightarrow v \,$
and one with $\, v \rightarrow u \,,\, u \rightarrow -v$) into eqn~%
(\ref{eq:SPEDCEE3}) and dividing by $\Phi(\alpha(q),u)$, we obtain
\begin{equation} \label{eq:SPIAC11}
  N_{\alpha,\beta}^{} \, M_k^{\alpha+\beta}~=~0 \qquad
  \mbox{for $\, \alpha \smin \tilde{\Delta}$, $\beta \smin \Delta_0$}~,
\end{equation}
which in turn reduces eqn~(\ref{eq:SPEDCEE2}) to
\begin{equation} \label{eq:SPIAC12}
 \alpha(K_\beta^{}) \, M_k^\alpha~
 =~\beta(M_\alpha) \, M_k^\beta \qquad
  \mbox{for $\, \alpha \smin \Delta_0$, $\beta \smin \tilde{\Delta}$}~.
\end{equation}
In the degenerate case, the same conclusion is reached along a slightly
different path, since in this case eqn~(\ref{eq:SPEDCEE3}) is void while
eqn~(\ref{eq:SPEDCEE2}) takes the form
\[
 \alpha(K_\beta) \, w(\beta(q)) \, M_k^\alpha \, + \,
 N_{\alpha,\beta}^{} \, M_k^{\alpha+\beta} \, w^\prime(\beta(q)) \, - \,
 \beta(M_\alpha^{}) \, w(\beta(q))~=~0~,
\]
which again leads to eqns~(\ref{eq:SPIAC11}) and~(\ref{eq:SPIAC12})
since $w$ and $w^\prime$ are functionally independent.

Before proceeding to the solution of the remaining equations, let us pause
to draw a few consequences of the algebraic constraints (\ref{eq:SPIAC01})--%
(\ref{eq:SPIAC04}), (\ref{eq:SPIAC07}/\ref{eq:SPIAC08}) and (\ref{eq:SPIAC11})
derived so far; this will help us considerably to simplify our further work.
For this purpose, we must distinguish between real and complex roots $\alpha$
in $\tilde{\Delta}$:
\begin{itemize}
 \item For real roots $\, \alpha \smin\, \Delta$ $(\theta\alpha = -\alpha$),
       eqns~(\ref{eq:SPCOEF5}) and~(\ref{eq:SPCOEF6}), together with the
       fact that $\, K_\alpha \smin\, \mathrm{i} \mathfrak{b}_0 \,$ and
       hence $\, \theta K_\alpha = K_\alpha \,$, imply that $\, M_{-\alpha}
       = - M_\alpha \,$, $K_{-\alpha} = K_\alpha \,$ and $\, \alpha(K_\alpha)
       = \alpha(\theta K_\alpha) = \theta\alpha(K_\alpha) = -\alpha(K_\alpha)$,
       so $\, \alpha(K_{\pm\alpha}) = 0$. Thus in this case, the last term and
       the second term in eqn~(\ref{eq:SPIAC08}) drop out, so we get
       \[
        \alpha(M_\alpha^{}) \, M_\alpha^{}~
        =~{\textstyle \frac{1}{2}} \, (H_\alpha^{})_{\mathfrak{a}}^{}~.
       \]
       Applying $\alpha$ to this relation and using that in this case,
       $(H_\alpha^{})_{\mathfrak{a}}^{} = H_\alpha^{}$, we conclude that
       \[
        2 \, \alpha(M_\alpha^{})^2~=~\alpha(H_\alpha^{})~=~(\alpha,\alpha)~,
       \]
       i.e.,
       \begin{equation} \label{eq:SPIAC13}
        \alpha(M_\alpha^{})~=~\frac{\epsilon_\alpha}{\sqrt{2}} \; |\alpha|
        \qquad \mbox{and} \qquad
        M_\alpha^{}~=~\frac{\epsilon_\alpha}{\sqrt{2} \, |\alpha|} \;
                      (H_\alpha^{})_{\mathfrak{a}}^{}~,
       \end{equation}
       where $\, \epsilon_\alpha^{} = \epsilon_{-\alpha}^{}
       = \epsilon_{\theta\alpha}^{} = \epsilon_{-\theta\alpha}^{} \,$
       is a sign factor ($\pm 1$).
 \item For complex roots $\, \alpha \smin\, \Delta$ $(\theta\alpha \neq
       \pm\alpha$), eqn~(\ref{eq:SPIAC01}) implies that $\, \alpha
       (K_{-\alpha}) = \alpha(M_{-\alpha}) \,$, so we may rewrite
       eqn~(\ref{eq:SPIAC08}) in the form
       \[
        {\textstyle \frac{1}{2}} \, (H_\alpha^{})_{\mathfrak{a}}^{} \, + \,
        2 \, \alpha(M_\alpha^+) \, M_\alpha^+ \, - \,
        2 \, \alpha(M_\alpha^-) \, M_\alpha^- \, + \,
        {\textstyle \frac{1}{2}} \,
        N_{\theta\alpha,-\alpha}^{} \, M_{\theta\alpha-\alpha}^{}~=~0~.
       \]
       But under the substitution $\, \alpha \rightarrow -\alpha$, the
       first three terms in this equation are odd while the last is even,
       which forces them to vanish separately. Now we must distinguish
       two cases:
 \begin{itemize}
  \item If $\theta\alpha - \alpha$ is not a root, the last term drops out and,
        according to eqn~(\ref{eq:SPIAC04}), so does the second. Thus we get
        \[
         \alpha(M_\alpha^-) \, M_\alpha^-~
         =~{\textstyle \frac{1}{4}} \, (H_\alpha^{})_{\mathfrak{a}}^{}~.
        \]
        Applying $\alpha$ to this relation, we conclude that
        \[
         4 \, \alpha(M_\alpha^-)^2~=~\alpha((H_\alpha^{})_{\mathfrak{a}}^{})~
         =~{\textstyle \frac{1}{2}} \, (\alpha,\alpha - \theta\alpha)
        \]
        which can be further simplified because $\theta\alpha + \alpha$
        is never a root~\cite[Ex.\ F.2, p.\ 530]{He} so that in this case,
        $\theta\alpha$ and $\alpha$ are (strongly) orthogonal, leading to
        \begin{equation} \label{eq:SPIAC14}
         \alpha(M_\alpha^-)~
         =~\frac{\epsilon_\alpha}{2 \, \sqrt{2}} \; |\alpha|
         \qquad \mbox{and} \qquad
         M_\alpha^-~=~\frac{\epsilon_\alpha}{\sqrt{2} \, |\alpha|} \;
                      (H_\alpha^{})_{\mathfrak{a}}^{}~,
        \end{equation}
        where $\, \epsilon_\alpha^{} = \epsilon_{-\alpha}^{}
        = \epsilon_{\theta\alpha}^{} = \epsilon_{-\theta\alpha}^{} \,$
        is a sign factor ($\pm 1$).
  \item If $\, \theta\alpha - \alpha \,$ is a root, we arrive at a
        contradiction, since in this case $\, \theta\alpha - \alpha \,$
        is a real root, so that according to the previous item,
        $M_{\theta\alpha-\alpha}$ cannot vanish. Therefore,
        this possibility must be excluded.
 \end{itemize}
\end{itemize}
Moreover, we see that $M_\alpha$ can never vanish, so the only way
to guarantee the validity of eqn~(\ref{eq:SPIAC11}) is to assume that
the sum of a root $\alpha$ in $\tilde{\Delta}$ and a root $\beta$ in
$\Delta_0$ is never a root. Since we may freely change the sign of
$\beta$, this forces all roots in $\tilde{\Delta}$ to be (strongly)
orthogonal to all roots in $\Delta_0$, which is only possible if
one of these two sets is empty, since $\mathfrak{g}$ is supposed
to be simple and hence $\Delta$ must be irreducible. This proves
the two restrictions on the action of $\theta$ on $\Delta$ stated
in the proposition, namely
\begin{equation} \label{eq:SPAC8}
 \Delta_0~=~\emptyset~~,~~\tilde{\Delta}~=~\Delta~,
\end{equation}
and, for any root $\alpha$ in $\Delta$,
\begin{equation} \label{eq:SPAC9}
 \theta\alpha \pm \alpha \,\nsmin\, \Delta
 \qquad \mbox{and} \qquad
 \mbox{either}~~\theta\alpha = -\alpha~~\mbox{or}~~\theta\alpha \perp \alpha~.
\end{equation}
Moroever, eqns~(\ref{eq:SPIAC13}) and~(\ref{eq:SPIAC14}) can be
unified into a single formula
\begin{equation} \label{eq:SPIAC15}
 M_\alpha^-~=~\frac{\epsilon_\alpha}{\sqrt{2} \, |\alpha|} \;
              (H_\alpha^{})_{\mathfrak{a}}^{}~,
\end{equation}
% with
% \begin{equation} \label{eq:SPIAC16}
%  \alpha(M_\alpha^-)~=~\frac{\epsilon_\alpha}{2 \, \sqrt{2}} \;
%                       (1 + \delta_{\theta\alpha,-\alpha}^{}) \, |\alpha|~,
% \end{equation}
where $\, \epsilon_\alpha^{} = \epsilon_{-\alpha}^{}
= \epsilon_{\theta\alpha}^{} = \epsilon_{-\theta\alpha}^{} \,$
is a sign factor ($\pm 1$).

\vspace{3mm}

Let us summarize the results obtained so far. With the exception of
eqn~(\ref{eq:SPEDCHE1}), the system of equations (\ref{eq:SPEDCHH1})--%
(\ref{eq:SPEDCEE8}) has been completely solved in terms of the algebraic
conditions (\ref{eq:SPAC8}), (\ref{eq:SPAC9}), the explicit formulae
(\ref{eq:SPDGPOT1})--(\ref{eq:SPEGPOT2}) for the gauge potential with
the explicit formula (\ref{eq:SPIAC15}) for the odd part $M_\alpha^-$
of the coefficient vectors $M_\alpha$ and the algebraic constraints
(\ref{eq:SPAC4}), (\ref{eq:SPIAC03}), (\ref{eq:SPIAC05}/\ref{eq:SPIAC06})
and (\ref{eq:SPIAC09}/\ref{eq:SPIAC10}). Thus we are left with the task of
verifying the implications of eqns~(\ref{eq:GPOT4}), (\ref{eq:GPOT5}) and
(\ref{eq:SPEDCHE1}).

\pagebreak

Beginning with eqn~(\ref{eq:GPOT4}), we use the functional equation~%
(\ref{eq:EFUNEQ2}) and eqns~(\ref{eq:SPAC8})--(\ref{eq:SPIAC15}) to
compute
\vspace{2mm}
\begin{eqnarray*}
\lefteqn{\partial_k^{\vphantom{\mathfrak{h}}} A_l^{\mathfrak{h}}(q,u) \, - \,
         \partial_l^{\vphantom{\mathfrak{h}}} A_k^{\mathfrak{h}}(q,u) \, + \,
         \sum_{\alpha \ssmin \Delta} \, A_k^\alpha(q,u) A_l^{-\alpha}(q,u) \,
         H_\alpha^{}}                                      \hspace{5mm} \\[1mm]
 &=&\!\! \sum_{\alpha \ssmin \Delta} \zeta^\prime(\alpha(q))
         \Bigl( \, \alpha_l^{} \, (H_k^{} \,, M_{-\alpha}^{}) \,
                   K_\alpha^{} \, - \,
                   \alpha_k^{} \, (H_l^{} \,, M_{-\alpha}^{}) \,
                   K_\alpha^{} \Bigr)                                   \\
 & & \mbox{} + \, \sum_{\alpha \ssmin \Delta} 
         \Phi(\alpha(q),u) \, \Phi(-\alpha(q),u) \;
         (H_k^{} \,, M_\alpha^{}) \, (H_l^{} \,, M_{-\alpha}^{}) \,
         H_\alpha^{}                                                    \\[4mm]
 &=&\!\! {\textstyle \frac{1}{2}} \;
         \sum_{\alpha \ssmin \Delta} \zeta^\prime(\alpha(q))
         \Bigl( \mbox{} + \, \alpha_l^{} \, (H_k^{} \,,
                             M_{-\alpha}^{}) \, K_\alpha^{} \,
                        - \, \alpha_l^{} \, (H_k^{} \,,
                             M_\alpha^{}) \, K_{-\alpha}^{}            \\[-3mm]
 & & \hspace{27.5mm} \mbox{} - \, \alpha_k^{} \, (H_l^{} \,,
                             M_{-\alpha}^{}) \, K_\alpha^{} \,
                        + \, \alpha_k^{} \, (H_l^{} \,,
                             M_\alpha^{}) \, K_{-\alpha}^{}             \\
 & & \hspace{27.5mm} \mbox{} + \, (H_k^{} \,, M_\alpha^{}) \, (H_l^{} \,,
                             M_{-\alpha}^{}) \, H_\alpha^{} \,
                        - \, (H_k^{} \,, M_{-\alpha}^{}) \, (H_l^{} \,,
                             M_\alpha^{}) \, H_\alpha^{} \Bigr)         \\[1mm]
 & & \mbox{} - \, \zeta^\prime(u) \, \sum_{\alpha \ssmin \Delta}
                  (H_k^{} \,, M_\alpha^{}) \,
                  (H_l^{} \,, M_{-\alpha}^{}) \, H_\alpha^{}            \\[4mm]
 &=&\!\! \sum_{\alpha \ssmin \Delta} \zeta^\prime(\alpha(q))
         \Bigl( \mbox{} - \, \alpha_l^{} \, (H_k^{} \,,
                             M_\alpha^-) \, K_\alpha^+ \,
                        + \, \alpha_l^{} \, (H_k^{} \,,
                             M_\alpha^+) \, K_\alpha^-                 \\[-3mm]
 & & \hspace{23.5mm} \mbox{} + \, \alpha_k^{} \, (H_l^{} \,,
                             M_\alpha^-) \, K_\alpha^+ \,
                        - \, \alpha_k^{} \, (H_l^{} \,,
                             M_\alpha^+) \, K_\alpha^-                  \\
 & & \hspace{23.5mm} \mbox{} + \, (H_k^{} \,, M_\alpha^-) \, (H_l^{} \,, 
                             M_\alpha^+) \, H_\alpha^{} \,
                        - \, (H_k^{} \,, M_\alpha^+) \, (H_l^{} \,,
                             M_\alpha^-) \, H_\alpha^{} \Bigr)          \\[1mm]
 & & \mbox{} - \, \zeta^\prime(u) \, \sum_{\alpha \ssmin \Delta}
                  (H_k^{} \,, M_\alpha) \, (H_l^{} \,, M_{-\alpha}) \,
                  (H_\alpha^{})_{\mathfrak{b}}^{}                       \\[4mm]
 &=&\!\! \sum_{\alpha \ssmin \Delta} \zeta^\prime(\alpha(q))
         \Bigl( \mbox{} - \, \frac{\epsilon_\alpha}{\sqrt{2} \, |\alpha|} \;
                             \alpha_l^{} \, \alpha_k^{} \, K_\alpha^+ \,
                        + \, \frac{\epsilon_\alpha}{\sqrt{2} \, |\alpha|} \;
                             \alpha_k^{} \, \alpha_l^{} \, K_\alpha^+ \Bigr)
                                                                        \\[1mm]
 & & \mbox{} + \,
         \sum_{\alpha \ssmin \Delta} \zeta^\prime(\alpha(q))
         \Bigl( \mbox{} + \, \alpha_l^{} \, (H_k^{} \,, M_\alpha^+) \,
                             K_\alpha^- \, - \,
                             \alpha_k^{} \, (H_l^{} \,, M_\alpha^+) \,
                             K_\alpha^-                                \\[-2mm]
 & & \hspace{30.5mm} \mbox{} + \,
                             \frac{\epsilon_\alpha}{\sqrt{2} \, |\alpha|} \;
                             \alpha_k^{} \, (H_l^{} \,, M_\alpha^+) \,
                             (H_\alpha^{})_{\mathfrak{b}}^{} \, - \,
                             \frac{\epsilon_\alpha}{\sqrt{2} \, |\alpha|} \;
                             \alpha_l^{} \, (H_k^{} \,, M_\alpha^+) \,
                             (H_\alpha^{})_{\mathfrak{b}}^{} \Bigr)     \\[2mm]
 & & \mbox{} - \, \zeta^\prime(u) \, \sum_{\alpha \ssmin \Delta}
                  (H_k^{} \,, M_\alpha) \, (H_l^{} \,, M_{-\alpha}) \,
                  (H_\alpha^{})_{\mathfrak{b}}^{}~.
\end{eqnarray*}
Obviously, the whole expression will vanish provided we assume that
\begin{equation} \label{eq:SPIAC17}
 K_\alpha^-~=~\frac{\epsilon_\alpha}{\sqrt{2} \, |\alpha|} \;
              (H_\alpha^{})_{\mathfrak{b}}^{}~,
\end{equation}
which is complementary to the condition (\ref{eq:SPIAC15}) derived previously
% , with
% \begin{equation} \label{eq:SPIAC18}
%  \alpha(K_\alpha^-)~=~\frac{\epsilon_\alpha}{2 \, \sqrt{2}} \;
%                       (1 - \delta_{\theta\alpha,-\alpha}^{}) \, |\alpha|~,
% \end{equation}
and that
\begin{equation} \label{eq:SPIAC19}
 \sum_{\alpha \ssmin \Delta}
 (H_\alpha^{})_{\mathfrak{b}}^{} \otimes M_\alpha^{} \otimes M_{-\alpha}^{}~
 =~0~,
\end{equation}
which is complementary to the condition (\ref{eq:SPIAC06}) derived previously.
Note that in the degenerate case, the same argument works, but eqn~%
(\ref{eq:SPIAC19}) is not needed. Note also that eqn~(\ref{eq:SPIAC03})
can now be eliminated because it follows from eqns~(\ref{eq:SPAC9}),
(\ref{eq:SPIAC15}) and~(\ref{eq:SPIAC17}).

For the proof of eqn~(\ref{eq:GPOT5}), the trick is to split the sum over
roots $\beta$ coming from the third and fourth term into various pieces:
the contribution with $\, \beta = \alpha \,$ and, if \linebreak $\theta\alpha
\neq - \alpha$, also the contribution with $\, \beta = \theta\alpha$,
which cancel mutually, the contribution with $\, \beta = - \alpha \,$
and, if $\, \theta\alpha \neq - \alpha$, also the contribution with
$\, \beta = - \theta\alpha$, which combine with the contributions
coming from the first and second term (transformed using the functional
equation (\ref{eq:EFUNEQ7})), and finally the remaining contributions
with $\, \beta \neq \pm\alpha \,$ and $\, \beta \neq \pm\theta\alpha$:
these can be complemented by terms that also cancel mutually (marked
by underlining) and then be combined with the contributions from the
fifth term (transformed using the functional equation (\ref{eq:EFUNEQ8})):
\vspace{2mm}
\begin{eqnarray*}
\lefteqn{
 \begin{array}{rcl}
  \partial_k^{\vphantom{\alpha}} A_l^\alpha(q,u) \, - \,
  \partial_l^{\vphantom{\alpha}} A_k^\alpha(q,u) \!\! &+& \!\!
  \alpha(A_k^{\mathfrak{h}}(q,u)) \,
  A_l^{\vphantom{\mathfrak{h}}\alpha}(q,u) \, - \,
  \alpha(A_l^{\mathfrak{h}}(q,u)) \,
  A_k^{\vphantom{\mathfrak{h}}\alpha}(q,u) \\[4mm]
  &+& \!\!\! {\displaystyle
              \sum_{\beta,\gamma \ssmin \Delta \atop \beta + \gamma = \alpha}
              N_{\beta,\gamma}^{} A_k^\beta(q,u) A_l^\gamma(q,u)}
 \end{array}
}                                                          \hspace{3mm} \\[5mm]
 &=&\!\! \Phi^\prime(\alpha(q),u) \,
         \alpha_k^{} \, (H_l^{} \,, M_\alpha^{}) \, - \,
         \Phi^\prime(\alpha(q),u) \,
         \alpha_l^{} \, (H_k^{} \,, M_\alpha^{})                        \\[4mm]
 & &\!\! \mbox{} + \, \Phi(\alpha(q),u) \,
         \Bigl( \, \sum_{\beta \ssmin \Delta}
                   \zeta(\beta(q)) \; \alpha(K_{-\beta}^{}) \,
                   (H_k^{} \,, M_\beta^{}) \, (H_l^{} \,, M_\alpha^{}) \, - \,
                   \zeta(u) \; \alpha_k^{} \, (H_l^{} \,, M_\alpha^{})
                   \Bigr)                                               \\
 & &\!\! \mbox{} - \, \Phi(\alpha(q),u) \,
         \Bigl( \, \sum_{\beta \ssmin \Delta}
                   \zeta(\beta(q)) \, \alpha(K_{-\beta}^{}) \,
                   (H_l^{} \,, M_\beta^{}) \, (H_k^{} \,, M_\alpha^{}) \, - \,
                   \zeta(u) \; \alpha_l^{} \, (H_k^{} \,, M_\alpha^{})
                   \Bigr)                                               \\
 & &\!\! \mbox{} + \,
         \sum_{\gamma,\delta \ssmin \Delta \atop
               \gamma + \delta = \alpha}
         N_{\gamma,\delta}^{} \, \Phi(\gamma(q),u) \, \Phi(\delta(q),u) \,
         (H_k^{} \,, M_\gamma^{}) \, (H_l^{} \,, M_\delta^{})           \\[3mm]
 &=&\!\! \Bigl( \Phi^\prime(\alpha(q),u) \, - \,
                \Phi(\alpha(q),u) \, \zeta(u) \Bigr) \,
         \Bigl( \, (H_k \,, H_\alpha) \, (H_l \,, M_\alpha) \, - \,
                   (H_l \,, H_\alpha) \, (H_k \,, M_\alpha) \Bigr)      \\[2mm]
 & &\!\! \mbox{} + \, \Phi(\alpha(q),u) \, \zeta(\alpha(q))             \\[1mm]
 & &\!\! \mbox{} \hspace{5mm} \times
         \Bigl( \mbox{} + \, \alpha(K_{-\alpha}) \,
                   (H_k \,, M_\alpha) \, (H_l \,, M_\alpha)            \\[-2mm]
 & &\!\! \mbox{} \hspace{7mm} \phantom{\times \Bigl(}
                \mbox{} - \, \alpha(K_\alpha) \,
                   (H_k \,, M_{-\alpha}) \, (H_l \,, M_\alpha)         \\[-2mm]
 & &\!\! \mbox{} \hspace{7mm} \phantom{\times \Bigl(}
                \mbox{} - \, (1 - \delta_{\theta\alpha,-\alpha}) \,
                   \alpha(K_{-\theta\alpha}) \,
                   (H_k \,, M_{\theta\alpha}) \, (H_l \,, M_\alpha)    \\[-2mm]
 & &\!\! \mbox{} \hspace{7mm} \phantom{\times \Bigl(}
                \mbox{} + \, (1 - \delta_{\theta\alpha,-\alpha}) \,
                   \alpha(K_{\theta\alpha}) \,
                   (H_k \,, M_{-\theta\alpha}) \, (H_l \,, M_\alpha)   \\[-2mm]
 & &\!\! \mbox{} \hspace{7mm} \phantom{\times \Bigl(}
                \mbox{} - \, \alpha(K_{-\alpha}) \,
                   (H_l \,, M_\alpha) \, (H_k \,, M_\alpha)            \\[-2mm]
 & &\!\! \mbox{} \hspace{7mm} \phantom{\times \Bigl(}
                \mbox{} + \, \alpha(K_\alpha) \,
                   (H_l \,, M_{-\alpha}) \, (H_k \,, M_\alpha)         \\[-2mm]
 & &\!\! \mbox{} \hspace{7mm} \phantom{\times \Bigl(}
                \mbox{} + \, (1 - \delta_{\theta\alpha,-\alpha}) \,
                   \alpha(K_{-\theta\alpha}) \,
                   (H_l \,, M_{\theta\alpha}) \, (H_k \,, M_\alpha)    \\[-2mm]
 & &\!\! \mbox{} \hspace{7mm} \phantom{\times \Bigl(}
                \mbox{} - \, (1 - \delta_{\theta\alpha,-\alpha}) \,
                   \alpha(K_{\theta\alpha}) \,
                   (H_l \,, M_{-\theta\alpha}) \, (H_k \,, M_\alpha) \, \Bigr)
                                                                        \\[1mm]
 & &\!\! \mbox{} + \, \Phi(\alpha(q),u) \!\!\!
         \sum_{\beta \ssmin \Delta \atop
               \beta \neq \pm\alpha,\beta \neq \pm \theta\alpha} \!\!\!\!\!
         \zeta(\beta(q)) \, (H_k^{} \,, M_\beta^{})
         \Bigl( \, \alpha(K_{-\beta}^{}) \, (H_l^{} \,, M_\alpha^{}) \, + \,
                   \underline{\beta(M_\alpha^{}) \, (H_l^{} \,, M_{-\beta}^{})}
                \, \Bigr)                                               \\
 & &\!\! \mbox{} - \, \Phi(\alpha(q),u) \!\!\!
         \sum_{\beta \ssmin \Delta \atop
               \beta \neq \pm\alpha,\beta \neq \pm \theta\alpha} \!\!\!\!\!
         \zeta(\beta(q)) \, (H_l^{} \,, M_\beta^{})
         \Bigl( \, \alpha(K_{-\beta}^{}) \, (H_k^{} \,, M_\alpha^{}) \, + \,
                   \underline{\beta(M_\alpha^{}) \, (H_k^{} \,, M_{-\beta}^{})}
                \, \Bigr)                                               \\
 & &\!\! \mbox{} + \, \Phi(\alpha(q),u)
         \sum_{\gamma,\delta \ssmin \Delta \atop
               \gamma + \delta = \alpha}
         \Bigl( \, \zeta(\gamma(q)) \, + \, \zeta(\delta(q)) \Bigr) \,
         N_{\gamma,\delta} \,
         (H_k \,, M_\gamma) \, (H_l \,, M_\delta)                       \\
 & &\!\! \mbox{} - \, \Phi(\alpha(q),u)
         \Bigl( \, \zeta(\alpha(q) + u) \, - \, \zeta(u) \Bigr)
         \sum_{\gamma,\delta \ssmin \Delta \atop
               \gamma + \delta = \alpha}
         N_{\gamma,\delta} \,
         (H_k \,, M_\gamma) \, (H_l \,, M_\delta)                       \\[3mm]
 &=&\!\! \Phi(\alpha(q),u) \,
         \Bigl( \, \zeta(\alpha(q)+u) \, - \, \zeta(\alpha(q)) \, - \,
                   \zeta(u) \Bigr)                                      \\
 & &\!\! \mbox{} \times \,
         \Bigl( \, (H_k^{} \,, H_\alpha^{}) \, (H_l^{} \,, M_\alpha^+) \, - \,
                   (H_k^{} \,, M_\alpha^+) \, (H_l^{} \,, H_\alpha^{}) \\[-1mm]
 & &\!\! \mbox{} \hspace{1cm}
         - \, 2 \, (2 - \delta_{\theta\alpha,-\alpha}) \, \alpha(K_\alpha) \,
              (H_k^{} \,, M_\alpha^-) \, (H_l^{} \,, M_\alpha^+)       \\[-1mm]
 & &\!\! \mbox{} \hspace{1cm}
         + \, 2 \, (2 - \delta_{\theta\alpha,-\alpha}) \, \alpha(K_\alpha) \,
              (H_k^{} \,, M_\alpha^+) \, (H_l^{} \,, M_\alpha^-) \Bigr) \\[2mm]
 & &\!\! \mbox{} + \, \Phi(\alpha(q),u) \,
         \Bigl( \, \zeta(\alpha(q)+u) \, - \, \zeta(u) \Bigr)           \\
 & &\!\! \mbox{} \hspace{6mm} \times \,
         \Bigl( \, + \; 2 \, (2 - \delta_{\theta\alpha,-\alpha}) \,
                        \alpha(K_\alpha) \, (H_k^{} \,, M_\alpha^-) \,
                        (H_l^{} \,, M_\alpha^+)                        \\[-1mm]
 & &\!\! \mbox{} \hspace{14mm}
                   -    2 \, (2 - \delta_{\theta\alpha,-\alpha}) \,
                        \alpha(K_\alpha) \, (H_k^{} \,, M_\alpha^+) \,
                        (H_l^{} \,, M_\alpha^-)                         \\
 & &\!\! \mbox{} \hspace{14mm} - \,
         \sum_{\beta,\gamma \ssmin \Delta \atop
               \beta + \gamma = \alpha}
         N_{\beta,\gamma} \,
         (H_k \,, M_\beta) \, (H_l \,, M_\gamma) \, \Bigr)              \\
 & &\!\! \mbox{} + \, \Phi(\alpha(q),u)
         \sum_{\beta \ssmin \Delta \atop
               \beta \neq \pm\alpha,\beta \neq \pm\theta\alpha}
         \zeta(\beta(q)) \, (H_k^{} \,, M_\beta^{})                    \\[-5mm]
 & &\!\! \mbox{} \hspace{42mm} \times
         \Bigl( H_l^{} \,, \alpha(K_{-\beta}^{}) \, M_\alpha^{} \, + \,
                           \beta(M_\alpha^{}) \, M_{-\beta}^{} \, + \,
                           N_{\beta,\alpha-\beta}^{} \,
                           M_{\alpha-\beta}^{} \Bigr)                   \\
 & &\!\! \mbox{} - \, \Phi(\alpha(q),u)
         \sum_{\beta \ssmin \Delta \atop
               \beta \neq \pm\alpha,\beta \neq \pm\theta\alpha}
         \zeta(\beta(q)) \, (H_l^{} \,, M_\beta^{})                    \\[-5mm]
 & &\!\! \mbox{} \hspace*{42mm} \times
         \Bigl( H_k^{} \,, \alpha(K_{-\beta}^{}) \, M_\alpha^{} \, + \,
                           \beta(M_\alpha^{}) \, M_{-\beta}^{} \, - \,
                           N_{\alpha-\beta,\beta}^{} \,
                           M_{\alpha-\beta}^{} \Bigr)                   \\
 & &\!\! \mbox{} - \, (1 - \delta_{\theta\alpha,-\alpha}) \,
         \Phi(\alpha(q),u) \, \zeta(\alpha(q)) \, \Bigl( \, + \,\,
                N_{\theta\alpha,\alpha-\theta\alpha}^{} \,
                (H_k^{} \,, M_{\theta\alpha}^{}) \,
                (H_l^{} \,, M_{\alpha-\theta\alpha}^{})                 \\
 & &\!\! \mbox{} \hspace{6.25cm} - \,
                N_{-\theta\alpha,\alpha+\theta\alpha}^{} \,
                (H_k^{} \,, M_{-\theta\alpha}^{}) \,
                (H_l^{} \,, M_{\alpha+\theta\alpha}^{})                 \\[1mm]
 & &\!\! \mbox{} \hspace{6.25cm} + \,
                N_{\alpha-\theta\alpha,\theta\alpha}^{} \,
                (H_l^{} \,, M_{\theta\alpha}^{}) \,
                (H_k^{} \,, M_{\alpha-\theta\alpha}^{})                \\[-1mm]
 & &\!\! \mbox{} \hspace{6.25cm} - \,
                N_{\alpha+\theta\alpha,-\theta\alpha}^{} \,
                (H_l^{} \,, M_{-\theta\alpha}^{}) \,
                (H_k^{} \,, M_{\alpha+\theta\alpha}^{}) \Bigr)~.
\end{eqnarray*}
The last term vanishes due to the condition (\ref{eq:SPAC9}), whereas the
previous two terms vanish due to eqn~(\ref{eq:SPIAC10}). The first term also
vanishes because, as already observed before, $M_\alpha^+ = 0 \,$ for real
roots $\alpha$ while, according to eqns~(\ref{eq:SPIAC03}), (\ref{eq:SPIAC04})
and~(\ref{eq:SPIAC14}),
\[
 4 \;\! \alpha(K_\alpha^{}) \, M_\alpha^-~
 =~4 \;\! \alpha(K_\alpha^-) \, M_\alpha^-~
 =~4 \;\! \alpha(M_\alpha^-) \, M_\alpha^-~
 =~(H_\alpha^{})_{\mathfrak{a}}^{}
\]
for complex roots $\alpha$. Finally, the same reasoning shows that the
second term will vanish provided we assume that
\begin{equation} \label{eq:SPIAC20}
 \sum_{\beta,\gamma \ssmin \Delta \atop \beta + \gamma = \alpha}
 N_{\beta,\gamma}^{} \, M_\beta^{} \otimes M_\gamma^{}~
 =~\frac{\epsilon_\alpha}{\sqrt{2}} \, |\alpha| \,
   \Bigl( M_\alpha^{} \otimes M_{-\alpha}^{} \, - \,
          M_{-\alpha}^{} \otimes M_\alpha^{} \Bigr)~,
\end{equation}
which is easily reduced to the second equation in eqn~(\ref{eq:SPEAC2}) by
noting that
\[
 {\textstyle \frac{1}{2}}
 \left( M_\alpha^{} \otimes M_{-\alpha}^{} \, - \,
        M_{-\alpha}^{}\otimes M_\alpha^{} \right) \,
 =~M_\alpha^- \otimes M_\alpha^+ \, - \, M_\alpha^+ \otimes M_\alpha^-~
 =~M_\alpha^- \otimes M_\alpha^{} \, - \, M_\alpha^{} \otimes
 M_\alpha^-
\]
and using eqn~(\ref{eq:SPAC5}). Note that in the degenerate case, the same
argument works, but eqn~(\ref{eq:SPIAC20}) is not needed.

The proof of eqn~(\ref{eq:SPEDCHE1}) proceeds along similar lines, using the
functional equations~(\ref{eq:EFUNEQ6})--(\ref{eq:EFUNEQ8}):
\vspace{2mm}
\begin{eqnarray*}
\lefteqn{\alpha_k^{} \, \Phi^\prime(\alpha(q),v) \, K_\alpha^{}}
                                                           \hspace{4mm} \\[2mm]
 & &\!\! \mbox{} + \, {\textstyle \frac{1}{2}} \,
         \Bigl( \, \zeta(u-v) + \zeta(u+v) \Bigr) \,
         A_k^\alpha(q,v) \, (H_\alpha^{})_{\mathfrak{a}}^{}             \\[1mm]
 & &\!\! \mbox{} + \, {\textstyle \frac{1}{2}} \,
         \Bigl( \, \zeta(u-v) - \zeta(u+v) + 2 \>\! \zeta(v) \Bigr) \,
         A_k^\alpha(q,v) \, (H_\alpha^{})_{\mathfrak{b}}^{}             \\[2mm]
 & &\!\! \mbox{} + \, \Phi(\alpha(q),v) \,
                      \alpha(A_k^{\mathfrak{h}}(q,v)) \, K_\alpha^{} \, -
                      \sum_{\gamma,\delta \ssmin \Delta
                            \atop \gamma + \delta = \alpha}
                      N_{\gamma,\delta}^{} \, \Phi(\gamma(q),v) \,
                      A_k^\delta(q,v) \, K_{\gamma}^{}                  \\[2mm]
 & &\!\! \mbox{} + \, {\textstyle \frac{1}{2}} \, \Phi(\alpha(q),v-u) \, 
                      A_k^\alpha(q,u) \, H_\alpha^{} \,
                 + \, {\textstyle \frac{1}{2}} \, \Phi(\alpha(q),v+u) \,
                      A_k^\alpha(q,-u) \, \theta H_\alpha^{}            \\[2mm]
 & &\!\! \mbox{} + \, {\displaystyle \sum_{j=1}^r} \, \alpha_j^{} \,
                      A_k^\alpha(q,v) A_j^{\mathfrak{h}}(q,u)           \\[4mm]
 &=&     \Phi^\prime(\alpha(q),v) \; \alpha_k^{} \, K_\alpha^{}         \\[2mm]
 & & \mbox{} + \, {\textstyle \frac{1}{2}} \, \Phi(\alpha(q),v) \;
     \Bigl( \, \zeta(u-v) + \zeta(u+v) \Bigr) \,
     (H_k^{} \,, M_\alpha^{}) \, (H_\alpha^{})_{\mathfrak{a}}^{}        \\[1mm]
 & & \mbox{} + \, {\textstyle \frac{1}{2}} \, \Phi(\alpha(q),v) \;
     \Bigl( \, \zeta(u-v) - \zeta(u+v) + 2 \>\! \zeta(v) \Bigr) \,
     (H_k^{} \,, M_\alpha^{}) \, (H_\alpha^{})_{\mathfrak{b}}^{}        \\[2mm]
 & & \mbox{} + \, \Phi(\alpha(q),v) \,
     \Bigl( \, \sum_{\beta \ssmin \Delta}
               \zeta(\beta(q)) \; \alpha(K_{-\beta}^{}) \,
               (H_k^{} \,, M_\beta^{}) \, K_\alpha^{} \, - \,
               \zeta(v) \; \alpha_k^{} \, K_\alpha^{} \Bigr)            \\[1mm]
 & & \mbox{} - \,
     \sum_{\gamma,\delta \ssmin \Delta \atop \gamma + \delta = \alpha}
     N_{\gamma,\delta}^{} \, \Phi(\gamma(q),v) \, \Phi(\delta(q),v) \,
     (H_k^{} \,, M_\delta^{}) \, K_\gamma^{}                            \\[1mm]
 & & \mbox{} + \, {\textstyle \frac{1}{2}} \, \Phi(\alpha(q),v-u) \,
     \Phi(\alpha(q),u) \, (H_k^{} \,, M_\alpha^{}) \,
     \Bigl( \, (H_\alpha^{})_{\mathfrak{a}}^{} \, + \,
               (H_\alpha^{})_{\mathfrak{b}}^{} \Bigr)                   \\[2mm]
 & & \mbox{} - \, {\textstyle \frac{1}{2}} \, \Phi(\alpha(q),v+u) \,
     \Phi(\alpha(q),-u) \, (H_k^{} \,, M_\alpha^{}) \,
     \Bigl( \, (H_\alpha^{})_{\mathfrak{a}}^{} \, - \,
               (H_\alpha^{})_{\mathfrak{b}}^{} \Bigr)                   \\[2mm]
 & & \mbox{} - \, \Phi(\alpha(q),v) \,
     \sum_{j=1}^r \sum_{\beta \ssmin \Delta}
     \zeta(\beta(q)) \; \alpha_j^{} \, (H_j^{} \,, M_{-\beta}^{}) \,
     (H_k^{} \,, M_\alpha^{}) \, K_\beta^{}                             \\
 & & \mbox{} - \, \Phi(\alpha(q),v) \, \zeta(u) \,
     \sum_{j=1}^r \alpha_j^{} \, (H_k^{} \,, M_\alpha^{}) \, H_j^{}     \\[4mm]
 &=& \Bigl( \Phi^\prime(\alpha(q),v) \, - \,
            \Phi(\alpha(q),v) \, \zeta(v) \Bigr) \,
            (H_k^{} \,, H_\alpha^{}) \, K_\alpha^{}                     \\[2mm]
 & & \mbox{} + \, {\textstyle \frac{1}{2}} \,
     \Bigl( \bigl( \zeta(u-v) + \zeta(u+v) - 2 \>\! \zeta(u) \bigr) \,
            \Phi(\alpha(q),v)                                           \\
 & & \mbox{} \hspace{12mm} + \,
     \Phi(\alpha(q),v-u) \, \Phi(\alpha(q),u)                           \\
 & & \mbox{} \hspace{12mm} - \,
     \Phi(\alpha(q),v+u) \, \Phi(\alpha(q),-u) \Bigr) \,
     (H_k^{} \,, M_\alpha^{}) \, (H_\alpha^{})_{\mathfrak{a}}^{}        \\[2mm]
 & & \mbox{} + \, {\textstyle \frac{1}{2}} \,
     \Bigl( \bigl( \zeta(u-v) - \zeta(u+v) + 2 \>\! \zeta(v) \bigr) \,
            \Phi(\alpha(q),v)                                           \\
 & & \mbox{} \hspace{12mm} + \,
     \Phi(\alpha(q),v-u) \, \Phi(\alpha(q),u)                           \\
 & & \mbox{} \hspace{12mm} + \,
     \Phi(\alpha(q),v+u) \, \Phi(\alpha(q),-u) \Bigr) \,
     (H_k^{} \,, M_\alpha^{}) \, (H_\alpha^{})_{\mathfrak{b}}^{}        \\[2mm]
 & & \mbox{} + \, \Phi(\alpha(q),v) \, \zeta(\alpha(q)) \,
     \Bigl( \mbox{} + \, \alpha(K_{-\alpha}) \,
                         (H_k \,, M_\alpha) \, K_\alpha \,
                    - \, \alpha(K_\alpha) \,
                         (H_k \,, M_{-\alpha}) \, K_\alpha              \\
 & & \hspace{41mm}
            \mbox{} - \, (1 - \delta_{\theta\alpha,-\alpha}) \,
                         \alpha(K_{-\theta\alpha}) \,
                         (H_k \,, M_{\theta\alpha}) \, K_\alpha         \\[1mm]
 & & \hspace{41mm}
            \mbox{} + \, (1 - \delta_{\theta\alpha,-\alpha}) \,
                         \alpha(K_{\theta\alpha}) \,
                         (H_k \,, M_{-\theta\alpha}) \, K_\alpha        \\[1mm]
 & & \hspace{41mm}
            \mbox{} - \, \alpha(M_{-\alpha}) \,
                         (H_k \,, M_\alpha) \, K_\alpha
            \mbox{} + \, \alpha(M_\alpha) \,
                         (H_k \,, M_\alpha) \, K_{-\alpha}              \\[1mm]
 & & \hspace{41mm}
            \mbox{} + \, (1 - \delta_{\theta\alpha,-\alpha}) \,
                         \alpha(M_{-\theta\alpha}) \,
                         (H_k \,, M_\alpha) \, K_{\theta\alpha}         \\
 & & \hspace{41mm}
            \mbox{} - \, (1 - \delta_{\theta\alpha,-\alpha}) \,
                         \alpha(M_{\theta\alpha}) \,
                         (H_k \,, M_\alpha) \, K_{-\theta\alpha} \Bigr) \\[1mm]
 & & \mbox{} + \, \Phi(\alpha(q),v)
     \sum_{\beta \ssmin \Delta \atop
           \beta \neq \pm\alpha,\beta \neq \pm\theta\alpha}
     \zeta(\beta(q)) \; (H_k^{} \,, M_\beta^{}) \,
     \Bigl( \alpha(K_{-\beta}^{}) \, K_\alpha^{} \, + \,
            \underline{\beta(K_\alpha^{}) \, K_{-\beta}^{}} \Bigr)    \\
 & & \mbox{} - \, \Phi(\alpha(q),v)
     \sum_{\beta \ssmin \Delta \atop
           \beta \neq \pm\alpha,\beta \neq \pm\theta\alpha}
     \zeta(\beta(q)) \;
     \Bigl( H_k^{} \,, \alpha(M_{-\beta}^{}) \, M_\alpha^{} \, + \,
            \underline{\beta(K_\alpha^{}) \, M_{-\beta}^{}} \Bigr) \,
     K_\beta^{}                                                         \\
 & & \mbox{} - \, \Phi(\alpha(q),v)
     \sum_{\gamma,\delta \ssmin \Delta \atop \gamma + \delta = \alpha}
     \Bigl( \, \zeta(\gamma(q)) \, + \, \zeta(\delta(q)) \, \Bigr)
     N_{\gamma,\delta} \, (H_k \,, M_\delta) \, K_\gamma                \\
 & & \mbox{} + \, \Phi(\alpha(q),v)
     \Bigl( \, \zeta(\alpha(q) + v) \, - \, \zeta(v) \, \Bigr)
     \sum_{\gamma,\delta \ssmin \Delta \atop \gamma + \delta = \alpha}
     N_{\gamma,\delta} \, (H_k \,, M_\delta) \, K_\gamma               %\\[3mm]
 \\[3cm]
 &=& \Phi(\alpha(q),v) \,
     \Bigl( \, \zeta(\alpha(q)+v) \, - \, \zeta(\alpha(q)) \, - \,
               \zeta(v) \Bigr)                                          \\
 & & \mbox{} \times \,
     \Bigl( \, (H_k^{} \,, H_\alpha^{}) \, K_\alpha^{} \, - \,
               (H_k^{} \,, M_\alpha^{}) \, (H_\alpha^{})_{\mathfrak{b}}^{}
                                                                       \\[-1mm]
 & & \mbox{} \hspace{1cm}
     - \, (2 - \delta_{\theta\alpha,-\alpha}) \, 
          (\alpha(K_{-\alpha}) - \alpha(M_{-\alpha})) \,
          (H_k^{} \,, M_\alpha) \, K_\alpha                             \\
 & & \mbox{} \hspace{1cm}
     + \, (2 - \delta_{\theta\alpha,-\alpha}) \, \alpha(K_\alpha) \,
          (H_k^{} \,, M_{-\alpha}) \, K_\alpha                         \\[-1mm]
 & & \mbox{} \hspace{1cm}
     - \, (2 - \delta_{\theta\alpha,-\alpha}) \, \alpha(M_\alpha) \,
          (H_k^{} \,, M_\alpha) \, K_{-\alpha} \Bigr)                   \\[2mm]
 & & \mbox{} + \, \Phi(\alpha(q),v) \,
     \Bigl( \, \zeta(\alpha(q)+v) \, - \, \zeta(v) \Bigr)               \\
 & & \mbox{} \hspace{6mm} \times \,
     \Bigl( \, + \; (2 - \delta_{\theta\alpha,-\alpha}) \,
                    (\alpha(K_{-\alpha}) - \alpha(M_{-\alpha})) \,
                    (H_k^{} \,, M_\alpha) \, K_\alpha                  \\[-1mm]
 & & \mbox{} \hspace{14mm}
               -    (2 - \delta_{\theta\alpha,-\alpha}) \, \alpha(K_\alpha) \,
                    (H_k^{} \,, M_{-\alpha}) \, K_\alpha                \\
 & & \mbox{} \hspace{14mm}
               +    (2 - \delta_{\theta\alpha,-\alpha}) \, \alpha(M_\alpha) \,
                    (H_k^{} \,, M_\alpha) \, K_{-\alpha}                \\
 & & \mbox{} \hspace{14mm} - \,
     \sum_{\beta,\gamma \ssmin \Delta \atop \beta + \gamma = \alpha}
     N_{\beta,\gamma} \, (H_k \,, M_\beta) \, K_\gamma \, \Bigr)        \\[2mm]
 & & \mbox{} + \, \Phi(\alpha(q),v) \!\!
     \sum_{\beta \ssmin \Delta \atop
           \beta \neq \pm\alpha,\beta \neq \pm\theta\alpha}
     \zeta(\beta(q))                                                   \\[-5mm]
 & & \hspace*{41mm} \times \,
     (H_k^{} \,, M_\beta^{})
     \Bigl( \alpha(K_{-\beta}^{}) \, K_\alpha^{} +
            \beta(K_\alpha^{}) \, K_{-\beta}^{} -
            N_{\alpha-\beta,\beta}^{} \, K_{\alpha-\beta}^{} \Bigr)     \\[2mm]
 & & \mbox{} - \, \Phi(\alpha(q),v) \!\!
     \sum_{\beta \ssmin \Delta \atop
           \beta \neq \pm\alpha,\beta \neq \pm\theta\alpha}
     \zeta(\beta(q))                                                   \\[-5mm]
 & & \hspace*{41mm} \times
     \Bigl( H_k^{} \,, \alpha(M_{-\beta}^{}) \, M_\alpha^{} +
                       \beta(K_\alpha^{}) \, M_{-\beta}^{} +
                       N_{\beta,\alpha-\beta}^{} \,
                       M_{\alpha-\beta}^{} \Bigr) \, K_\beta^{}         \\[4mm]
 & & \mbox{} + \, (1 - \delta_{\theta\alpha,-\alpha}) \,
     \Phi(\alpha(q),v) \, \zeta(\alpha(q)) \,
     \Bigl( \, + \,\, N_{\alpha-\theta\alpha,\theta\alpha}^{} \,
                      (H_k^{} \,, M_{\theta\alpha}^{}) \,
                      K_{\alpha-\theta\alpha}^{}                        \\
 & & \mbox{} \hspace{6.25cm}
               - \,   N_{\alpha+\theta\alpha,-\theta\alpha}^{} \,
                      (H_k^{} \,, M_{-\theta\alpha}^{}) \,
                      K_{\alpha+\theta\alpha}^{}                        \\[1mm]
 & & \mbox{} \hspace{6.25cm}
               + \,   N_{\theta\alpha,\alpha-\theta\alpha}^{} \,
                      (H_k^{} \,, M_{\alpha-\theta\alpha}^{}) \,
                      K_{\theta\alpha}^{}                              \\[-1mm]
 & & \mbox{} \hspace{6.25cm}
               - \,   N_{-\theta\alpha,\alpha+\theta\alpha}^{} \,
                      (H_k^{} \,, M_{\alpha+\theta\alpha}^{}) \,
                      K_{-\theta\alpha}^{} \Bigr)~.
\end{eqnarray*}

\noindent
The last term vanishes due to the condition (\ref{eq:SPAC9}), whereas the
previous two terms vanish due to eqn~(\ref{eq:SPIAC10}) and provided we
impose the relation
\begin{equation} \label{eq:SPIAC21}
 \begin{array}{c}
  \alpha(K_\beta^{}) \, K_\alpha^{} \, - \, {\textstyle \frac{1}{2}} \,
  \Bigl( N_{\alpha,\beta}^{} \, K_{\alpha+\beta}^{} \, + \,
         N_{\theta\alpha,\beta}^{} \, K_{\theta\alpha+\beta}^{} \Bigr) \, - \,
  \beta(K_\alpha^{}) \, K_\beta^{}~=~0 \\[4mm]
  \mbox{for $\, \alpha,\beta \smin \Delta \,$ such that
  $\, \beta \neq \pm \alpha \,,\, \beta \neq \pm \theta\alpha$}~,
 \end{array}
\end{equation}
which is complementary to it. The first term also vanishes because,
according to eqns~(\ref{eq:SPIAC15}) and (\ref{eq:SPIAC17}),

\pagebreak

\begin{eqnarray*}
\lefteqn{(H_k^{} \,, H_\alpha^{}) \, K_\alpha^{} \, - \,
         (H_k^{} \,, M_\alpha^{}) \, (H_\alpha^{})_{\mathfrak{b}}^{}}
                                                           \hspace{5mm} \\[2mm]
 &=&\!\! (H_k^{} \,, H_\alpha^{}) \, K_\alpha^+ \, - \,
         (H_k^{} \,, M_\alpha^+) \, (H_\alpha^{})_{\mathfrak{b}}^{}     \\[1mm]
 &=&\!\! \epsilon_\alpha \, \sqrt{2} \, |\alpha| \,
         \Bigl( (H_k^{} \,, M_\alpha^-) \, K_\alpha^+ \, - \,
                (H_k^{} \,, M_\alpha^+) \, K_\alpha^- \Bigr)            \\[1mm]
 &=&\!\! \frac{\epsilon_\alpha}{\sqrt{2}} \, |\alpha| \,
         \Bigl( (H_k^{} \,, M_\alpha^{}) \, K_{-\alpha}^{} \, - \,
                (H_k^{} \,, M_{-\alpha}^{}) \, K_\alpha^{} \Bigr)~,
\end{eqnarray*}
whereas for real roots $\, \alpha \smin\, \Delta$ $(\theta\alpha = -\alpha)$,
we have $\, K_{-\alpha} = K_\alpha$, $M_{-\alpha} = - M_\alpha$,
$\alpha(K_{\pm\alpha}) = 0$ and hence by eqn~(\ref{eq:SPIAC13})
\begin{eqnarray*}
\lefteqn{\begin{array}{l}
          \mbox{} - \, (2 - \delta_{\theta\alpha,-\alpha}^{}) \, 
          (\alpha(K_{-\alpha}^{}) - \alpha(M_{-\alpha}^{})) \,
          (H_k^{} \,, M_\alpha^{}) \, K_\alpha^{} \\[1mm]
          \mbox{} + \, (2 - \delta_{\theta\alpha,-\alpha}^{}) \,
          \alpha(K_\alpha^{}) \,
          (H_k^{} \,, M_{-\alpha}^{}) \, K_\alpha^{} \\[1mm]
          \mbox{} - \, (2 - \delta_{\theta\alpha,-\alpha}^{}) \,
          \alpha(M_\alpha^{}) \,
          (H_k^{} \,, M_\alpha^{}) \, K_{-\alpha}^{}
         \end{array}}                                      \hspace{5mm} \\[3mm]
 &=&\!\! \alpha(M_{-\alpha}^{}) \, (H_k^{} \,, M_\alpha^{}) \,
         K_\alpha^{} \, - \,
         \alpha(M_\alpha^{}) \, (H_k^{} \,, M_\alpha^{}) \,
         K_{-\alpha}^{}                                                 \\[2mm]
 &=&\!\! \mbox{} - \, \frac{\epsilon_\alpha}{\sqrt{2}} \, |\alpha| \,
         \Bigl( (H_k^{} \,, M_\alpha^{}) \, K_{-\alpha}^{} \, - \,
                (H_k^{} \,, M_{-\alpha}^{}) \, K_\alpha^{} \Bigr)~,
\end{eqnarray*}
while for complex roots $\, \alpha \smin\, \Delta$ $(\theta\alpha \neq \pm
\alpha$), we have $\, \alpha(K_{\pm\alpha}^{}) = \alpha(M_{\pm\alpha}^{}) \,$
and hence by eqn~(\ref{eq:SPIAC14})
\begin{eqnarray*}
\lefteqn{\begin{array}{l}
          \mbox{} - \, (2 - \delta_{\theta\alpha,-\alpha}^{}) \, 
          (\alpha(K_{-\alpha}^{}) - \alpha(M_{-\alpha}^{})) \,
          (H_k^{} \,, M_\alpha^{}) \, K_\alpha^{} \\[1mm]
          \mbox{} + \, (2 - \delta_{\theta\alpha,-\alpha}^{}) \,
          \alpha(K_\alpha^{}) \,
          (H_k^{} \,, M_{-\alpha}^{}) \, K_\alpha^{} \\[1mm]
          \mbox{} - \, (2 - \delta_{\theta\alpha,-\alpha}^{}) \,
          \alpha(M_\alpha^{}) \,
          (H_k^{} \,, M_\alpha^{}) \, K_{-\alpha}^{}
         \end{array}}                                      \hspace{5mm} \\[3mm]
 &=&\!\! 2 \, \alpha(M_\alpha^{}) \, (H_k^{} \,, M_{-\alpha}^{}) \,
         K_\alpha^{} \, - \,
         2 \, \alpha(M_\alpha^{}) \, (H_k^{} \,, M_\alpha^{}) \,
         K_{-\alpha}^{}                                                 \\[2mm]
 &=&\!\! \mbox{} - \, \frac{\epsilon_\alpha}{\sqrt{2}} \, |\alpha| \,
         \Bigl( (H_k^{} \,, M_\alpha^{}) \, K_{-\alpha}^{} \, - \,
                (H_k^{} \,, M_{-\alpha}^{}) \, K_\alpha^{} \Bigr)~.
\end{eqnarray*}
Finally, the same reasoning shows that the second term will vanish provided
we assume that
\begin{equation} \label{eq:SPIAC22}
 \sum_{\beta,\gamma \ssmin \Delta \atop \beta + \gamma = \alpha}
 N_{\beta,\gamma}^{} \, M_\beta^{} \otimes K_\gamma^{}~
 =~\frac{\epsilon_\alpha}{\sqrt{2}} \, |\alpha| \,
   \Bigl( M_\alpha^{} \otimes K_{-\alpha}^{} \, - \,
          M_{-\alpha}^{} \otimes K_\alpha^{} \Bigr)
\end{equation}
which is complementary to the condition (\ref{eq:SPIAC20}) derived previously
and is easily reduced to the first equation in eqn~(\ref{eq:SPEAC2}) by
noting that
\[
 {\textstyle \frac{1}{2}}
 \left( M_\alpha^{} \otimes K_{-\alpha}^{} \, - \,
        M_{-\alpha}^{}\otimes K_\alpha^{} \right) \,
 =~M_\alpha^- \otimes K_\alpha^+ \, - \, M_\alpha^+ \otimes K_\alpha^-~
 =~M_\alpha^- \otimes K_\alpha^{} \, - \, M_\alpha^{} \otimes K_\alpha^-
\]
and using eqn~(\ref{eq:SPAC5}). Note that in the degenerate case, the same
argument works, but eqn~(\ref{eq:SPIAC22}) is not needed.
\hspace*{\fill} \raisebox{-2mm}{$\Box$} \vspace{4mm}

Having concluded the proof of Proposition~2, we pass to analyzing the
implications of the algebraic constraints that we have derived. The first
thing that suggests itself is to combine the generators $K_\alpha$ and
$M_\alpha$ into generators
\begin{equation} \label{eq:LASPG1}
 F_\alpha~=~K_\alpha + M_\alpha
\end{equation}
which, according to eqns~(\ref{eq:SPCOEF5}) and~(\ref{eq:SPCOEF6}), define
a $\theta$-covariant map from $\Delta$ to $\mathfrak{h}_{\mathbb{R}}$:
\begin{equation} \label{eq:LASPG2}
 \theta F_\alpha~=~F_{\theta\alpha}~.
\end{equation}
Then eqn~(\ref{eq:SPAC4}) becomes equivalent to eqn~(\ref{eq:LAAC4})
and eqn~(\ref{eq:SPAC5}) becomes equivalent to eqn~(\ref{eq:LAAC5}).
The relation between eqn~(\ref{eq:SPAC6}) and eqn~(\ref{eq:LAAC6}),
however, is more intricate. To approach this question, note that
eqn~(\ref{eq:LAAC6}) decomposes naturally into a component along
$\mathrm{i} \mathfrak{b}_0$,
\begin{equation} \label{eq:SPAC10}
 \begin{array}{c}
  \alpha(F_\beta^{}) \, K_\alpha^{} \, - \, \beta(F_\alpha^{}) \, K_\beta^{}~
  =~N_{\alpha,\beta}^{} \, K_{\alpha+\beta}^{} \\[2mm]
  \mbox{for $\, \alpha,\beta \smin\, \Delta \,$ such that
  $\, \beta \neq \pm \alpha \,,\, \beta \neq \pm \theta\alpha$}~,
 \end{array}
\end{equation}
and a component along $\mathfrak{a}_0$,
\begin{equation} \label{eq:SPAC11}
 \begin{array}{c}
  \alpha(F_\beta^{}) \, M_\alpha^{} \, - \, \beta(F_\alpha^{}) \, M_\beta^{}~
  =~N_{\alpha,\beta}^{} \, M_{\alpha+\beta}^{} \\[2mm]
  \mbox{for $\, \alpha,\beta \smin\, \Delta \,$ such that
  $\, \beta \neq \pm \alpha \,,\, \beta \neq \pm \theta\alpha$}~.
 \end{array}
\end{equation}
Now observe that eqn~(\ref{eq:SPCOEF5}) can be used to show that the second
equation in eqn~(\ref{eq:SPAC6}) is equivalent to eqn~(\ref{eq:SPAC11}).
Indeed, antisymmetrizing eqn~(\ref{eq:SPAC6}) with respect to the exchange
of $\alpha$ and $\beta$ eliminates one of the two terms containing structure
constants and leads to eqn~(\ref{eq:SPAC11}), and conversely, substituting
$\alpha$ by $\theta\alpha$ in eqn~(\ref{eq:SPAC11}) and subtracting the
result, we are led back to the second equation in eqn~(\ref{eq:SPAC6}).
On the other hand, using eqn~(\ref{eq:SPCOEF6}) and applying the same
argument, the first equation in eqn~(\ref{eq:SPAC6}) turns out to be
a consequence of eqn~(\ref{eq:SPAC10}) but is apparently weaker.
Similarly, eqn~(\ref{eq:SPEAC1}) is part of eqn~(\ref{eq:LAEAC1}), from
which it can be obtained by projecting from $\mathfrak{h}_{\mathbb{R}}$
onto $\mathfrak{a}_0$ in the third tensor factor, that is, by applying
the operator $\, 1 \otimes 1 \otimes \frac{1}{2} (1 - \theta)$, and
eqn~(\ref{eq:SPEAC2}) is part of eqn~(\ref{eq:LAEAC2}), from which
it can be obtained by projecting from $\mathfrak{h}_{\mathbb{R}}$
onto $\mathfrak{a}_0$ in the first tensor factor, that is, by applying
the operator $\, \frac{1}{2} (1 - \theta) \otimes 1$.

Although the conditions stated in Proposition~2 thus seem to be weaker than
those stated in Proposition~1, it turns out that they are still sufficiently
strong to allow for a complete classification of all possible solutions.
As a by-product, we shall be able to reduce eqn~(\ref{eq:SPAC5}) to the
form given in eqn~(\ref{eq:SPAC7}). The arguments employed to achieve this
are essentially the same as the ones in the previous section. First, we
argue that, as before, the signs $\epsilon_\alpha$ that appear in eqn~%
(\ref{eq:SPAC5}) may without loss of generality be assumed to be independent
of $\alpha$. Next, writing down the system obtained from eqn~(\ref{eq:SPAC11})
upon replacing $\alpha$ by $-\alpha$ and $\beta$ by $-\beta$, adding the
resulting four equations, inserting eqn~(\ref{eq:SPAC5}) and separating the
coefficients of $(H_\alpha)_{\mathfrak{a}}$ and $(H_\beta)_{\mathfrak{a}}$,
we arrive at the same formula as in the previous section, eqn~%
(\ref{eq:LAIAC15}). Once again, it is to be noted that this derivation
is only valid when $\, \beta \neq \pm \alpha \,, \beta \neq \pm \theta\alpha$,
as stated in eqn~(\ref{eq:SPAC11}): this supplementary condition is also needed
to guarantee that $(H_\alpha)_{\mathfrak{a}}$ and $(H_\beta)_{\mathfrak{a}}$
are linearly independent but can in fact be eliminated from eqn~%
(\ref{eq:LAIAC15}) since this formula is automatically satisfied
when $\, \beta = \pm \alpha \,$ or $\, \beta = \pm \theta\alpha$.
(Indeed, for $\, \beta = \pm \alpha \,$ or $\, \beta = \pm \theta\alpha \,$
the rhs is understood to vanish since $2 \>\! \alpha$, $0$ and $\, \alpha
\pm \theta\alpha \,$ do not belong to the root system~$\Delta$, whereas the
lhs vanishes as a consequence of eqn~(\ref{eq:SPAC4}).)
\begin{quote}
 The statement that for $\, \beta \neq \pm \alpha \,$ and $\, \beta
 \neq \pm \theta\alpha$, the generators $(H_\alpha)_{\mathfrak{a}}$
 and $(H_\beta)_{\mathfrak{a}}$ are linearly independent, used in the
 derivation of eqn~(\ref{eq:LAIAC15}) given here, can be proved indirectly,
 as follows. Suppose that for some pair of roots $\, \alpha,\beta \smin\,
 \Delta \,$ satisfying $\, \beta \neq \pm \alpha \,$ and $\, \beta \neq
 \pm \theta\alpha$, these generators were linearly dependent. Since
 $\Delta_0$ is empty so that $(H_\alpha)_{\mathfrak{a}}$ and
 $(H_\beta)_{\mathfrak{a}}$ are both non-zero, this amounts to
 assuming that there exists a non-zero real number $\lambda\,$ such that
 $\, (H_\beta)_{\mathfrak{a}} = \lambda \, (H_\alpha)_{\mathfrak{a}}$, or
 equivalently,
 \begin{equation} \label{eq:SPLD1}
  \beta - \theta\beta~=~\lambda \, (\alpha - \theta\alpha)~.
 \end{equation}
 Obviously, if both roots are real, eqn~(\ref{eq:SPLD1}) reduces to
 $\, \beta = \lambda \alpha$, with $\, \lambda = \pm 1$, a contradiction.
 Similarly, if one of the two roots is complex while the other is real,
 we also get a contradiction since if, for example, $\alpha$ is complex
 and $\beta$ is real, eqn~(\ref{eq:SPLD1}) becomes $\, \beta = \frac{1}{2}
 \lambda (\alpha-\theta\alpha) \,$ which is excluded since $\alpha$
 and~$\theta\alpha$ being strongly orthogonal implies that the only
 linear combinations of $\alpha$ and $\theta\alpha$ which are roots
 are $\pm \alpha$ and $\pm \theta\alpha$. To handle the case where
 both roots are complex and hence $\alpha$ and $\theta\alpha$ as well as
 $\beta$ and $\theta\beta$ are strongly orthogonal, we begin by noting 
 that $\beta$ cannot be orthogonal to both $\alpha$ and~$\theta\alpha$
 since otherwise, $\theta\beta$ would be so as well and hence $\beta
 - \theta\beta$ would be orthogonal to $\alpha - \theta\alpha$, which
 contradicts eqn~(\ref{eq:SPLD1}). Exchanging $\alpha$ with~$\theta\alpha$
 and $\beta$ with $\theta\beta$ if necessary, we may assume without loss
 of generality that $\alpha$ is not orthogonal to $\beta$ and that the
 factor $\lambda$ in eqn~(\ref{eq:SPLD1}) is positive. \linebreak
 With these conventions, taking the scalar product of eqn~(\ref{eq:SPLD1})
 with $\alpha$ and with $\beta$ gives
 \begin{equation} \label{eq:SPLD2}
  \begin{array}{c}
   (\beta \,, \alpha) - (\theta\beta \,, \alpha)~
   =~\lambda (\alpha \,, \alpha)~, \\[1mm]
   (\beta \,, \beta)~
   =~\lambda (\beta \,, \alpha) - \lambda (\beta \,, \theta\alpha)~,
  \end{array}
 \end{equation}
 implying
 \[
  (\beta \,, \beta)~=~\lambda^2 (\alpha \,, \alpha)~.
 \]
 In the root system of an arbitrary simple complex Lie algebra, this
 forces $\lambda^2$ to be $1$, $2$, $3$, $\frac{1}{2}$ or $\frac{1}{3}$.
 But eqn~(\ref{eq:SPLD1}) excludes the possibility of $\lambda^2$ being
 different from $1$ since the root system of any simple complex Lie
 algebra is contained in an appropriate lattice formed by the integer
 linear combinations of vectors $\frac{1}{2} e_i$ where the $e_i$
 are an orthonormal basis of $\mathbb{R}^n$, so an equation of the
 form~(\ref{eq:SPLD1}) with an irrational value of $\lambda$ can only
 hold if both sides vanish, which is impossible since $\Delta_0$ is
 empty. Thus we conclude that $\, \lambda = 1$, so $\alpha$, $\beta$,
 $\theta\alpha$ and $\theta\beta$ all have the same length and
 eqn~(\ref{eq:SPLD1}) becomes
 \begin{equation} \label{eq:SPLD3}
  \beta - \theta\beta~=~\alpha - \theta\alpha~.
 \end{equation}
 This allows us to determine the $\alpha$-string through $\beta$.
 First, $\beta - \alpha$ cannot be a root since if it were, it would
 belong to~$\Delta_0$ which is empty. Second, $\beta + \alpha$ must
 therefore be a root, since $\alpha$ and $\beta$ are not orthogonal.
 Third, $\beta + 2\alpha$ cannot be a root since if it were, we would
 have $\, 2 \, (\beta \,, \alpha) / (\alpha \,, \alpha) \leqslant -2$,
 \linebreak
 implying $\, |\beta + \alpha|^2 \leqslant 0$, which is absurd. Hence
 $\alpha$ and $\beta$ generate a root system of type $A_2$ for which
 they act as simple roots; in particular, $2 \, (\beta \,, \alpha) /
 (\alpha \,, \alpha) = -1$. Inserting this conclusion back into
 eqn~(\ref{eq:SPLD2}), we see that $\, 2 \, (\theta\beta \,, \alpha) /
 (\alpha \,, \alpha) = -3$, which is only possible if the $\alpha$-string
 through $\theta\beta$ consists of four roots, namely $\, \theta\beta$,
 $\theta\beta + \alpha$, $\theta\beta + 2\alpha \,$ and $\, \theta\beta
 + 3\alpha$ (recall that any root string has length at most $4$).
 But this requires the angle between $\theta\beta$ and $\alpha$ to
 be $- 150^{\mathrm{o}}$ and forces $\theta\beta$ and $\alpha$ to
 have different length, contrary to a conclusion reached before.
\end{quote}
In this way, we arrive once again at the conclusion that the simple
complex Lie algebra $\mathfrak{g}$ must belong to the $A$-series.
Moreover, the automorphism $\theta$ that defines the symmetric pair
$(\mathfrak{g},\theta)$ is further restricted by various additional
constraints. The first such condition is that the root generators
$E_\alpha$ in $\mathfrak{g}$ can be chosen so that $\, \theta E_\alpha
= E_{\theta\alpha} \,$ for all $\, \alpha \smin\, \Delta$, which
according to the erratum of Ref.~\cite{FW1} is not only sufficient
but also necessary to guarantee that the proof of integrability given
in Ref.~\cite{FW1} really works: this excludes the symmetric pairs of
the $A\,I\;\!$-\,series $\, SL(n,\mathbb{R}) / SO(n) \,$ for which all
roots are real and $\, \theta E_\alpha = -E_{\theta\alpha} \,$ for all
$\, \alpha \smin\, \Delta$. The second condition is that there should
be no imaginary roots: this excludes the symmetric pairs of the
$A\,II\;\!$-\,series $\, SL(n,\mathbb{H}) / Sp\;\!(n) \,$ as well
as the symmetric pairs of the $A\,III\;\!$-\,series of complex
Grassmannians $\, SU(p,q) / \mbox{$S(U(p) \times U(q))$}$ \linebreak
with $\, |p-q| \geqslant 1$. The third and final condition is that
for all complex roots $\alpha$, $\theta\alpha$ should be strongly
orthogonal to $\alpha$: this excludes the symmetric pairs of the
$A\,III\;\!$-\,series of complex Grassmannians $\, SU(p,q) /
S(U(p) \times U(q)) \,$ with $\, p \neq q$. On the other hand,
it is clear that the symmetric pairs associated with the Grass%
mannians $\, SU(n,n) / S(U(n) \times U(n)) \,$ do provide a non-%
trivial solution: explicitly, we have in the notation employed at
the end of the previous section (with $n$ replaced by $2n$) and
in Sect.~3.2 of Ref.~\cite{FW1}
\begin{equation} \label{eq:SPSOL1}
 \begin{array}{c}
  {\displaystyle
   K_{ab}^+~= \; - \, \frac{1}{4} \, ( E_{aa} + E_{bb} +
                                       E_{\theta(a) \, \theta(a)} +
                                       E_{\theta(b) \, \theta(b)}) \,
                 + \, \frac{1}{2n} \, \mathbf{1}_{2n}^{}~,} \\[4mm]
  {\displaystyle
   M_{ab}^+~= \; - \, \frac{1}{4} \, ( E_{aa} + E_{bb} -
                                       E_{\theta(a) \, \theta(a)} -
                                       E_{\theta(b) \, \theta(b)})~,}
 \end{array}
\end{equation}
and
\begin{equation} \label{eq:SPSOL2}
 \begin{array}{c}
  {\displaystyle
   K_{ab}^-~=~\frac{\epsilon}{4} \, ( E_{aa} - E_{bb} +
                                      E_{\theta(a) \, \theta(a)} -
                                      E_{\theta(b) \, \theta(b)})~,} \\[4mm]
  {\displaystyle
   M_{ab}^-~=~\frac{\epsilon}{4} \, ( E_{aa} - E_{bb} -
                                      E_{\theta(a) \, \theta(a)} +
                                      E_{\theta(b) \, \theta(b)})~,}
 \end{array}
\end{equation}

\pagebreak

\noindent
implying that
\begin{equation} \label{eq:SPSOL3}
 K_{ab}^{}~= \; - \, \frac{1}{2} \left( E_{bb} \, + \,
                                        E_{\theta(b) \, \theta(b)} \right) + \,
                     \frac{1}{2n} \, \mathbf{1}_{2n}^{}~~,~~
 M_{ab}^{}~= \; - \, \frac{1}{2} \left( E_{bb} \, - \,
                                        E_{\theta(b) \, \theta(b)} \right)
\end{equation}
when $\, \epsilon = +1$, while
\begin{equation} \label{eq:SPSOL4}
 K_{ab}^{}~= \; - \, \frac{1}{2} \left( E_{aa} \, + \,
                                        E_{\theta(a) \, \theta(a)} \right) + \,
                     \frac{1}{2n} \, \mathbf{1}_{2n}^{}~~,~~
 M_{ab}^{}~= \; - \, \frac{1}{2} \left( E_{aa} \, - \,
                                        E_{\theta(a) \, \theta(a)} \right)
\end{equation}
when $\, \epsilon = -1$.

\section{Conclusions and Outlook}

Our analysis of the question whether the known dynamical $R$-matrices
for integrable Calogero models can be gauge transformed to numerical
$R$-matrices has revealed that this is possible in some cases but not
in all~-- a conclusion that could definitely not be reached by looking
at the standard model associated with the root system of the $A$-series
alone. In fact, it had been known from previous work that a)~the Calogero
models associated with the root systems of simple complex Lie algebras
$\mathfrak{g}$ are integrable, in the sense of admitting a Lax
representation with a dynamical $R$-matrix, if and only if
$\, \mathfrak{g} = \mathfrak{sl}(n,\mathbb{C})$~\cite{FW1}
and b)~that this dynamical $R$-matrix can be gauge transformed
to a numerical one~\cite{HW,FP}. The results reported in this
paper show that for the Calogero models associated with the root
systems of symmetric pairs $(\mathfrak{g},\theta)$, the situation
is more intricate. First of all, there is still no complete answer
to the question which of these models are integrable, in the sense
of admitting a Lax representation with a dynamical $R$-matrix:
the only case that has been analyzed completely is that of the
$A\,III\;\!$-\,series of complex Grassmannians $\, SU(p,q) /
S(U(p) \times U(q))$, where integrability has been shown to occur
if and only if $|p-q|$ is either $0$ or $1$. We strongly suspect
that this is in fact the only class of symmetric spaces where
integrability prevails, but a rigorous proof of this conjecture
is still missing. What is shown in this paper is that dynamical
$R$-matrices of integrable Calogero models associated with non-%
Grassmannian symmetric pairs~-- should they exist~-- cannot be
gauge transformed to numerical $R$-matrices and, more importantly,
that the dynamical $R$-matrices of the Grassmannian Calogero models
can be gauge transformed to numerical $R$-matrices if $p=q$ but not
if $|p-q|=1$. The first case includes the $C_n$ and $D_n$ models,
whereas the second case includes the $B_n$ and $BC_n$ models.

In summary, our results show that the question which originally
motivated our work on integrability of the Calogero models, namely
the search for an understanding of the mathematical nature and role
of dynamical $R$-matrices, is still far from a definite answer, since
the attempt to reduce them to numerical $R$-matrices via gauge
transformations is only partially successful.

Accepting the fact that the role of dynamical $R$-matrices for our
understanding of integrable systems can apparently not be reduced to
that of numerical $R$-matrices in disguise, there are many questions
that gain new impetus. Continuing to use the Calogero models as a
guideline, we believe that there are several directions in which
future work will be capable of providing new insights into the problem.
One of them is the question of what should be the algebro-differential
constraints to be satisfied by a truly dynamical $R$-matrix, or in
other words, what is the real mathematical status and interpretation
of the dynamical Yang-Baxter equation. A remarkable fact is that,
as will be shown in a separate publication~\cite{FW3}, there is a
natural candidate which is gauge invariant. Another promising
direction for research is a further clarification of the relation
between Calogero models and the geodesic flow on symmetric spaces
subjected to Marsden-Weinstein phase space reduction: this relation
should also shed new light on the role of the recently introduced
spin Calogero models~\cite{LX}.

\end{document}